\documentclass[12pt]{article}
\usepackage{epsf,epsfig,here}
\begin{document}
\newcommand{\Kstarr}{\mathrm{K^{*\pm}}}
\newcommand{\Kstarm}{\mathrm{K^{*-}}}
\newcommand{\Kstarmn}{\mathrm{K^{*-}}}
\newcommand{\Km}{\mathrm{K^{-}}}
\newcommand{\Dsl}{{\mathrm{ D_s}}\ell}
\newcommand{\bt}{\begin{tabular}}
\newcommand{\et}{\end{tabular}}
\newcommand{\Mecqt}{{$\mathrm{MeV/c^2}$}}
\newcommand{\Mecqm}{{\mathrm{MeV/c^2}}}
\newcommand{\Gecqt}{{$\mathrm{GeV/c^2}$}}
\newcommand{\Gecqm}{{\mthrm{GeV/c^2}}}
\newcommand{\DB}{\Delta B}
\newcommand{\as}{\alpha_{ s}}
\newcommand{\ep}{\varepsilon}
\newcommand{\rp}{\tau({\rm B^+})/\tau({\rm B^0_d})}
\newcommand{\rs}{\tau({\rm B^0_s})/\tau({\rm B^0_d})}
\newcommand{\rl}{\tau({\rm \Lambda_b})/\tau({\rm B^0_d})}
\newcommand{\dgs}{\Delta \Gamma_{\Bs}}
\newcommand{\dgbs}{\Delta \Gamma_{\rm \Bs}/\Gamma_{\rm \Bs}}
\newcommand{\tbs}{\tau_{\Bs}}
\newcommand{\tbd}{\tau_{\Bd}}
\newcommand{\Gs}{\Gamma_{{\rm B^0_s}}}
\newcommand{\Gd}{\Gamma_{{\rm B^0_d}}}
\newcommand{\rhobar}{\overline {\rho}}
\newcommand{\etabar}{\overline{\eta}}
\newcommand{\epsilonk}{\varepsilon_K}
\newcommand{\vubovcb}{\left | \frac{V_{ub}}{V_{cb}} \right |}
\newcommand{\vubsvcb}{\left | V_{ub}/V_{cb}  \right |}
\newcommand{\vtdovts}{\left | \frac{V_{td}}{V_{ts}} \right |}
\newcommand{\epsp}{\frac{\varepsilon^{'}}{\varepsilon}}
\newcommand{\dmd}{\Delta m_d}
\newcommand{\dms}{\Delta m_s}
\newcommand{\Do}{{\rm D}^0}
\newcommand{\pimora}{\pi^{\ast}}
\newcommand{\Bstar}{{\rm B}^{\ast}}
\newcommand{\Bstarstar}{{\rm B}^{\ast \ast}}
\newcommand{\Dstar}{{\rm D}^{\ast}}
\newcommand{\Dstars}{{\rm D}^{\ast +}_s}
\newcommand{\Dstaro}{{\rm D}^{\ast 0}}
\newcommand{\Dstarp}{{\rm D}^{\ast +}}
\newcommand{\Dstarstar}{{\rm D}^{\ast \ast}}
\newcommand{\pistar}{\pi^{\ast}}
\newcommand{\pisstar}{\pi^{\ast \ast}}
\newcommand{\bptre}{\rm b^{+}_{3}}
\newcommand{\Vcb}{\left | {\rm V}_{cb} \right |}
\newcommand{\Vub}{\left | {\rm V}_{ub} \right |}
\newcommand{\Vcs}{\left | {\rm V}_{cs} \right |}
\newcommand{\Vtd}{\left | {\rm V}_{td} \right |}
\newcommand{\Vts}{\left | {\rm V}_{ts} \right |}
\newcommand{\bp}{\rm b^{+}_{1}}
\newcommand{\bo}{\rm b^0}
\newcommand{\bos}{\rm b^0_s}
\newcommand{\bss}{\rm b^s_s}
\newcommand{\qq}{\rm q \overline{q}}
\newcommand{\cc}{\rm c \overline{c}}
\newcommand{\BsDmX}{{B_{s}^{0}} \rightarrow D \mu X}
\newcommand{\BsDsm}{{B_{s}^{0}} \rightarrow D_{s} \mu X}
\newcommand{\BsDsX}{{B_{s}^{0}} \rightarrow D_{s} X}
\newcommand{\BDsX}{B \rightarrow D_{s} X}
\newcommand{\BDomX}{B \rightarrow D^{0} \mu X}
\newcommand{\BDpmX}{B \rightarrow D^{+} \mu X}
\newcommand{\Dsfmn}{D_{s} \rightarrow \phi \mu \nu}
\newcommand{\Dsfipi}{D_{s} \rightarrow \phi \pi}
\newcommand{\DsfX}{D_{s} \rightarrow \phi X}
\newcommand{\DpfX}{D^{+} \rightarrow \phi X}
\newcommand{\DofX}{D^{0} \rightarrow \phi X}
\newcommand{\DfX}{D \rightarrow \phi X}
\newcommand{\DsD}{B \rightarrow D_{s} D}
\newcommand{\DsmX}{D_{s} \rightarrow \mu X}
\newcommand{\DmX}{D \rightarrow \mu X}
\newcommand{\Zbb}{Z^{0} \rightarrow \rm b \overline{b}}
\newcommand{\Zcc}{Z^{0} \rightarrow \rm c \overline{c}}
\newcommand{\Rbb}{\frac{\Gamma_{Z^0 \rightarrow \rm b \overline{b}}}
{\Gamma_{Z^0 \rightarrow Hadrons}}}
\newcommand{\Rcc}{\frac{\Gamma_{Z^0 \rightarrow \rm c \overline{c}}}
{\Gamma_{Z^0 \rightarrow Hadrons}}}
\newcommand{\bb}{b \overline{b}}
\newcommand{\str}{\rm s \overline{s}}
\newcommand{\Bs}{\rm{B^0_s}}
\newcommand{\Bsb}{\overline{\rm{B}^0_s}}
\newcommand{\Bp}{\rm{B^{+}}}
\newcommand{\Bm}{\rm{B^{-}}}
\newcommand{\Bo}{\rm{B^{0}}}
\newcommand{\Bob}{\overline{\rm{B}^{0}}}
\newcommand{\Bd}{\rm{B^{0}_{d}}}
\newcommand{\Bdb}{\overline{\rm{B^{0}_{d}}}}
\newcommand{\Lb}{\Lambda^0_b}
\newcommand{\Lbb}{\overline{\Lambda^0_b}}
\newcommand{\Kstar}{\rm{K^{\star 0}}}
\newcommand{\phim}{\rm{\phi}}
\newcommand{\Ds}{\rm{D}_s}
\newcommand{\Dsp}{\rm{D}_s^+}
\newcommand{\Dsm}{\rm{D}_s^-}
\newcommand{\Dp}{\rm{D}^+}
\newcommand{\Dn}{\rm{D}^0}
\newcommand{\Dsb}{\overline{\rm{D}_s}}
\newcommand{\Dm}{\rm{D}^-}
\newcommand{\Dnb}{\overline{\rm{D}^0}}
\newcommand{\Lc}{\Lambda_c}
\newcommand{\Lcb}{\overline{\Lambda_c}}
\newcommand{\Dstarm}{\rm{D}^{\ast -}}
\newcommand{\Dsstarp}{\rm{D}_s^{\ast +}}
\newcommand{\Pb}{P_{b-baryon}}
\newcommand{\KKpi}{\rm{ K K \pi }}
\newcommand{\GeV}{\rm{GeV}}
\newcommand{\MeV}{\rm{MeV}}
\newcommand{\nb}{\rm{nb}}
\newcommand{\Zzero}{{\rm Z}^0}
\newcommand{\MZ}{\rm{M_Z}}
\newcommand{\MW}{\rm{M_W}}
\newcommand{\GF}{\rm{G_F}}
\newcommand{\Gm}{\rm{G_{\mu}}}
\newcommand{\MH}{\rm{M_H}}
\newcommand{\MT}{\rm{m_{top}}}
\newcommand{\GZ}{\Gamma_{\rm Z}}
\newcommand{\Afb}{\rm{A_{FB}}}
\newcommand{\Afbs}{\rm{A_{FB}^{s}}}
\newcommand{\sigmaf}{\sigma_{\rm{F}}}
\newcommand{\sigmab}{\sigma_{\rm{B}}}
\newcommand{\NF}{\rm{N_{F}}}
\newcommand{\NB}{\rm{N_{B}}}
\newcommand{\Nnu}{\rm{N_{\nu}}}
\newcommand{\RZ}{\rm{R_Z}}
\newcommand{\rhob}{\rho_{eff}}
\newcommand{\Gammanz}{\rm{\Gamma_{Z}^{new}}}
\newcommand{\Gammani}{\rm{\Gamma_{inv}^{new}}}
\newcommand{\Gammasz}{\rm{\Gamma_{Z}^{SM}}}
\newcommand{\Gammasi}{\rm{\Gamma_{inv}^{SM}}}
\newcommand{\Gammaxz}{\rm{\Gamma_{Z}^{exp}}}
\newcommand{\Gammaxi}{\rm{\Gamma_{inv}^{exp}}}
\newcommand{\rhoZ}{\rho_{\rm Z}}
\newcommand{\thw}{\theta_{\rm W}}
\newcommand{\swsq}{\sin^2\!\thw}
\newcommand{\swsqmsb}{\sin^2\!\theta_{\rm W}^{\overline{\rm MS}}}
\newcommand{\swsqbar}{\sin^2\!\overline{\theta}_{\rm W}}
\newcommand{\cwsqbar}{\cos^2\!\overline{\theta}_{\rm W}}
\newcommand{\swsqb}{\sin^2\!\theta^{eff}_{\rm W}}
\newcommand{\ee}{{e^+e^-}}
\newcommand{\eeX}{{e^+e^-X}}
\newcommand{\gaga}{{\gamma\gamma}}
\newcommand{\mumu}{\ifmmode {\mu^+\mu^-} \else ${\mu^+\mu^-} $ \fi}
\newcommand{\eeg}{{e^+e^-\gamma}}
\newcommand{\mumug}{{\mu^+\mu^-\gamma}}
\newcommand{\tautau}{{\tau^+\tau^-}}
\newcommand{\qqb}{{q\overline{q}}}
\newcommand{\eegg}{e^+e^-\rightarrow \gamma\gamma}
\newcommand{\eeggg}{e^+e^-\rightarrow \gamma\gamma\gamma}
\newcommand{\eeee}{e^+e^-\rightarrow e^+e^-}
\newcommand{\eeeeee}{e^+e^-\rightarrow e^+e^-e^+e^-}
\newcommand{\eeeeg}{e^+e^-\rightarrow e^+e^-(\gamma)}
\newcommand{\eeeegg}{e^+e^-\rightarrow e^+e^-\gamma\gamma}
\newcommand{\eeeg}{e^+e^-\rightarrow (e^+)e^-\gamma}
\newcommand{\eemumu}{e^+e^-\rightarrow \mu^+\mu^-}
\newcommand{\eetautau}{e^+e^-\rightarrow \tau^+\tau^-}
\newcommand{\eehad}{e^+e^-\rightarrow {\rm hadrons}}
\newcommand{\eettg}{e^+e^-\rightarrow \tau^+\tau^-\gamma}
\newcommand{\eell}{e^+e^-\rightarrow l^+l^-}
\newcommand{\Ztopig}{{\rm Z}^0\rightarrow \pi^0\gamma}
\newcommand{\Ztogg}{{\rm Z}^0\rightarrow \gamma\gamma}
\newcommand{\Ztoee}{{\rm Z}^0\rightarrow e^+e^-}
\newcommand{\Ztoggg}{{\rm Z}^0\rightarrow \gamma\gamma\gamma}
\newcommand{\Ztomumu}{{\rm Z}^0\rightarrow \mu^+\mu^-}
\newcommand{\Ztotautau}{{\rm Z}^0\rightarrow \tau^+\tau^-}
\newcommand{\Ztoll}{{\rm Z}^0\rightarrow l^+l^-}
\newcommand{\Ztocc}{{\rm Z^0\rightarrow c \overline c}}
\newcommand{\Lamp}{\Lambda_{+}}
\newcommand{\Lamm}{\Lambda_{-}}
\newcommand{\Pt}{\rm P_{t}}
\newcommand{\Gee}{\Gamma_{ee}}
\newcommand{\Gpig}{\Gamma_{\pi^0\gamma}}
\newcommand{\Ggg}{\Gamma_{\gamma\gamma}}
\newcommand{\Gggg}{\Gamma_{\gamma\gamma\gamma}}
\newcommand{\Gmumu}{\Gamma_{\mu\mu}}
\newcommand{\Gtautau}{\Gamma_{\tau\tau}}
\newcommand{\Ginv}{\Gamma_{\rm inv}}
\newcommand{\Ghad}{\Gamma_{\rm had}}
\newcommand{\Gnu}{\Gamma_{\nu}}
\newcommand{\GnuSM}{\Gamma_{\nu}^{\rm SM}}
\newcommand{\Gll}{\Gamma_{l^+l^-}}
\newcommand{\Gff}{\Gamma_{f\overline{f}}}
\newcommand{\Gtot}{\Gamma_{\rm tot}}
\newcommand{\Rb}{\mbox{R}_b}
\newcommand{\Rc}{\mbox{R}_c}
\newcommand{\al}{a_l}
\newcommand{\vl}{v_l}
\newcommand{\af}{a_f}
\newcommand{\vf}{v_f}
\newcommand{\ael}{a_e}
\newcommand{\ve}{v_e}
\newcommand{\amu}{a_\mu}
\newcommand{\vmu}{v_\mu}
\newcommand{\atau}{a_\tau}
\newcommand{\vtau}{v_\tau}
\newcommand{\ahatl}{\hat{a}_l}
\newcommand{\vhatl}{\hat{v}_l}
\newcommand{\ahate}{\hat{a}_e}
\newcommand{\vhate}{\hat{v}_e}
\newcommand{\ahatmu}{\hat{a}_\mu}
\newcommand{\vhatmu}{\hat{v}_\mu}
\newcommand{\ahattau}{\hat{a}_\tau}
\newcommand{\vhattau}{\hat{v}_\tau}
\newcommand{\vtildel}{\tilde{\rm v}_l}
\newcommand{\avsq}{\ahatl^2\vhatl^2}
\newcommand{\Ahatl}{\hat{A}_l}
\newcommand{\Vhatl}{\hat{V}_l}
\newcommand{\Afer}{A_f}
\newcommand{\Ael}{A_e}
\newcommand{\Aferb}{\overline{A_f}}
\newcommand{\Aelb}{\overline{A_e}}
\newcommand{\AVsq}{\Ahatl^2\Vhatl^2}
\newcommand{\Iwk}{I_{3l}}
\newcommand{\Qch}{|Q_{l}|}
\newcommand{\roots}{\sqrt{s}}
\newcommand{\pT}{p_{\rm T}}
\newcommand{\mt}{m_t}
\newcommand{\Rechi}{{\rm Re} \left\{ \chi (s) \right\}}
\newcommand{\up}{^}
\newcommand{\abscosthe}{|cos\theta|}
\newcommand{\sint}{\mbox{$\sin\theta$}}
\newcommand{\cost}{\mbox{$\cos\theta$}}
\newcommand{\mcost}{|\cos\theta|}
\newcommand{\epair}{\mbox{$e^{+}e^{-}$}}
\newcommand{\mupair}{\mbox{$\mu^{+}\mu^{-}$}}
\newcommand{\taupair}{\mbox{$\tau^{+}\tau^{-}$}}
\newcommand{\gamgam}{\mbox{$e^{+}e^{-}\rightarrow e^{+}e^{-}\mu^{+}\mu^{-}$}}
\newcommand{\fullskip}{\vskip 16cm}
\newcommand{\halfskip}{\vskip  8cm}
\newcommand{\quarskip}{\vskip  6cm}
\newcommand{\abitskip}{\vskip 0.5cm}
\newcommand{\ba}{\begin{array}}
\newcommand{\ea}{\end{array}}
\newcommand{\bc}{\begin{center}}
\newcommand{\ec}{\end{center}}
\newcommand{\be}{\begin{eqnarray}}
\newcommand{\eeq}{\end{eqnarray}}
\newcommand{\bes}{\begin{eqnarray*}}
\newcommand{\ees}{\end{eqnarray*}}
\newcommand{\Kz}{\ifmmode {\rm K^0_s} \else ${\rm K^0_s} $ \fi}
\newcommand{\Zz}{\ifmmode {\rm Z^0} \else ${\rm Z^0 } $ \fi}
\newcommand{\qqbar}{\ifmmode {\rm q\overline{q}} \else ${\rm q\overline{q}} $ \fi}
\newcommand{\ccbar}{\ifmmode {\rm c\overline{c}} \else ${\rm c\overline{c}} $ \fi}
\newcommand{\bbbar}{\ifmmode {\rm b\overline{b}} \else ${\rm b\overline{b}} $ \fi}
\newcommand{\xxbar}{\ifmmode {\rm x\overline{x}} \else ${\rm x\overline{x}} $ \fi}
\newcommand{\rphi}{\ifmmode {\rm R\phi} \else ${\rm R\phi} $ \fi}
\renewcommand\topfraction{1.}
\newcommand{\BK}{B_K}
\newcommand{\nubar}{\overline{\nu_{\ell}}}
\newcommand{\snb}{\sin{2\beta}}
\newcommand{\sna}{\sin{2\alpha}}
\newcommand{\vcb}{\left | {\rm V}_{cb} \right |}
\newcommand{\vub}{\left | {\rm V}_{ub} \right |}
\newcommand{\vus}{\left | V_{us} \right |}
\newcommand{\vud}{\left | V_{ud} \right |}
\newcommand{\vtd}{\left | {V}_{td} \right |}
\newcommand{\vts}{\left | { V}_{ts} \right |}
\newcommand{\fbdsqbd}{f_{B_d} \sqrt{\hat B_{B_d}}}
\newcommand{\fbssqbs}{f_{B_s} \sqrt{\hat B_{B_s}}}
\newcommand{\snbg}{\sin(2\beta + \gamma)}

\renewcommand{\arraystretch}{1.2}
\pagestyle{empty}
\pagenumbering{arabic}
\vskip  1.5 cm
\begin{center}
{\LARGE {\bf  CKM-TRIANGLE ANALYSIS:}}
\vskip .4 cm 
{ \Large {\bf  Updates and Novelties for Summer 2004}}
\vskip .1 cm 
\end{center}

\begin{figure}[htb!]
\begin{center}
\epsfig{figure=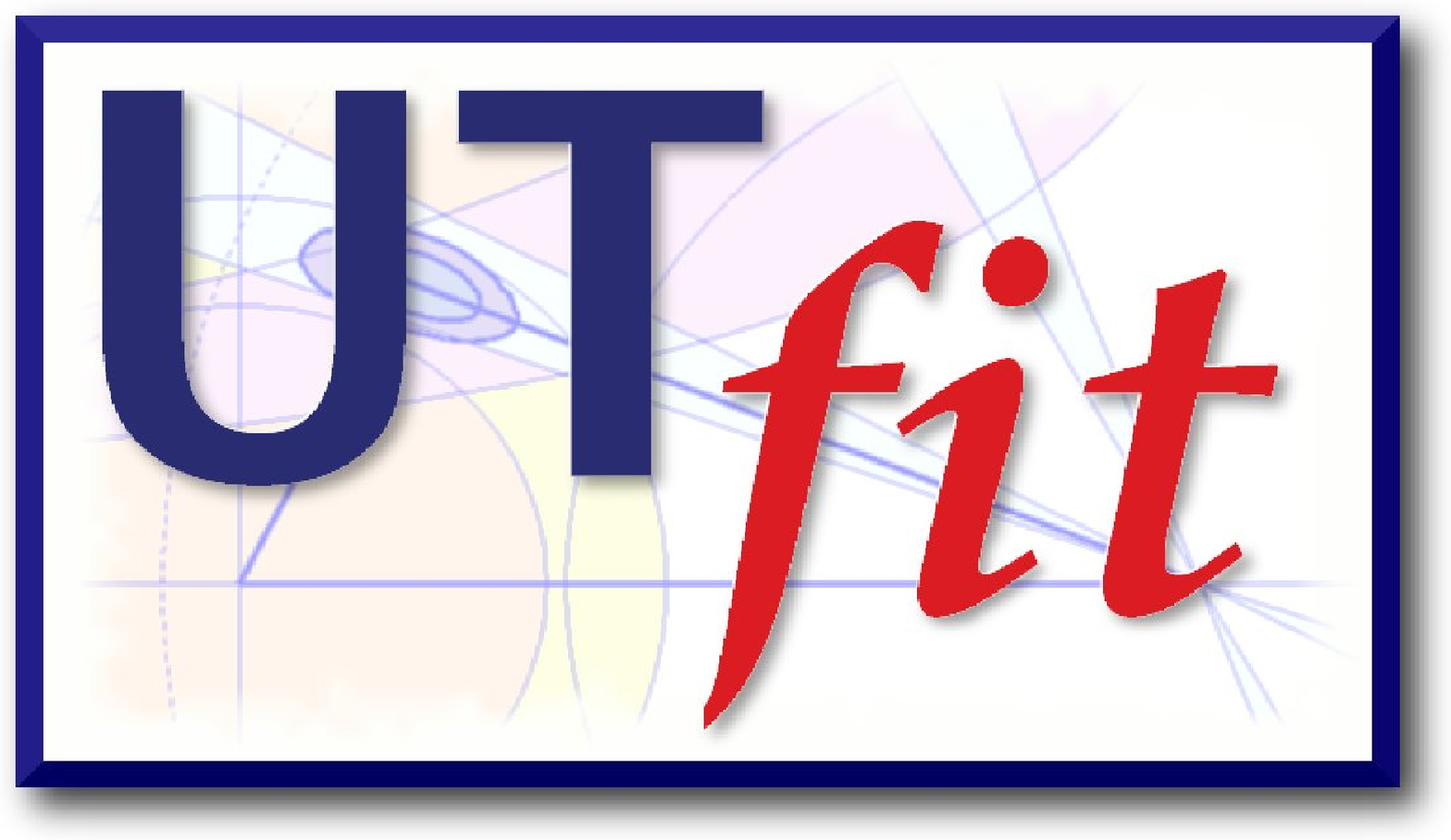,width=2.5cm}
\end{center}
\end{figure}

\vspace*{-0.5cm}
\begin{center}
{\Large {\bf UT}}{\large {\it fit}}{\large ~ Collaboration :} \\ 
\end{center}
\begin{center}
{\bf \large M.~Bona$^{(a)}$, M.~Ciuchini$^{(b)}$, E.~Franco$^{(c)}$,
V.~Lubicz$^{(b)}$, } \\
{\bf \large  G. Martinelli$^{(c)}$, F. Parodi$^{(d)}$, M. Pierini$^{(c)}$, P.
Roudeau$^{(e)}$, }\\
{\bf \large C. Schiavi$^{(d)}$, L.~Silvestrini$^{(c)}$ and A. Stocchi$^{(e)}$}
\end{center}

\vspace*{0.3cm}
\begin{center}
\noindent
{\small
\noindent
{\bf $^{(a)}$   INFN,  Sezione di Torino,}\\  
\hspace*{0.5cm}{Via P. Giuria 1, I-10125  Torino, Italy}\\
{\bf $^{(b)}$   Universit{\`a} di Roma Tre
 and INFN,  Sezione di Roma III,}\\  
\hspace*{0.5cm}{Via della Vasca Navale 84, I-00146 Roma, Italy}\\
\noindent
{\bf $^{(c)}$ Universit\`a di Roma ``La Sapienza'' and INFN, Sezione di Roma,}\\
\hspace*{0.5cm}{Piazzale A. Moro 2, 00185 Roma, Italy}\\
\noindent
{\bf $^{(d)}$ Dipartimento di Fisica, Universit\`a di Genova and INFN}\\
\hspace*{0.5cm}{Via Dodecaneso 33, 16146 Genova, Italy}\\
\noindent
{\bf $^{(e)}$ Laboratoire de l'Acc\'el\'erateur Lin\'eaire}\\
\hspace*{0.5cm}{IN2P3-CNRS et Universit\'e de Paris-Sud, BP 34, 
F-91898 Orsay Cedex}}\\
\noindent
\end{center}

\vspace*{0.5cm}
\begin{abstract}
Using the most recent determinations of several theoretical and experimental 
parameters, we update the Unitarity Triangle analysis in the Standard Model.
The basic experimental constraints come from the measurements of $\epsilonk$,
$\vubsvcb$, and $\dmd$, the limit on $\dms$, and the measurement of the CP
asymmetry in the $B$ sector through the $J/\psi K^0$ channel.  
In addition we also include in our analysis the direct determination of 
$\sin{2\alpha}$, $\gamma$, and $\sin(2\beta + \gamma)$ from the measurements
of new CP-violating quantities, recently coming from the B-Factories.
We also discuss the opportunities offered by improving the precision 
of the measurements of the various physical quantities entering in the 
determination of the Unitarity Triangle parameters. \\
The results and the plots presented in this paper can be also found at the
URL {\tt http://www.utfit.org}, where they are continuously updated with the
newest 
experimental and theoretical results~\cite{ref:pageweb}.
\end{abstract}

\vspace*{.5cm}
\vspace{\fill}
\begin{center}
  Submitted to the 32$^{\rm nd}$ International Conference on High-Energy Physics, ICHEP 04,\\
  16 August---22 August 2004, Beijing, China
\end{center}
\vspace{\fill}

\pagestyle{plain}

\newpage

\section{Introduction}
\label{sec:intro}

The analysis of the Unitarity Triangle (UT) and CP violation
represents one of the most stringent tests of the Standard Model (SM)
and, for this reason, it is also an interesting window on New Physics
(NP).  The most precise determination of the parameters governing this
phenomenon is obtained using B decays, $\Bo-\Bob$ oscillations and CP
asymmetries in the kaon and in the B sectors.

Up to now, the standard analysis~\cite{ref:noi,ref:loro} relies on the
following measurements: $\left | V_{ub} \right |/\left | V_{cb} \right
|$, $\Delta {m_d}$, the limit on $\Delta {m_s}$, and the measurements
of CP-violating quantities in the kaon ($\epsilonk$) and in the B
($\snb$) sectors.  Inputs to this analysis constitute a large body of
both experimental measurements and theoretically determined
parameters, where Lattice QCD calculations play a central role.  A
careful choice (and a continuous update) of the values of these
parameters is a prerequisite in this study.  The values and errors
attributed to these parameters are summarized in
Table~\ref{tab:inputs} (Section \ref{sec:inputs}). \\
The results of the analysis and the determination of the UT parameters
are presented and discussed in Section~\ref{sec:results} which is an
update of similar analyses
performed in~\cite{ref:noi} to which the readers can refer for more details. \\
New CP-violating quantities have been recently measured by the
B-Factories, allowing for the determination of several combinations of
UT angles. The measurements of sin2$\alpha$, $\gamma$, and
$\sin(2\beta + \gamma)$ are now available using B decays into $\pi
\pi$ and $\rho \rho$, D$^{(*)}$K and D$^{(*)}\pi$ final states,
respectively.  These measurements are presented in
Section~\ref{sec:newinputs} and their effect on
the UT fit is discussed in Section~\ref{sec:newresults}. \\
Finally we also discuss the perspectives opened by improving the
precision in the measurements of various physical quantities entering
the UT analysis. In particular, we investigate to which extent future
and improved determinations of the experimental constraints, such as
sin2$\beta$, $\dms$ and $\gamma$, could allow us to invalidate the SM,
thus signaling the presence of NP effects.

\section{Inputs used for the ``Standard'' analysis}
\label{sec:inputs}

The values and errors of the relevant quantities entering the standard
analysis of the CKM parameters (corresponding to the constraints from
$\left | V_{ub} \right |/\left | V_{cb} \right |$, $\Delta {m_d}$,
$\Delta {m_s}/\Delta {m_d}$, $\epsilonk$ and $\snb$) are summarized in
Table~\ref{tab:inputs}. \\
The novelties here are the final LEP/SLD likelihood from $\Delta m_s$,
the value of $|V_{ub}|$ from inclusive semileptonic decays
\cite{ref:hfag}, the new value of $\snb$ and a new treatment of the
non-perturbative QCD parameters as explained in the following section
\ref{sec:usefbs}.
 
\begin{table*}[htbp!]
{\footnotesize
\begin{center}
\begin{tabular}{@{}llll}
\hline\hline
\\
         Parameter                          &  Value                            
     & Gaussian ($\sigma$)      &   Uniform             \\
                                            &                                   
     &                          & (half-width)          \\ \hline\hline
         $\lambda$                          &  0.2265                           
     &  0.0020                  &    -                  \\ \hline
$\left |V_{cb} \right |$(excl.)             & $ 42.1 \times 10^{-3}$            
     & $2.1 \times 10^{-3}$     & -                     \\
$\left |V_{cb} \right |$(incl.)             & $ 41.4 \times 10^{-3}$            
     & $0.7 \times 10^{-3}$     & $0.6 \times 10^{-3}$  \\ 
$\left |V_{ub} \right |$(excl.)             & $ 33.0  \times 10^{-4}$           
     & $2.4 \times 10^{-4}$     & $4.6 \times 10^{-4}$  \\  
$\left |V_{ub} \right |$(incl.-LEP)         & $ 40.9  \times 10^{-4}$           
     & $6.2 \times 10^{-4}$     & $4.7 \times 10^{-4}$  \\
$\left |V_{ub} \right |$(incl.-HFAG)        & $ 45.7  \times 10^{-4}$           
     & $6.1 \times 10^{-4}$     &        -              \\ \hline
$\Delta m_d$                                & $0.503~\mbox{ps}^{-1}$            
     & $0.006~\mbox{ps}^{-1}$   &        -              \\
$\Delta m_s$                                & $>$ 14.5 ps$^{-1}$ at 95\% C.L.   
     & \multicolumn{2}{c}{sensitivity 18.3 ps$^{-1}$}   \\ \hline 
$m_t$                                       & $167$ GeV                        
     & $ 5$ GeV                 &          -            \\
$\fbssqbs$                                  & $276$ MeV                        
     & $38$ MeV                 &          -            \\
$\xi=\frac{\fbssqbs}{\fbdsqbd}$             & 1.24                              
     & 0.04                     & $\pm$ 0.06            \\
$\eta_b$                                    & 0.55                              
     & 0.01                     &          -            \\ \hline
$\hat B_K$                                  & 0.86                              
     & 0.06                     &     0.14              \\
$\epsilonk$                                 & $2.280 \times 10^{-3}$            
     & $0.013 \times 10^{-3}$   &          -            \\
$\eta_1$                                    & 1.38                              
     & 0.53                     &          -            \\
$\eta_2$                                    & 0.574                             
     & 0.004                    &          -            \\
$\eta_3$                                    & 0.47                              
     & 0.04                     &          -            \\
$f_K$                                       & 0.159 GeV                         
     & \multicolumn{2}{c}{fixed}                        \\
$\Delta m_K$                                & 0.5301 $\times 10^{-2}
~\mbox{ps}^{-1}$ & \multicolumn{2}{c}{fixed}                        \\ \hline
$\snb$                                      &  0.739                            
     &  0.048                   &          -            \\ \hline
$m_b$                                       & 4.21 GeV                          
     & 0.08 GeV                 &          -            \\
$m_c$                                       & 1.3 GeV                           
     & 0.1 GeV                  &          -            \\
$\alpha_s$                                  & 0.119                             
     & 0.03                     &          -            \\
$G_F $                                      & 1.16639 $\times 10^{-5} \GeV^{-2}$
     & \multicolumn{2}{c}{fixed}                        \\
$ m_{W}$                                    & 80.22 GeV                         
     & \multicolumn{2}{c}{fixed}                        \\
$ m_{B^0_d}$                                & 5.279 GeV                         
     & \multicolumn{2}{c}{fixed}                        \\
$ m_{B^0_s}$                                & 5.375 GeV                         
     & \multicolumn{2}{c}{fixed}                        \\
$ m_K$                                      & 0.493677 GeV                      
     & \multicolumn{2}{c}{fixed}                        \\ \hline\hline
\end{tabular} 
\end{center}
}
\caption {\it {Values of the relevant quantities used in the fit of the CKM
parameters.
In the third and fourth columns the Gaussian and the flat contributions to the
uncertainty are given
respectively (for details on the statistical treatment see~\cite{ref:noi}). 
The central values and errors are those adopted at the end of the ``CKM
Unitarity Triangle'' Workshops (\cite{ref:ckm1},~\cite{ref:ckm2}) and by
the HFAG~\cite{ref:hfag}.}}
\label{tab:inputs} 
\end{table*}

\subsection{Use of $\xi$, $f_{B_s}\sqrt{\hat B_{B_s}}$ and $f_{B_d}\sqrt{ \hat
B_{B_d}}$ in 
$\Delta m_s$ and $\Delta m_d$ constraints}
\label{sec:usefbs}

One of the important differences with respect to previous studies is
in the use of the information from non-perturbative QCD parameters
entering the expressions of $\Delta m_s$ and $\Delta m_d$.  The ${\rm
  B}^0_s-\overline{\rm {B}^0_s}$ time oscillation frequency, which can
be related to the mass difference between the light and heavy mass
eigenstates of the ${\rm B}^0_s-\overline{{\rm B}^0_s}$ system, is
proportional to the square of the $|V_{ts}|$ element.  Up to Cabibbo
suppressed corrections, $|V_{ts}|$ is independent of $\rhobar$ and
$\etabar$. As a consequence, the measurement of $\Delta m_s$ would
provide a strong constraint
on the non-perturbative QCD parameter $f_{B_s}^2 \hat B_{B_s}$.  \\
For this reason we propose a new and more appropriate way of treating
the constraints coming from the measurements of $\Delta m_s$ and
$\Delta m_d$. In previous analyses, these constraints were implemented
using the following equations:
\begin{eqnarray}
 \Delta m_d &\propto& [(1-\rhobar)^2+\etabar^2]  f_{B_d}^2 \hat B_{B_d}         
          \\ \nonumber
 \Delta m_s &\propto& f_{B_s}^2 \hat B_{B_s}  =  f_{B_d}^2 \hat B_{B_d} \times
\xi^2 
\end{eqnarray}
where $\xi=f_{B_s}\sqrt{\hat B_{B_s}}/f_{B_d}\sqrt{ \hat B_{B_d}}$. In
this case the input quantities are $f_{B_d}\sqrt{ \hat B_{B_d}}$ and
$\xi$.  The constraints from $\Delta m_s$ and the knowledge of $\xi$
are used to improve the knowledge on $f_{B_d} \sqrt{\hat B_{B_d}}$
which thus makes the constraint on $\Delta m_d$ more effective.  The
main problem of this method is that the quantity that we know best
from Lattice calculations is $f_{B_s}^2 \hat B_{B_s}$, whereas $\xi^2$
and $f_{B_d}^2 \hat B_{B_d}$ are affected by large uncertainties
coming from chiral extrapolations.  We thus suggest to use a different
method which consists in writing the constraints in the following way:
\begin{eqnarray}
 \Delta m_d &\propto& [(1-\rhobar)^2+\etabar^2] \frac{f_{B_s}^2 \hat
B_{B_s}}{\xi^2} \\ \nonumber
 \Delta m_s &\propto&  f_{B_s}^2 \hat B_{B_s}  
\end{eqnarray}
At present, this new parameterization does not have a large effect on
final results but, in the future, the measurement of $\Delta m_s$ will
allow the elimination of a further theoretical parameter, $f_{B_s}^2
\hat B_{B_s}$, from the UT fits. To obtain a more effective constraint
on $\Delta m_d$, also the error on $\xi$ should be improved.

\section{Determination of the Unitarity Triangle parameters}
\label{sec:results}

In this section, assuming the validity of the Standard Model, we give
the results for the quantities defining the Unitarity Triangle:
$\rhobar$, $\etabar$, $\snb$, $\sna$, $\gamma$, sin$(2\beta +\gamma)$
as well as other quantities such as $\dms$, $\fbssqbs$, $\hat{B}_K$
and $\xi$.  The inputs are summarized in Table~\ref{tab:inputs}.

\subsection{Fundamental test of the Standard Model in the fermion sector}
\label{sec:lati}

\begin{figure}[htb!]
\begin{center}
\includegraphics[width=16cm]{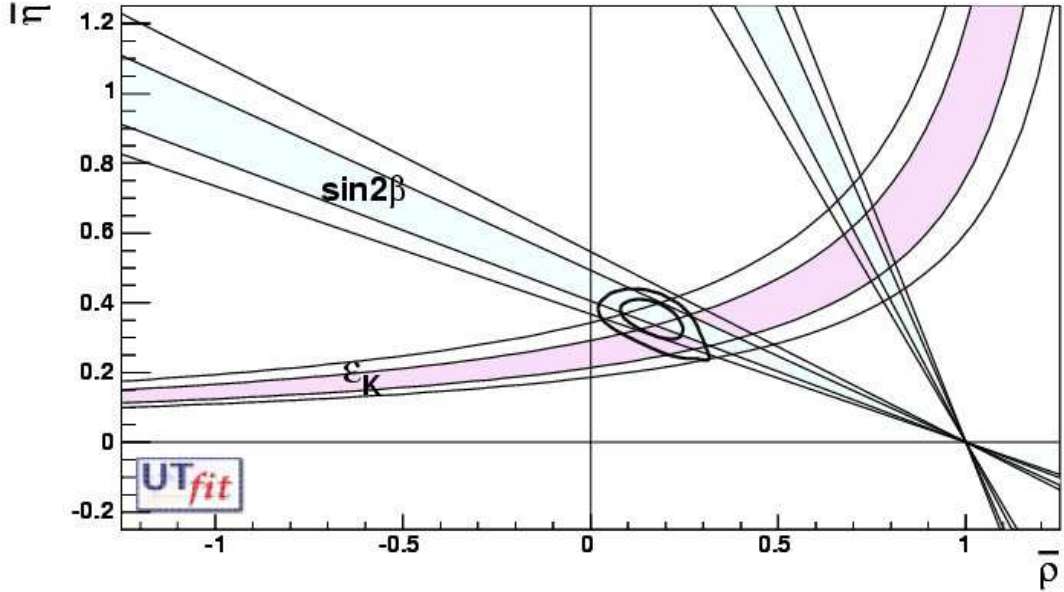}
\caption{\it {The allowed regions for $\overline{\rho}$ and $\overline{\eta}$
    (contours at 68\%, 95\% probability ranges), as selected by the
    measurements of $\left | V_{ub} \right |/\left | V_{cb} \right |$,
    $\Delta {m_d}$, and by the limit on $\Delta {m_s}/\Delta {m_d}$,
    are compared with the bands (at 68\% and 95\% probability ranges)
    from the measurements of CP-violating quantities in the kaon
    ($\epsilonk$) and in the B ($\snb$) sectors.}}
\label{fig:testcp}
\end{center}
\end{figure}

The most crucial test consists in the comparison between the
($\rhobar-\etabar$) region selected by the measurements which are
sensitive only to the sides of the Unitarity Triangle (semileptonic B
decays and ${\rm B}^0-\overline{{\rm B}^0}$ oscillations) and the
regions selected by the direct measurements of CP violation in the
kaon ($\epsilonk$) or in the B ($\snb$) sectors.  This test is shown
in Figure~\ref{fig:testcp}.  It can be translated quantitatively
through the comparison between the value of $\snb$ obtained from the
measurement of the CP asymmetry in $J/\psi K^0$ decays and the one
determined from ``sides" measurements:
\begin{eqnarray}
\snb = & 0.724 \pm 0.049 ~[0.613;0.803]~{\rm at} ~95\%~ C.L.  & ~~\rm {sides~
only}     \nonumber \\
\snb = & 0.739 \pm 0.048 ~[0.681;0.787]~{\rm at} ~95\%~ C.L.  & ~~\rm J/\psi
K^0. 
\label{eq:sin2beta}
\end{eqnarray}

The spectacular agreement between these values illustrates the
consistency of the Standard Model in describing CP violation phenomena
in terms of one single parameter $\etabar$.  It is also an important
test of the Operator Product Expansion (OPE), the Heavy Quark
Effective Theory (HQET) and Lattice QCD (LQCD) which have been used to
extract the CKM parameters.  It has to be noted that this test is even
more significant because the errors on $\snb$ from the two
determinations are comparable\footnote{In the following, for
  simplicity, we will denote as ``direct'' (``indirect'') the
  determination of any given quantity from a direct measurement (from
  the UT fit without using the measurement under consideration).}.

\begin{figure}[htb!]
\begin{center}
{\includegraphics[height=15cm]{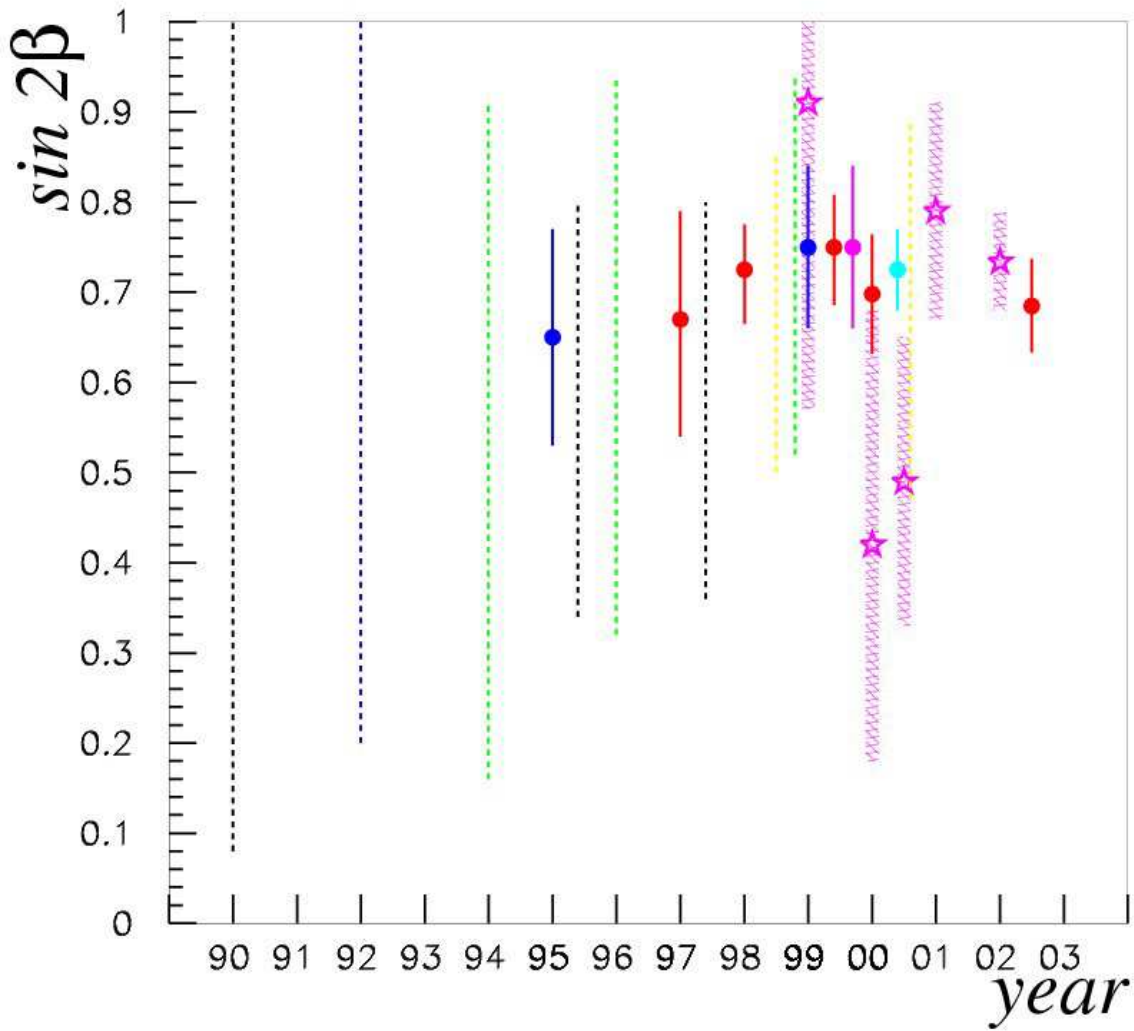}}
\caption{\it {Evolution of the ``indirect'' determination of $\snb$ over the
    years (until 2003).  From left to right, they correspond to the
    following papers~\cite{ref:loro,ref:allrhoeta}: DDGN90, LMMR92, AL94,
    CFMRS95, BBL95, AL96, PPRS97, BF97, BPS98, PS98, AL99, CFGLM99,
    CPRS99, M99, CDFLMPRS00, B.et.al.00, HLLL00 and
    CFLPSS~\cite{ref:noi}. The dotted lines correspond to the 95$\%$
    C.L. regions (the only information given in those papers). The
    larger bands (from year '99) correspond to values of $\snb$ from
    direct measurements ($\pm 1 \sigma$).}}
\label{fig:storiasin2beta}
\end{center}
\end{figure}

As a matter of fact, the value of $\snb$ was predicted, before its
first direct measurement was obtained, by using all other available
constraints, ($\left | V_{ub} \right |/\left | V_{cb} \right |$,
$\epsilonk$, $\Delta m_d$ and $\Delta m_s$). The ``indirect''
determination has improved regularly over the years.
Figure~\ref{fig:storiasin2beta} shows this evolution for the
``indirect'' determination of sin2$\beta$ which is compared with the
recent determinations of $\snb$ from direct measurements.

\subsection{Determination of the Unitarity Triangle parameters}
\label{sec:fit}

\begin{figure}[htbp!]
\begin{center}
{\includegraphics[height=7.cm]{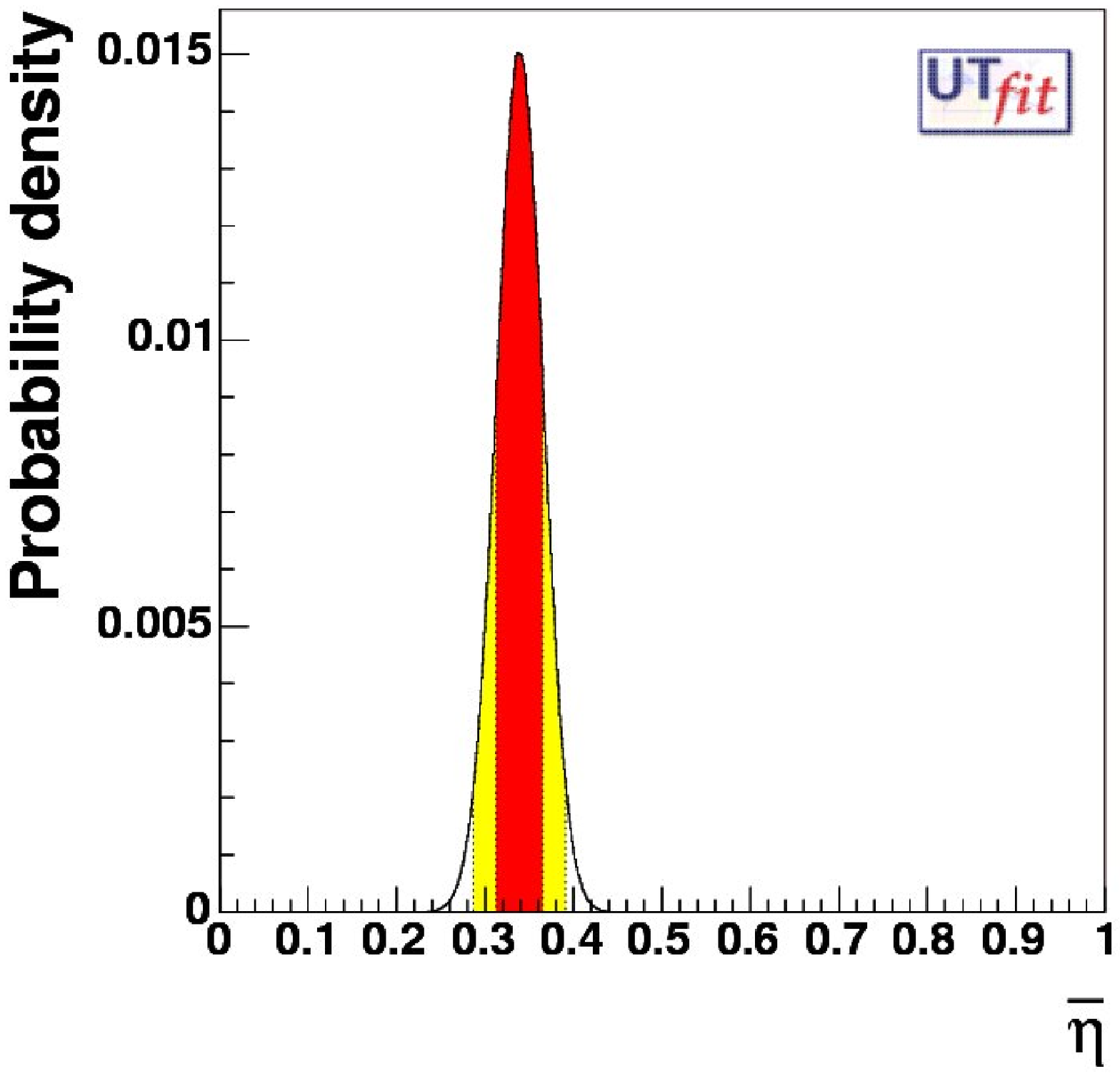}}
{\includegraphics[height=7.cm]{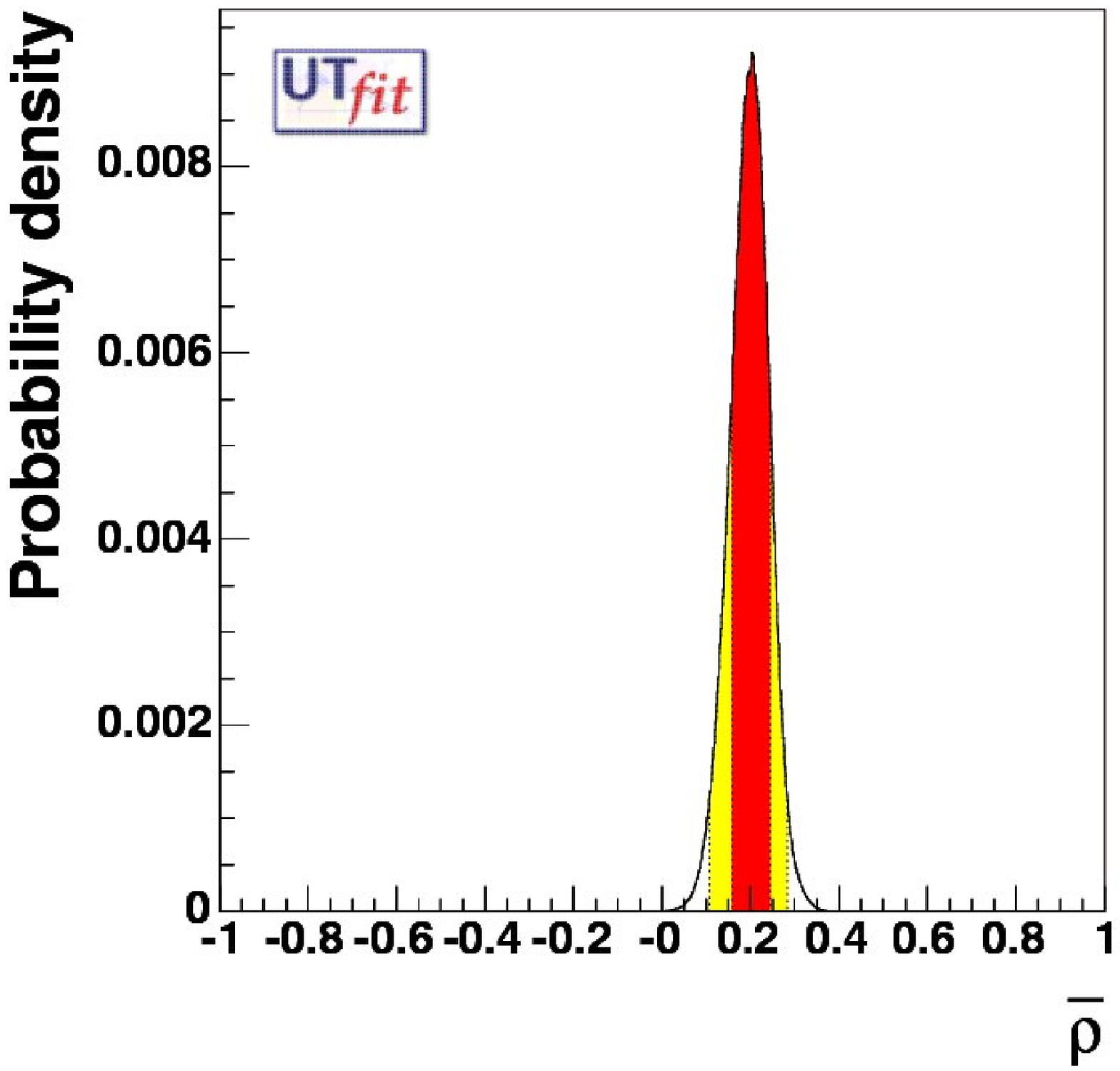}} \\
{\includegraphics[height=7.cm]{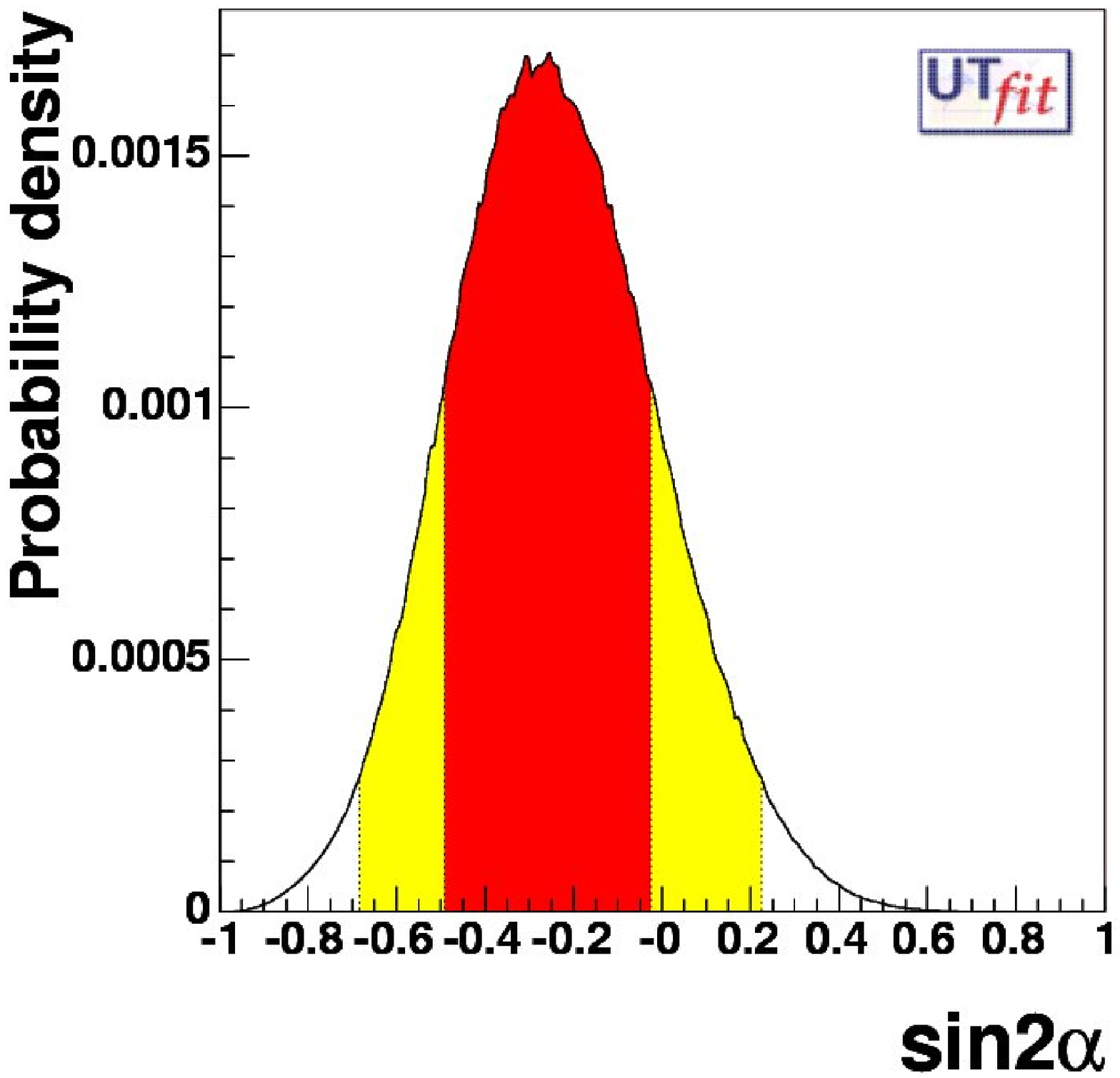}}
{\includegraphics[height=7.cm]{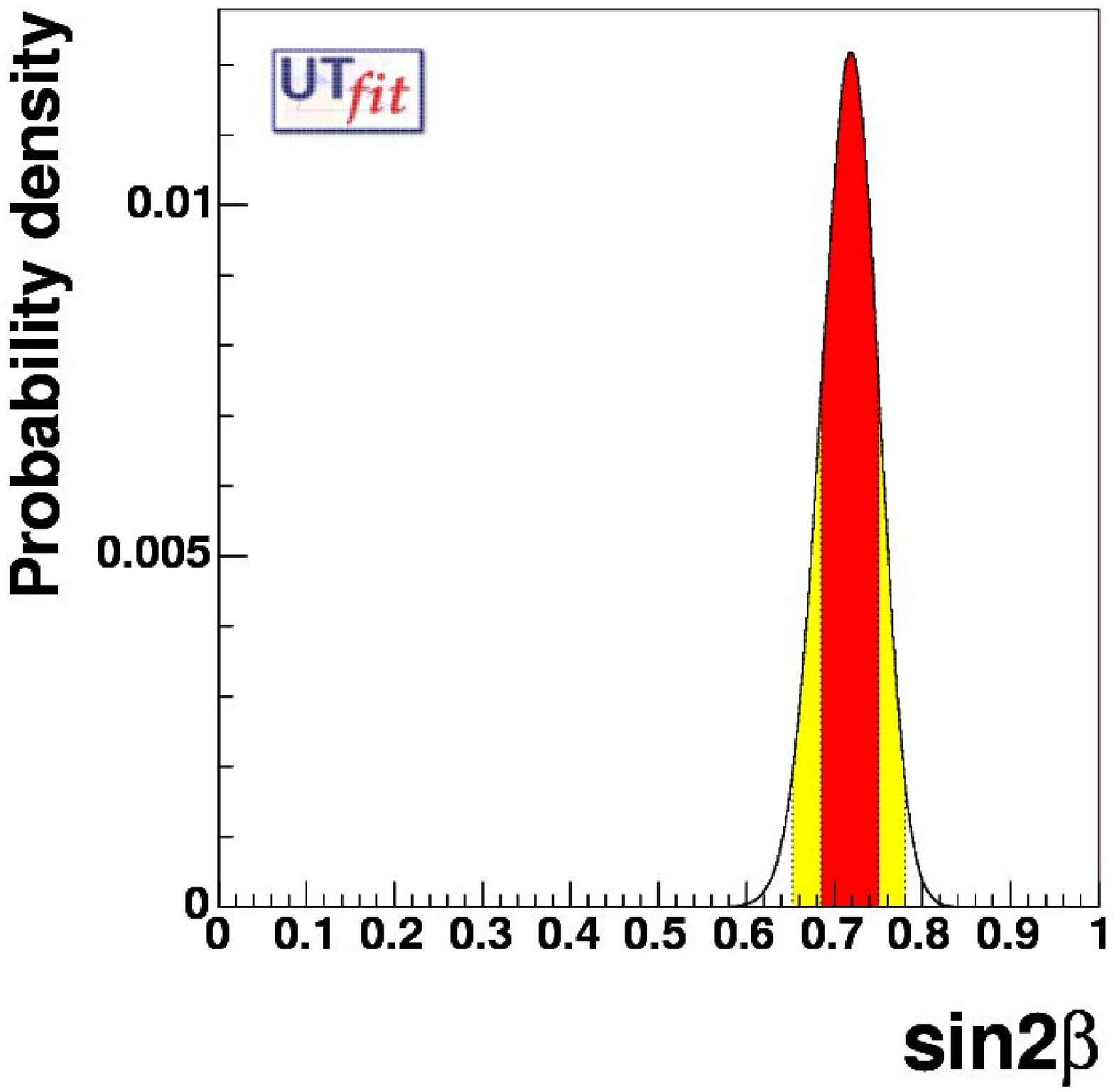}} \\
{\includegraphics[height=7.cm]{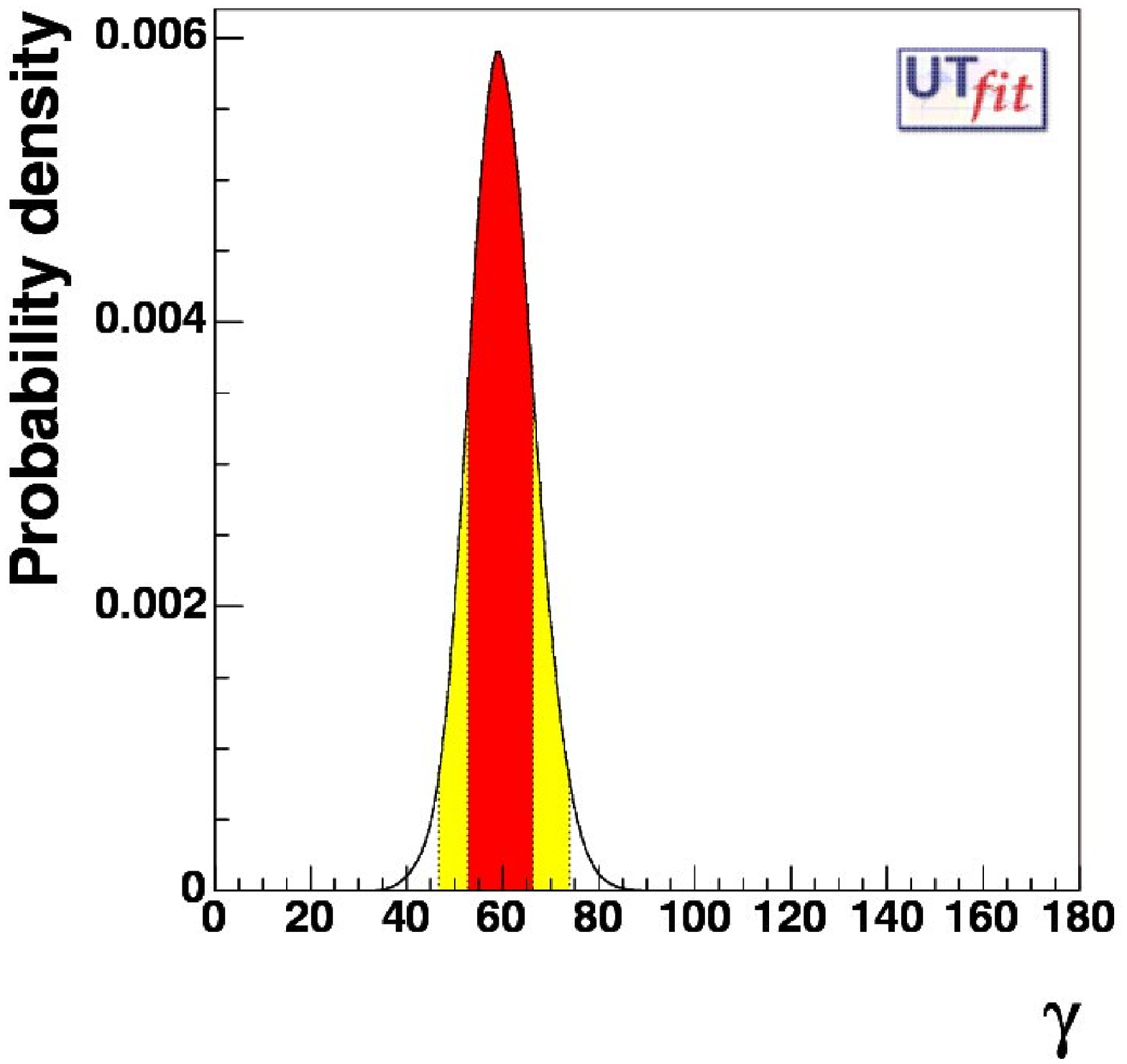}}
{\includegraphics[height=7.cm]{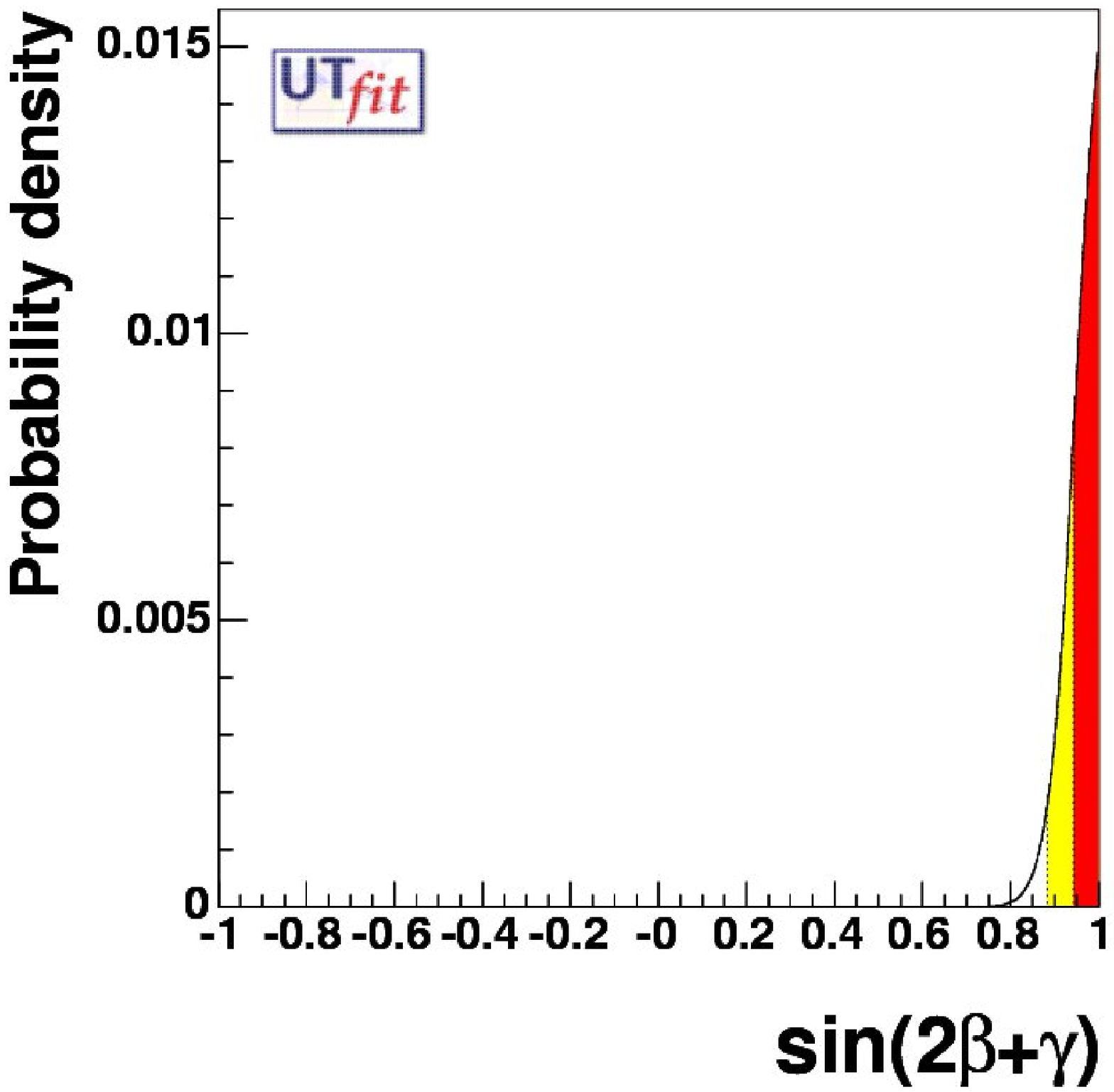}}
\caption{\it {From top left to bottom, the p.d.f.'s for $\etabar$, $\rhobar$,
    $\sna$, $\snb$, $\gamma$ and $\snbg$. The red (darker) and the
    yellow (lighter) zones correspond respectively to 68\% and 95\% of
    the normalised area. The following contraints have been used:
    $\left | V_{ub} \right |/\left | V_{cb} \right |$, $\epsilonk$,
    $\Delta m_d$, $\Delta m_s$ and $\snb$ from the measurement of the
    CP asymmetry in the $J/\psi K^0$ decays.}}
\label{fig:1dim}
\end{center}
\end{figure}

\begin{figure}[htb!]
\begin{center}
\includegraphics[width=16cm]{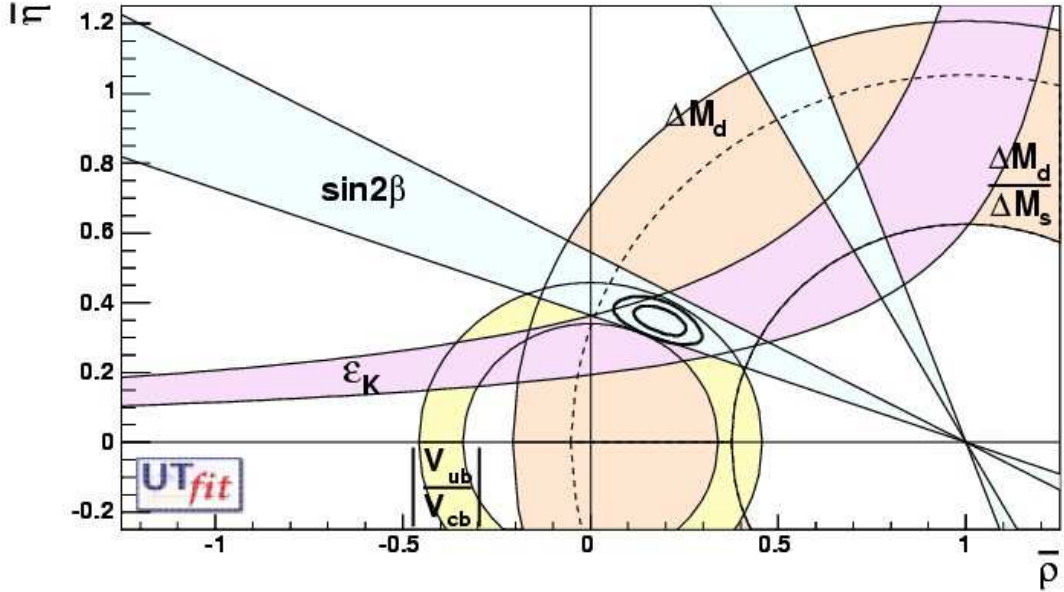}
\caption{ \it {Allowed regions for $\rhobar$ and $\etabar$ using the parameters
    listed in Table~\ref{tab:inputs}.  The closed contours at 68\% and
    95\% probability are shown. The full lines correspond to 95\%
    probability regions for the constraints, given by the measurements
    of $\left | V_{ub} \right |/\left | V_{cb} \right |$, $\epsilonk$,
    $\Delta m_d$, $\Delta m_s$ and $\snb$ from the measurement of the
    CP asymmetry in the $J/\psi K^0$ decays.}}
\label{fig:rhoeta}
\end{center}
\end{figure}

Using the constraints from $\left | V_{ub} \right |/\left | V_{cb}
\right |$, $\Delta {m_d}$, $\Delta {m_s}/\Delta {m_d} $, $\epsilonk$
and $\snb$, we obtain the results given in Table~\ref{tab:1dim}.

\begin{table*}[h]
\begin{center}
\begin{tabular}{@{}llllll}
\hline\hline
         Parameter          &   ~~~~~68$\%$        & ~~~~~95$\%$     &
~~~~~99$\%$   \\ \hline 
~~~~~$\overline {\eta}$     & 0.348  $\pm$ 0.028   & [0.293;0.403]   &
[0.275;0.418] \\
~~~~~$\overline {\rho}$     & 0.172  $\pm$ 0.047   & [0.082;0.270]   &
[0.051;0.302] \\
          ~$\snb$           & 0.725  $\pm$ 0.033   & [0.645;0.772]   &
[0.627;0.793] \\
          ~$\sna$           & -0.16  $\pm$ 0.26    & [-0.62;0.35]    &
[-0.75;0.48]  \\
~~~~$\gamma[^{\circ}$]      & 61.5   $\pm$ 7.0     & [47.5;76.6]     &
[43.3;81.6]   \\
          ~$\snbg$          & $> 0.94$             &  $> 0.88$       & $>0.84$  
    \\ 
$Im {\lambda}_t$[$10^{-5}$] & 13.5   $\pm$ 1.0     & [11.5;15.3]     &
[10.8;15.9]   \\
\hline\hline
\end{tabular} 
\end{center}
\caption {\it {Values and probability ranges for the Unitarity Triangle
parameters obtained by using  
the following constraints: $\left | V_{ub} \right |/\left | V_{cb} \right |$, 
$\Delta {m_d}$, $\Delta {m_s}/\Delta {m_d} $, $\epsilonk$ and $\snb$.}}
\label{tab:1dim} 
\end{table*}

Figures~\ref{fig:1dim} and \ref{fig:rhoeta} show, respectively, the
p.d.f.'s for the main Unitarity Triangle parameters and the selected
region in the $\rhobar-\etabar$ plane.

\subsection{Determination of other important quantities}

In the previous sections we have shown that it is possible to obtain
the p.d.f.'s for all the various UT parameters. It is instructive to
remove from the fitting procedure the external information on the
value of one (or more) of the constraints.

In this section we study the distributions of $\dms$ and of the
hadronic parameters.  For instance, in the case of the hadronic
parameters, it is interesting to remove from the fit the constraints
on their values coming from lattice calculations and use them as one
of the free parameters of the fit.  In this way we may compare the
uncertainty obtained on a given quantity through the UT fit to the
present theoretical error on the same quantity.

\subsubsection{The expected distribution for \boldmath$\dms$}

\begin{figure}[htb!]
\begin{center}
\includegraphics[height=7cm]{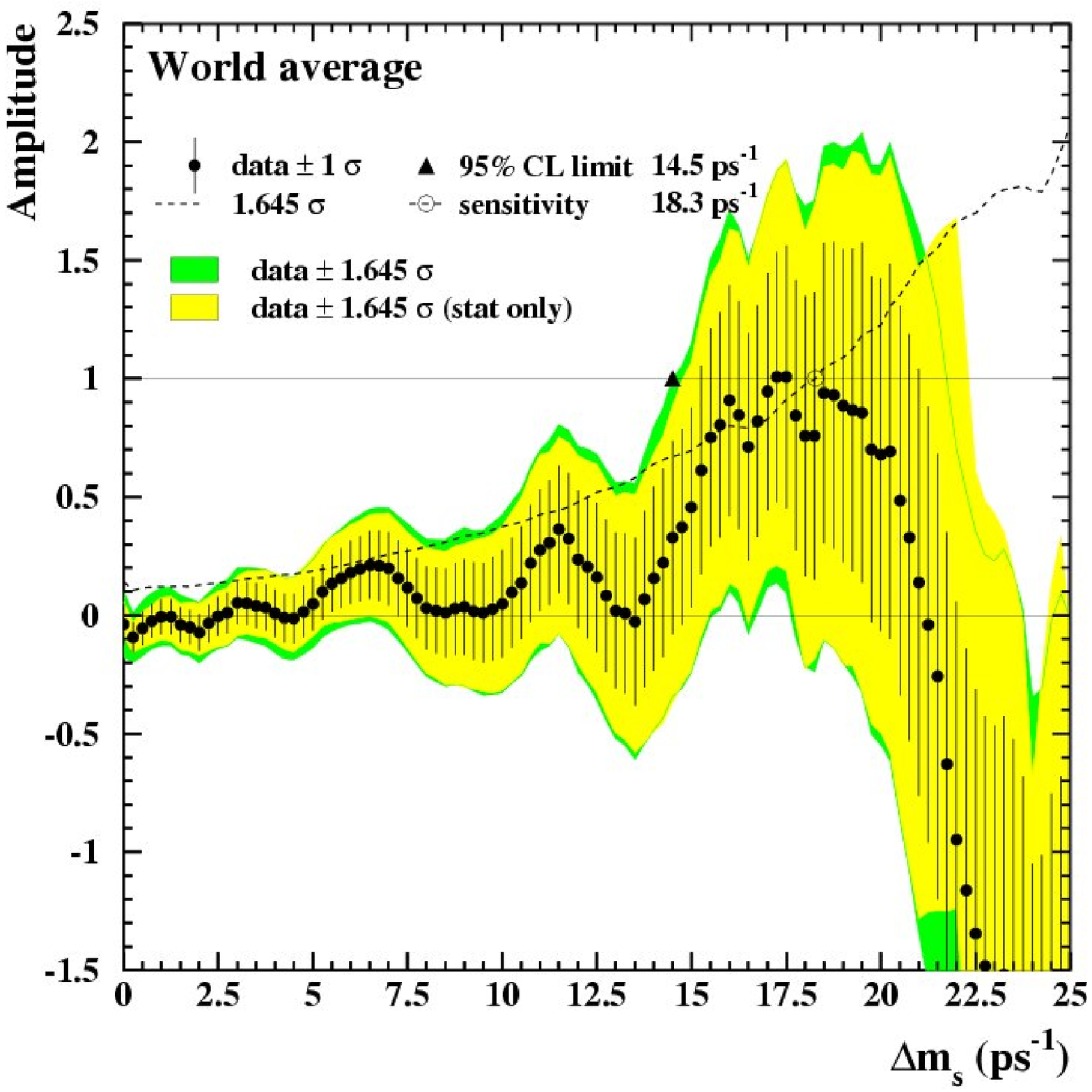} 
\includegraphics[height=8cm]{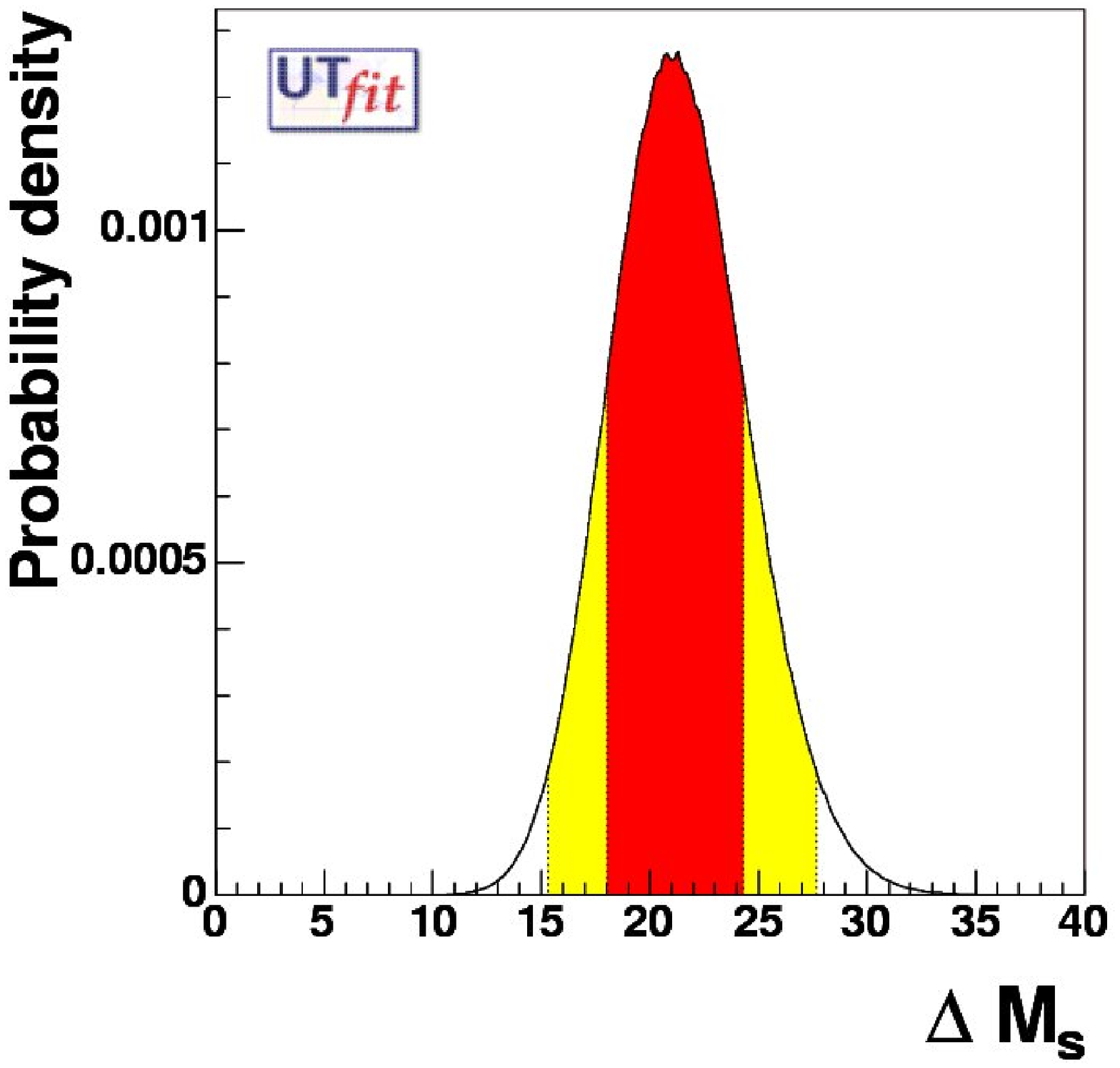} 
\end{center}
\caption{ \it {
    Left plot: combined results~\cite{ref:hfag} from all analyses on
    the oscillation amplitude, as a function of $\Delta {m}_s$. The
    points with error bars are the data; the lines show the 95\% C.L.
    curves (in dark the systematics have been included). The dotted
    curve corresponds to the sensitivity.  Right plot: $\Delta m_s$
    probability distributions, obtained without using the information
    from ${\rm B}^0_s-\overline{{\rm B}^0_s}$ mixing.}}
\label{fig:dms}
\end{figure}

Removing the constraint coming from $\dms$, the probability
distribution for $\dms$ itself can be extracted as shown in
Figure~\ref{fig:dms}.  The results of this exercise are given in
Table~\ref{tab:dmsresults}.  Present analyses at LEP/SLD have
established a sensitivity of 18.3~ps$^{-1}$ and they show a higher
probability region for a positive signal (see left plot in
Fig.~\ref{fig:dms}: a ``signal bump'' appears around 17.5 ps$^{-1}$)
well compatible with the range of the $\dms$ distribution from the UT
fit (see right plot in Fig.~\ref{fig:dms}).  Accurate measurements of
$\dms$ are expected from the TeVatron in the next future.

\begin{table*}[h]
\begin{center}
\begin{tabular}{@{}lllll}
\hline\hline
      ~~~~~~~~~~~~~~Parameter                              & ~~68$\%$          
& ~~~~95$\%$   & ~~~~99$\%$   \\ 
\hline
~~~~$\dms$(\rm{including ~$\dms$}) [{\rm ps}$^{-1}$]    &    18.3 $\pm$ 1.6 
& (15.4-23.1)  & (15.1-27.0)  \\
$\dms$(\rm{without~$\dms$}) [{\rm ps}$^{-1}$]  &    21.1 $\pm$ 3.1 
& (15.3-27.7)  & (13.7-29.9)  \\
\hline\hline
\end{tabular} 
\end{center}
\caption {\it Central values and ranges for $\dms$ corresponding to defined
levels of probability, obtained by including or not including the information
from the experimental amplitude spectrum ${\cal A}(\dms)$.}
\label{tab:dmsresults} 
\end{table*}

\subsubsection{Determination of \boldmath$\fbssqbs$, $\hat{B}_K$ and $\xi$}
\label{sec:2dpdfs}

\begin{figure}[htbp!]
\begin{center}
{\includegraphics[height=5.8cm]{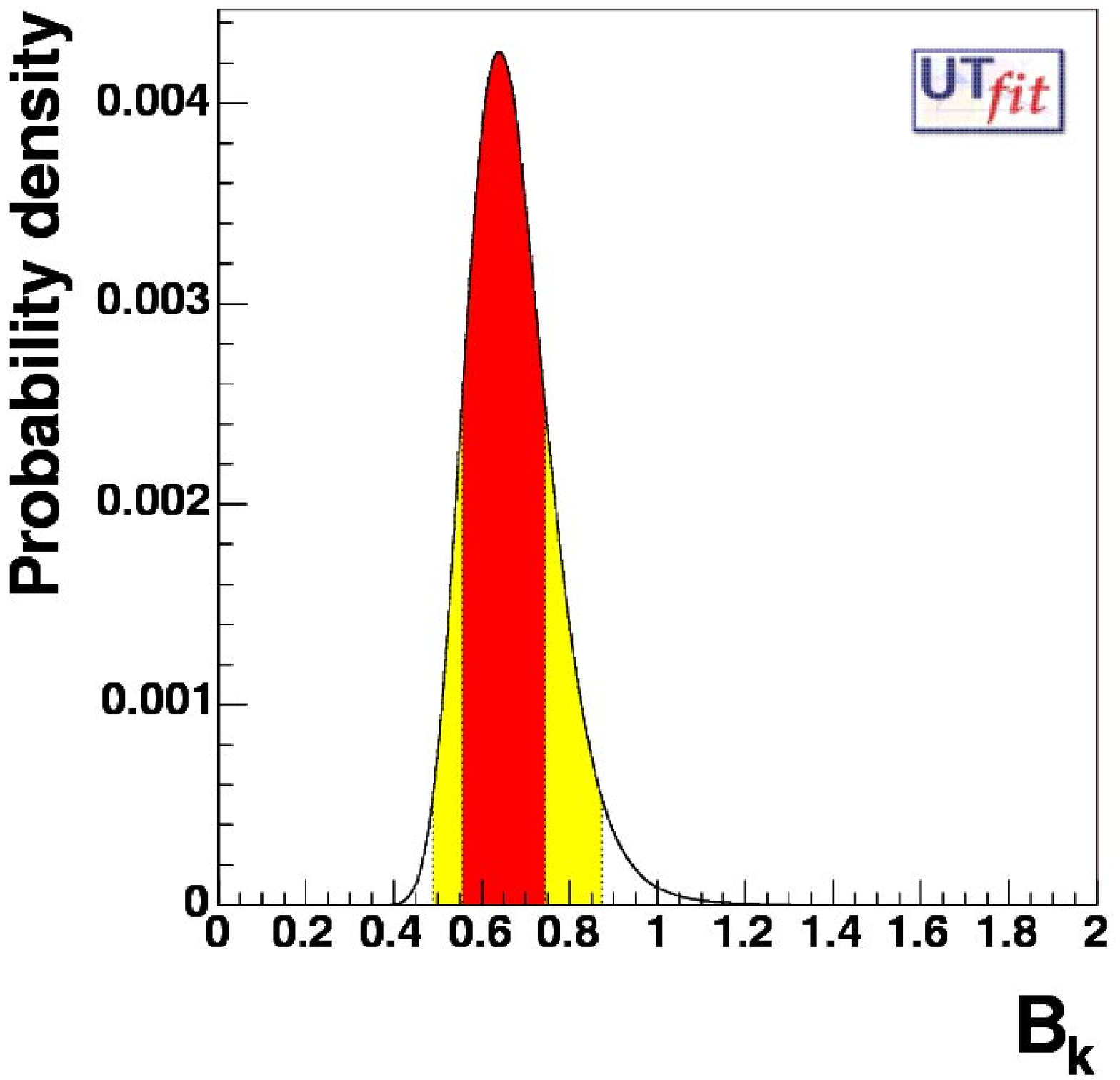}}
{\includegraphics[height=5.3cm]{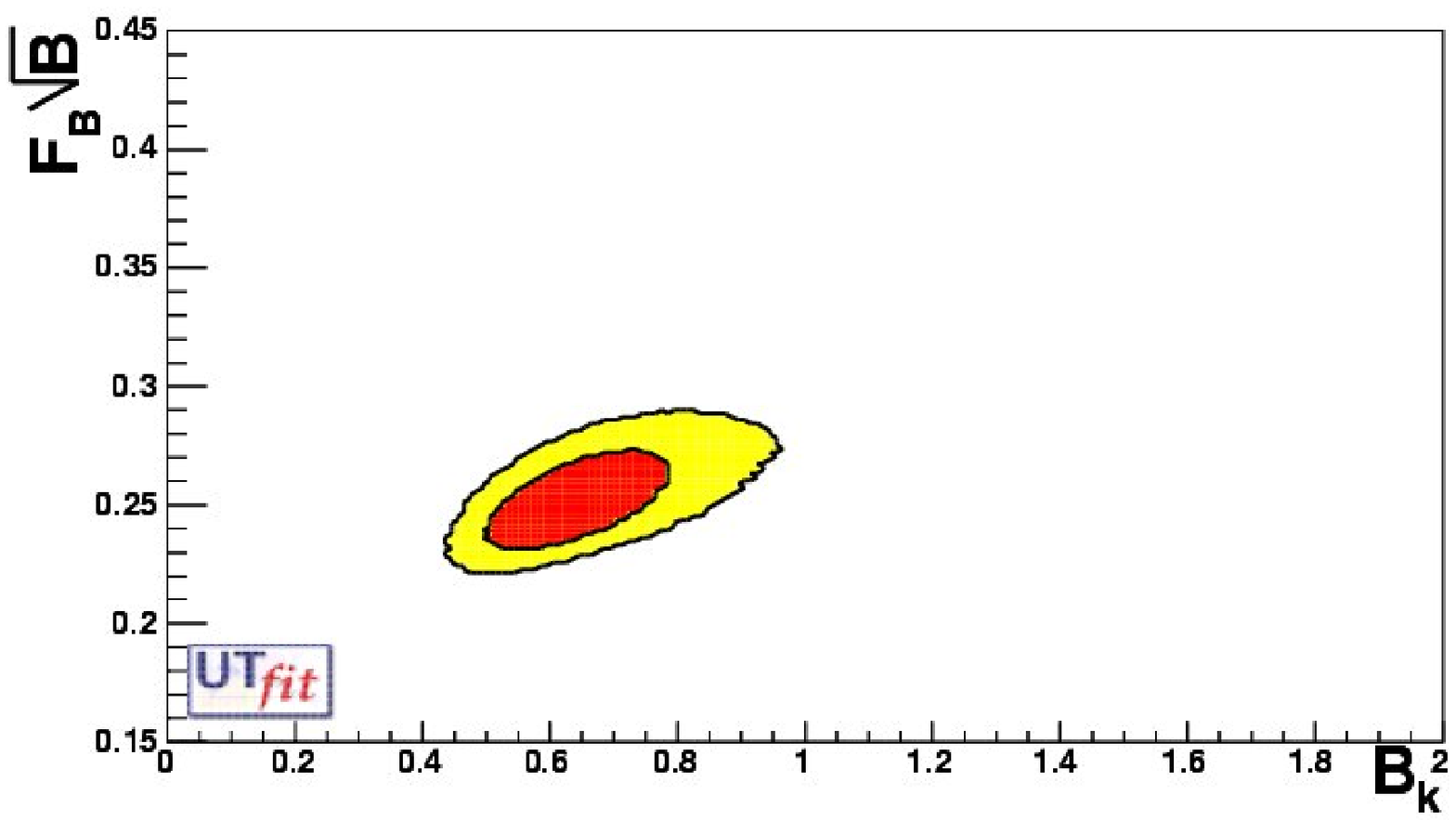}}  \\
{\includegraphics[height=5.8cm]{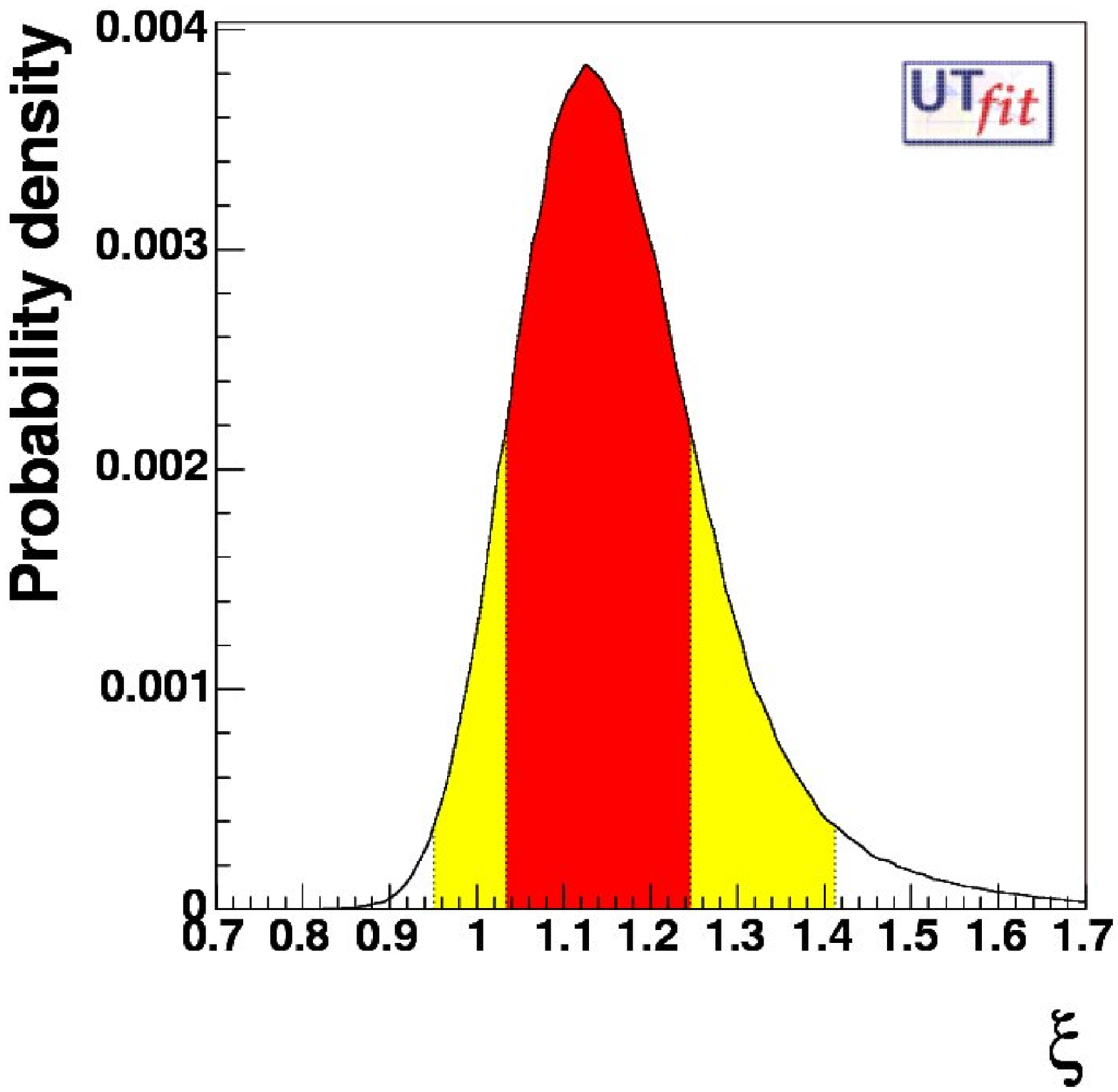}} 
{\includegraphics[height=5.3cm]{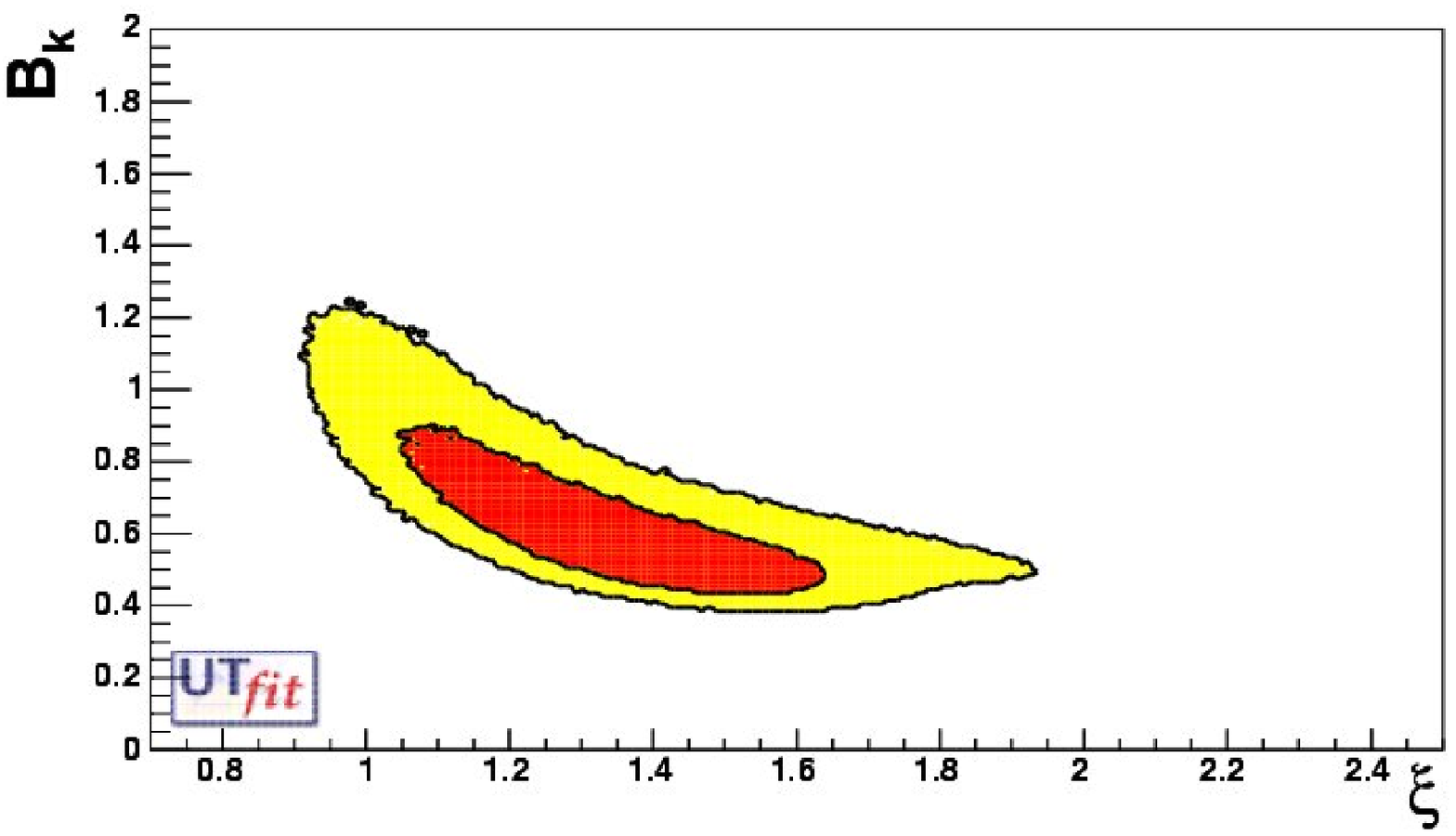}}  \\
{\includegraphics[height=5.8cm]{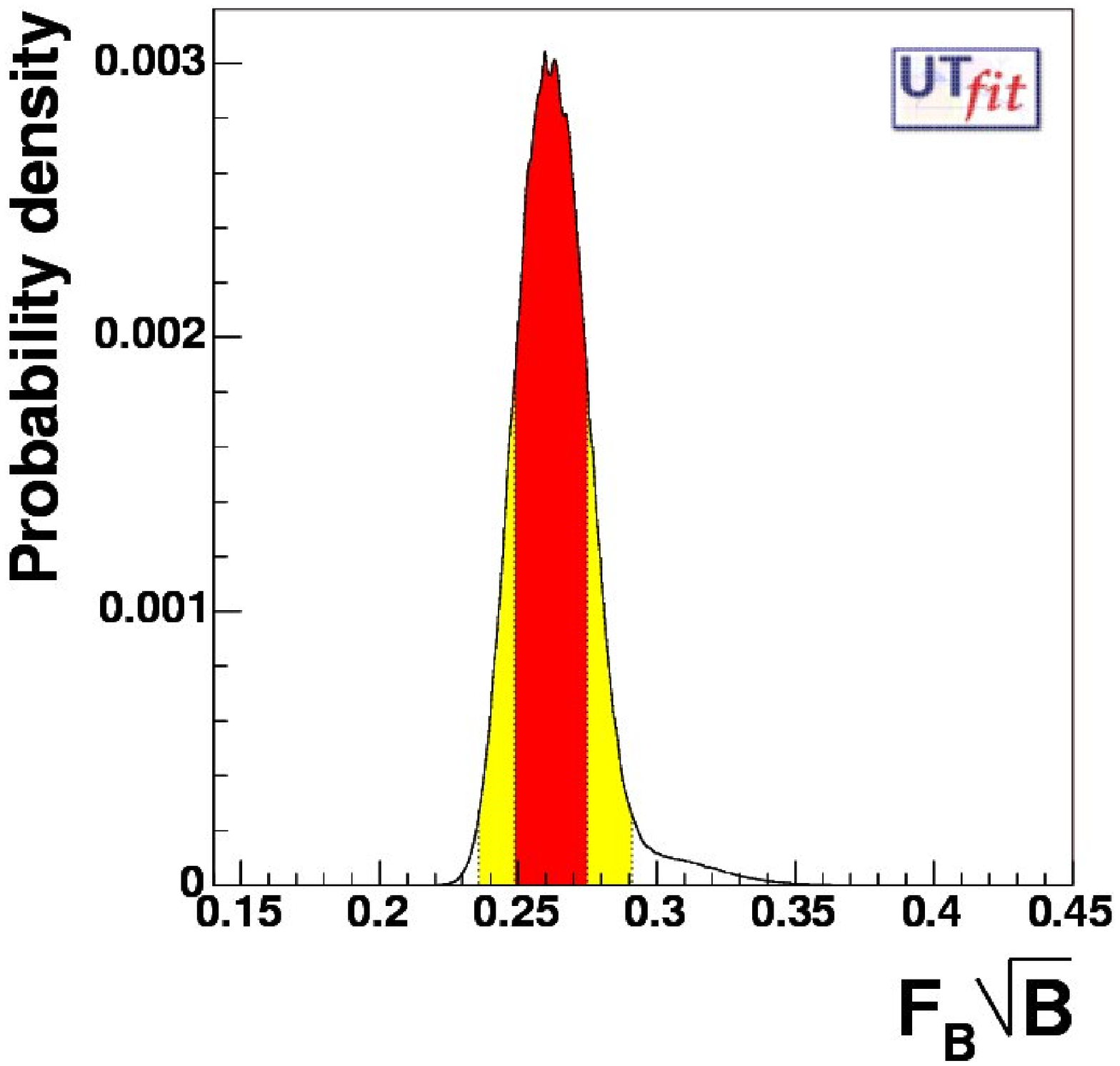}}      
{\includegraphics[height=5.3cm]{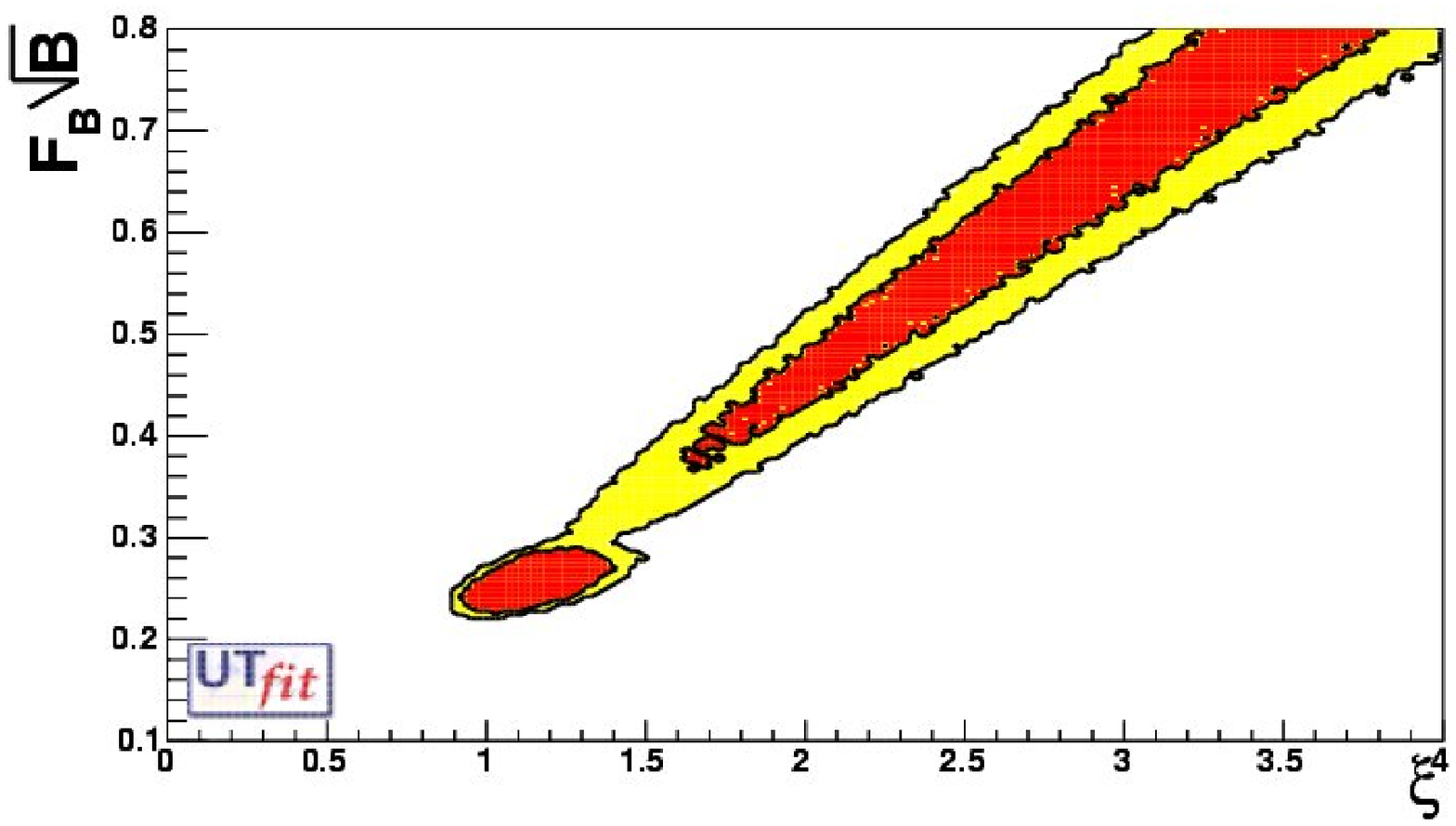}}  \\
\caption{\it {One- and two-dimensional 68\% and 95\% probability distributions
for $\hat B_K$, $\xi$ and $\fbssqbs$ (see Sec. \protect\ref{sec:2dpdfs}).}} 
\label{fig:bkfb}
\end{center}
\end{figure}

To obtain the p.d.f. for a given quantity, we perform the UT fit
imposing as input a uniform distribution of the quantity itself in a
range much wider than the expected interval of values assumed by the
parameter.  Table~\ref{tab:nonptsumm} shows the results of the UT fit
when one parameter at the time is taken out of the fit with this
procedure (see Figure \ref{fig:bkfb}).  The central value and the
error of each of these quantities has to be compared to the current
evaluation from lattice QCD, given in Table~\ref{tab:inputs}.

\begin{table}[h]
\begin{center}
\begin{tabular}{c|c|c|c}
\hline\hline
    Parameter         &     68$\%$             &      95$\%$  &  99$\%$     \\
\hline
$\xi$                 & 1.13$^{+0.12}_{-0.09}$ &  [0.95;1.41] & [0.92;1.57] \\
$\fbssqbs$(MeV)       & 263  $\pm$ 14          &  [236;290]   & [231;320]   \\
$\hat{B}_K$           & 0.65 $\pm$ 0.10        &  [0.49;0.87] & [0.45;0.99] \\
\hline\hline
\end{tabular} 
\caption {\it Values and probability ranges for the non-perturbative QCD
  parameters, if the external information (input) coming from the
  theoretical calculation of these parameters is not used in the CKM
  fits.}
\label{tab:nonptsumm} 
\end{center}
\end{table}

Some conclusions can be drawn.  The precision on $\fbssqbs$ obtained
from the fit has an accuracy which is better than the current
evaluation from lattice QCD.  This proves that the standard CKM fit
is, in practice, weakly dependent on the assumed theoretical
uncertainty on $\fbssqbs$.

The result on $\hat{B}_K$ indicates that values of $\hat{B}_K$ smaller
than $0.45$ are excluded at $99\%$ probability, while large values of
$\hat{B}_K$ are compatible with the prediction coming from the UT fit
using the other constraints.  The present estimate of $\hat{B}_K$ from
lattice QCD, which has a 15$\%$ relative error
(Table~\ref{tab:inputs}), is still more precise than the indirect
determination from the UT fit.  Likewise, the present best
determination of the parameter $\xi$ comes from lattice QCD.

In the above exercise we have removed from the UT fit individual
quantities one by one.  It is also interesting to see what can be
obtained taking out two of them simultaneously.
Figures~\ref{fig:bkfb} show the regions selected in the planes
($\fbssqbs$,~$\hat B_K$), ($\xi$,~$\hat B_K$) and ($\fbssqbs$,~$\xi$).
The corresponding results are summarized in
Table~\ref{tab:nonptsumm-two}.

\begin{table}[h]
\begin{center}
\begin{tabular}{c|c|c|c}
\hline\hline
    Parameter         &     68$\%$        &      95$\%$  &  99$\%$     \\ 
$\fbssqbs$(MeV)       & 252 $\pm$ 13     &  [228;283]   & [219;325]   \\
$\hat{B}_K$           & 0.63$\pm$ 0.10   &  [0.48;0.87] & [0.44;1.02] \\ \hline
$\fbssqbs$(MeV)       &        -         &  $>0.23$     & $>0.22$   \\
$\xi$                 &        -         &  $>0.98$     & $>0.92$ \\ \hline

$\hat{B}_K$           & $0.53^{+0.19}_{-0.07}$ & [0.41,0.99]  & [0.39,1.18] \\ 
$\xi$                 & 1.33 $\pm$ 0.20        & [0.99,1.63]  & [0.95,1.70] \\
\hline
\hline\hline
\end{tabular} 
\caption {\it Values and probability ranges for the non-perturbative QCD
  parameters, if two external pieces of information (inputs) coming
  from the theoretical calculation of these parameters are not used in
  the CKM fits.}
\label{tab:nonptsumm-two}
\end{center}
\end{table}

\section{New Constraints from UT angle measurements}
\label{sec:newinputs}

The values for $\sin{2\alpha}$, $\gamma$, and $\sin(2\beta +\gamma)$
given in Table~\ref{tab:1dim} have to be taken as predictions for
future measurements. A strong message is given for instance for the
angle $\gamma$. Its indirect determination is known with an accuracy
of about $10\%$.  It has to be stressed that, with present
measurements, the probability for $\gamma$ to be greater than
$90^{\circ}$ is only $0.0055\%$.

Thank to the huge statistics collected at the B-Factories, new
CP-violating quantities have been recently measured allowing for the
direct determination of $\sin{2\alpha}$, $\gamma$, and $\sin(2\beta
+\gamma)$.  In the following we present the UT fit results including
these new measurements and their impact on the $\rhobar-\etabar$
plane.

\subsection{Determination of the angle $\gamma$ using DK events}
\label{ref:gamma}

Various methods using $B \rightarrow DK$ decays have been used to
determine the Unitarity Triangle angle $\gamma$~\cite{ref:DKmethods}.
The basic idea in these methods is the following. A charged $B^-$ can
decay into a $D^0(\overline{D}^0) K^-$ final state via a
$V_{cb}$($V_{ub})$ mediated process. CP violation occurs if the $D^0$
and the $\overline{D}^0$ decay in the same final state.  The
measurement of the direct CP violation is thus sensitive to the
$V_{ub}$ and $V_{cb}$ phase difference, $\gamma$. The same argument
can be applied
to $B \rightarrow D^{*}K$ decays.\\
The most important aspect of these decays is that they proceed only
via tree-level diagrams, implying that the determination of $\gamma$
is not affected by possible New Physics loop contributions.

One of these methods is the ``GLW method'' and it consists in
reconstructing $D^0$ mesons in a specific CP (even/odd) mode. The
``ADS method'' is, instead, based on the fact that
$D^0(\overline{D}^0)$ decays can reach the same final state through
Doubly Cabibbo Suppressed (DCS) (Cabibbo Allowed (CA)) processes.  The
following observables are defined in these two methods:
\begin{eqnarray}
R_{C\!P^+} &=&      1 + r_B^2 -2 r_B \sin \gamma \sin \delta_B  \\ \nonumber
R_{C\!P^-} &=&      1 + r_B^2 -2 r_B \sin \gamma \cos \delta_B  \\ \nonumber
A_{C\!P^+} &=&      2 r_B \sin \gamma \sin \delta_B/ R_{C\!P^+}  \\ \nonumber
A_{C\!P^-} &=&     -2 r_B \sin \gamma \sin \delta_B/ R_{C\!P^-}  \\ \nonumber
R_{ADS}  &=&      r_{DCS}^2 + r_B^2 + 2 r_B r_{DCS} \cos \gamma \cos (\delta_B
~+~ \delta_D)
\end{eqnarray}
where
\begin{equation}
 r_B =  \frac{{\cal A} (B^- \rightarrow D^0 K^-)}{{\cal A} (B^- \rightarrow
\overline{D}^0 K^-)},
\label{eq:rbdef}
\end{equation}
$r_{DCS} = \sqrt{\frac{D^0 \rightarrow K^+ \pi^-}{D^0 \rightarrow K^-
    \pi^+}}$ and $\delta_B$ ($\delta_D$) is the difference between the
strong phases of
the two amplitudes in the B system (B and D systems). \\
In~\cite{ref:dalitz} a new method based on the Dalitz analysis of
three-body decays has been proposed and recent results using this new
technique applied to the $D^0 \rightarrow K_s \pi^- \pi^{+}$ decays
have been published by the Belle Collaboration~\cite{ref:belle}.  The
advantage of this method is that the full sub-resonance structure of
the three-body decay is considered, including interferences such as
those used for GLW and ADS methods plus additional interferences
because of the overlap of broad resonances in certain regions of the
Dalitz plot.  The same analysis can also be performed using $B
\rightarrow D^{*0}K$ decays.  The technique is identical to the one
used in $B \rightarrow D^{0}K$ decays but the values for $r_B$ and
$\delta_B$ are different so they will be indicated as $r^*_B$ and
$\delta^*_B$ in the following.  It is also interesting to note that
the Dalitz analysis has only a two-fold discrete ambiguity ($\gamma
+\pi,\gamma -\pi$) and not a four-fold ambiguity as in case of the GLW
and ADS methods.  It has to be noted that experimental likelihoods
have been used to
correctly implement the measurement of $R_{ADS}$ and the Dalitz result. \\
A summary of the experimental results is given in Table~\ref{tab:DK}.
\begin{table}[htbp!]
\begin{center}
\begin{tabular}{cc} \hline\hline
Observable & World Average \\
\hline
$A_{C\!P^+}$ &  0.07 $\pm$ 0.13     \\    
$A_{C\!P^-}$ & -0.19 $\pm$ 0.18     \\  
$R_{C\!P^+}$ &  1.09 $\pm$ 0.16     \\  
$R_{C\!P^-}$ &  1.30 $\pm$ 0.25     \\  \hline  
$R_{ADS}$  & 0.0054 $\pm$ 0.0124  \\   \hline
$r_{B} (r_{B}^*)$        &  $0.26^{+0.10}_{-0.14} \pm 0.03 \pm 0.04$
($0.20^{+0.19}_{-0.17} \pm 0.02 \pm 0.04$)  \\ 
$\gamma$                 &  $77^{+19}_{-17} \pm 13 \pm 11$ (Belle Dalitz)       
                               \\ \hline \hline
\end{tabular}
\caption{ \it {Summary of the results obtained with $B \rightarrow D^{(*)}K$
    decays and using the GLW and the ADS methods.  The last two lines
    give the results from the Dalitz analysis~\cite{ref:dalitz}, for
    which the ambiguity $\gamma \to \pi - \gamma$ is implicit.}}
\label{tab:DK}
\end{center}
\end{table}

\begin{figure}[htb!]
\begin{center}
\begin{tabular}{cc}
\includegraphics[height=5.3cm]{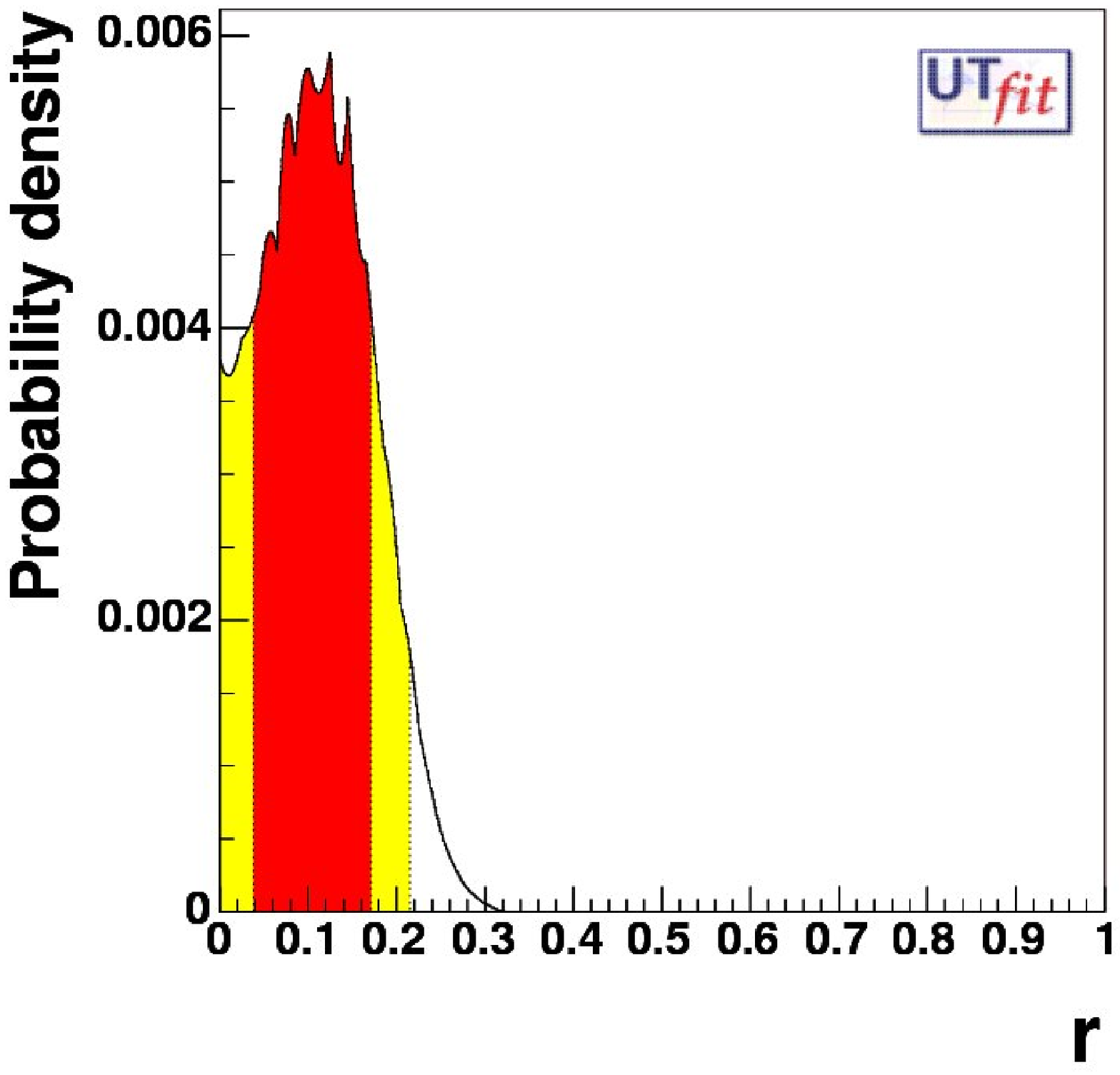}
\includegraphics[height=5.3cm]{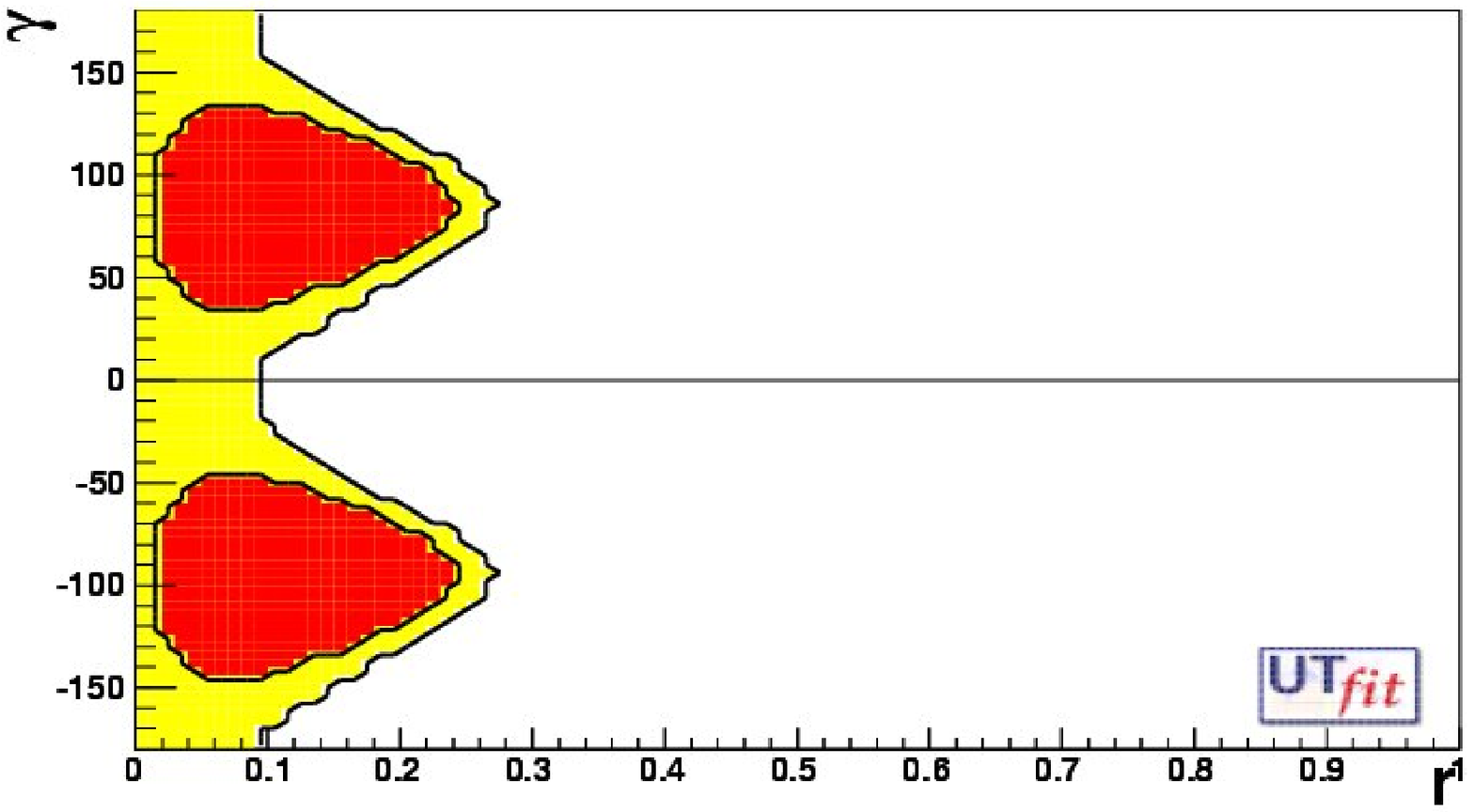} \\
\includegraphics[height=5.3cm]{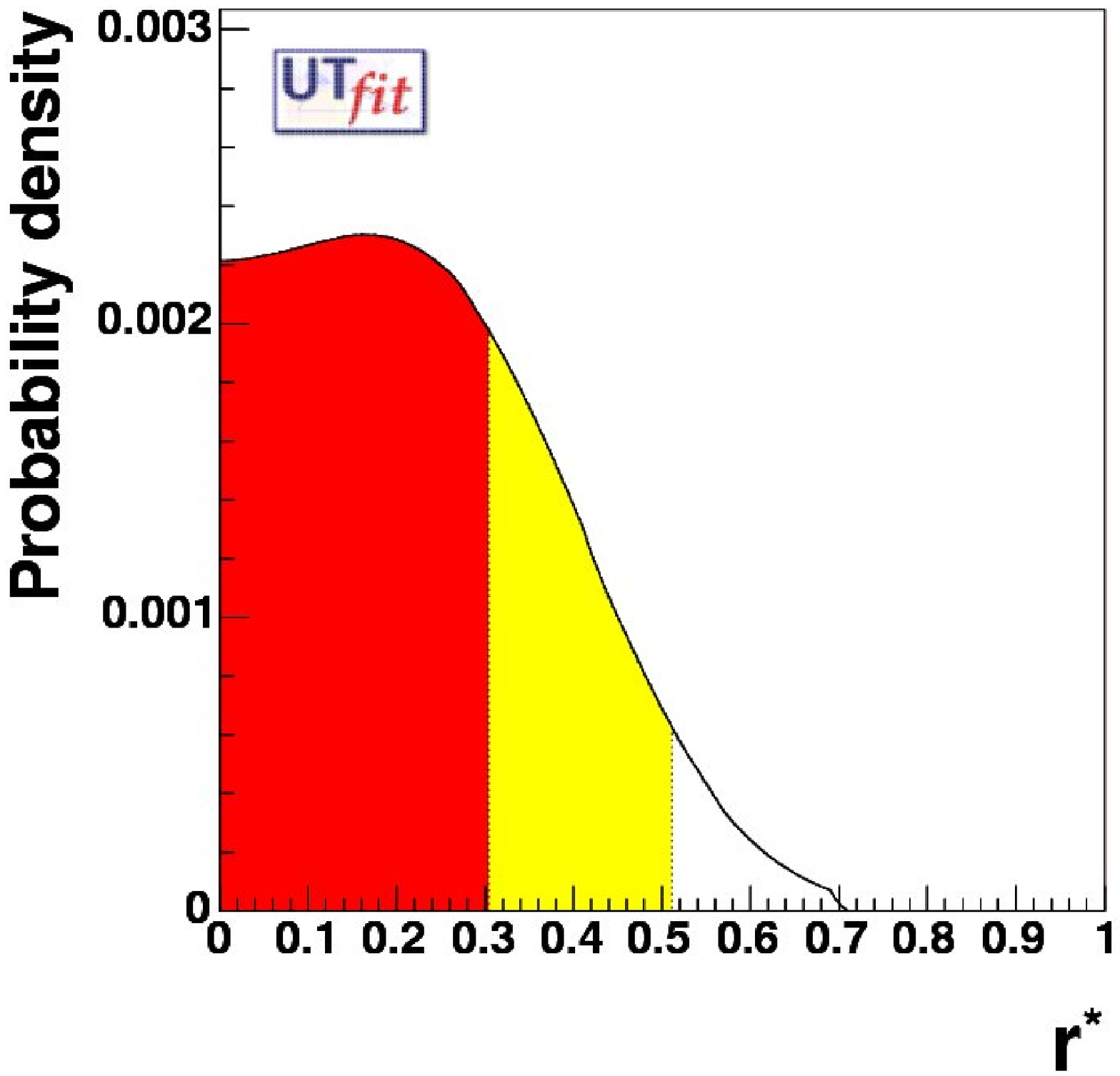} 
\includegraphics[height=5.3cm]{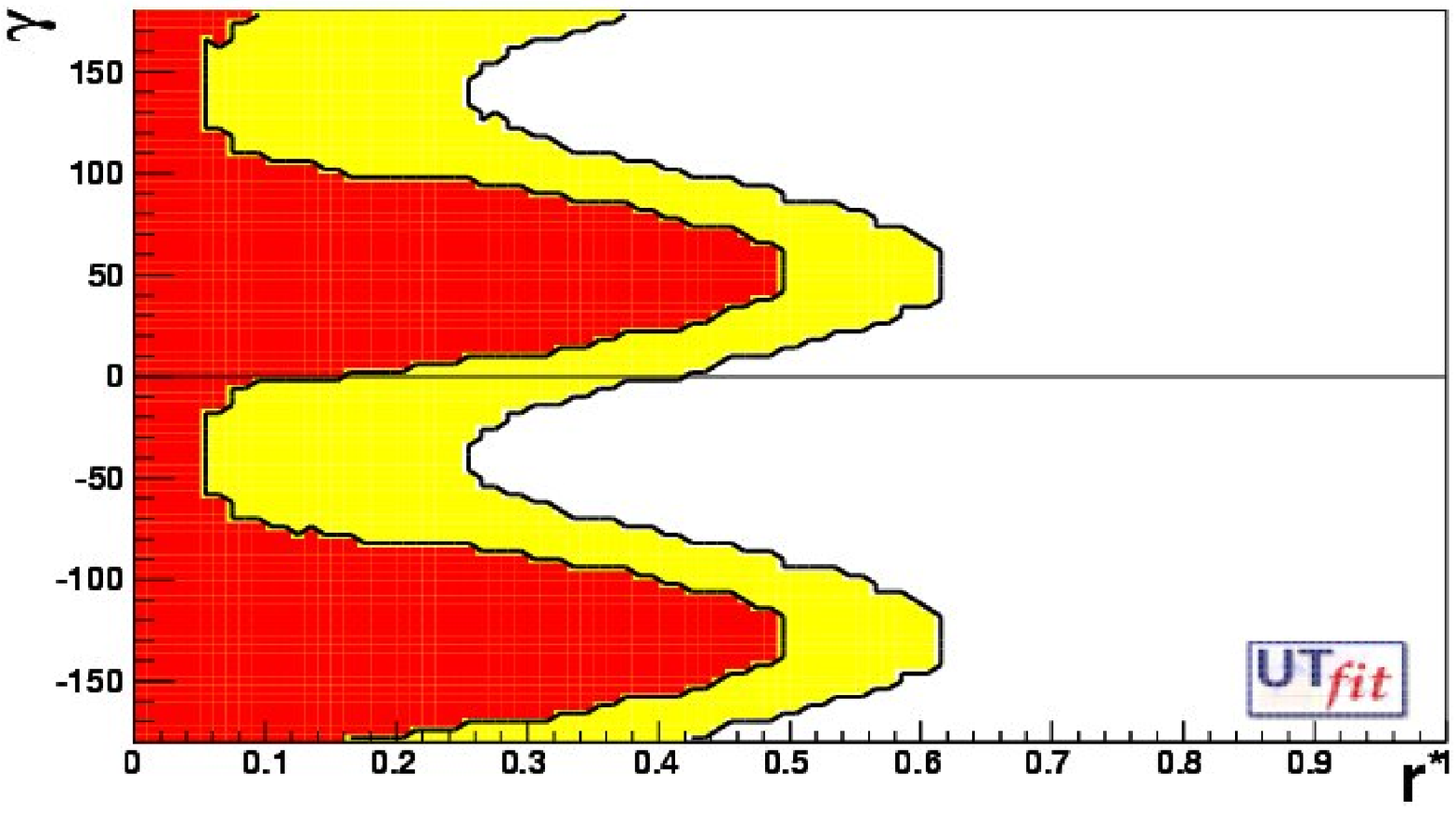} \\
\includegraphics[height=5.3cm]{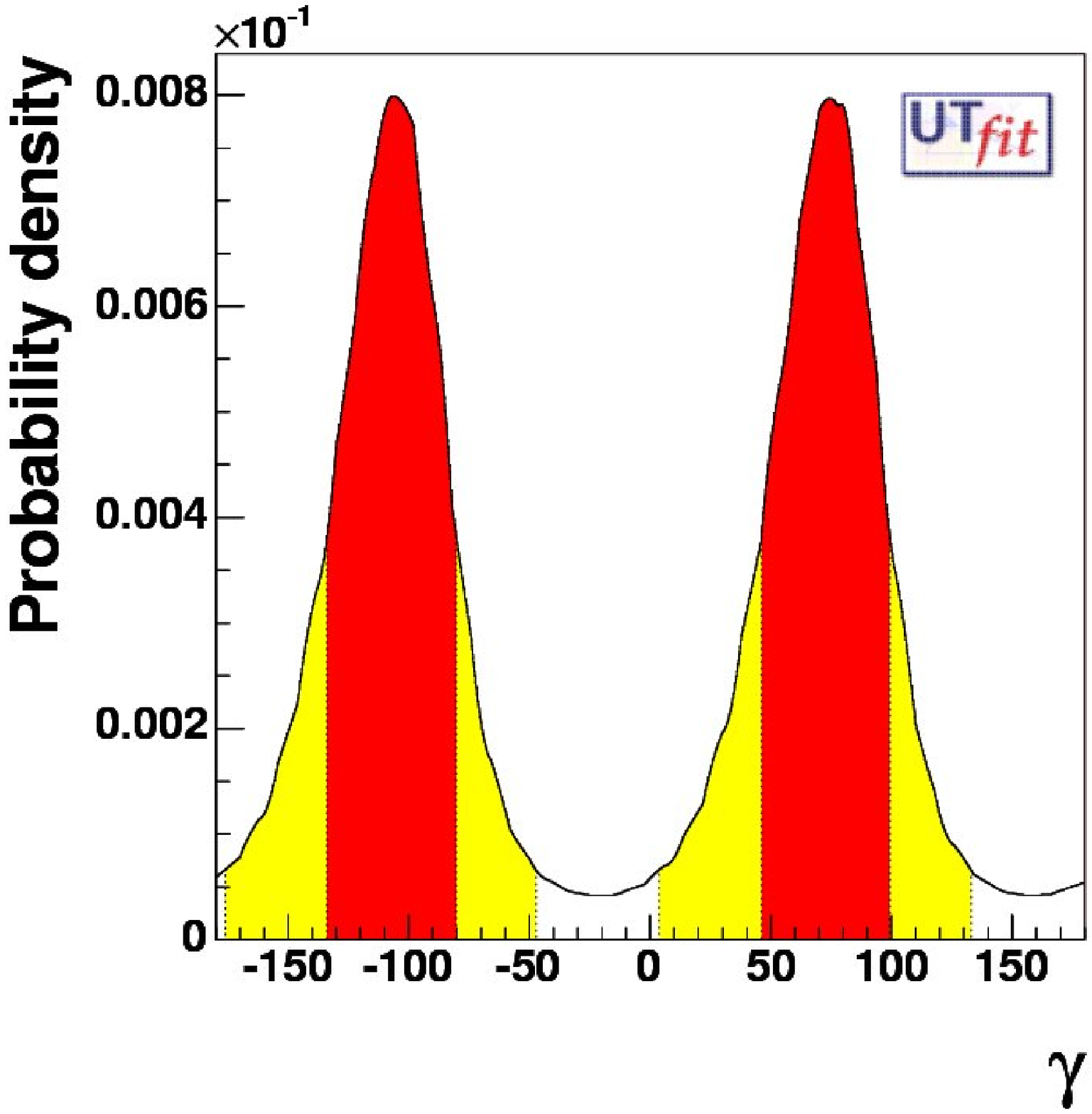} 
\includegraphics[height=5.3cm]{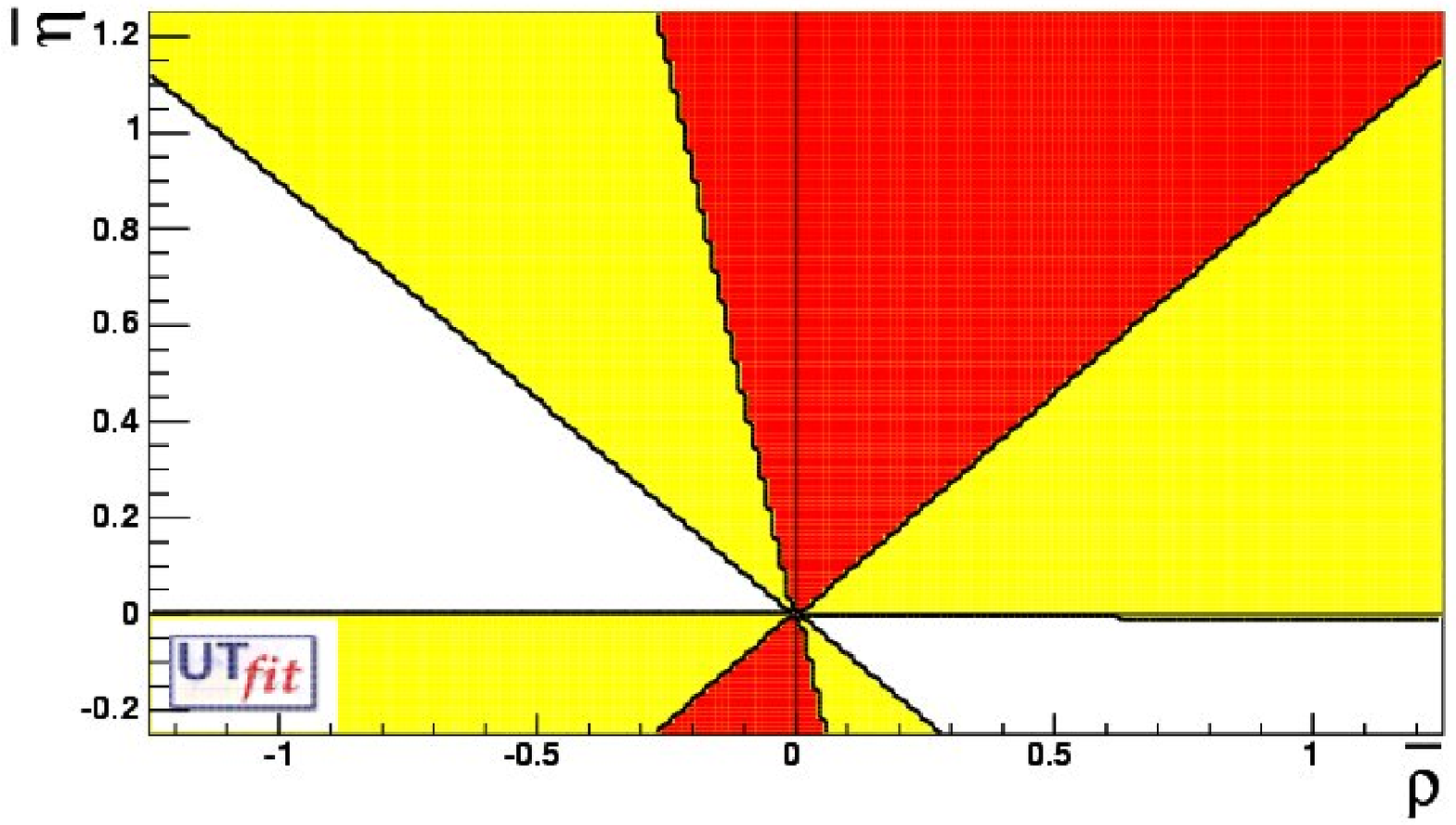} 
\end{tabular}
\end{center}
\caption{ {\it P.d.f.'s for $r_b$ from DK (top-left), for $r^*_b$ from D$^*$K
    (center-left) and for $\gamma$ (bottom-left) using the results
    from $B \rightarrow D^{(*)}K$ decays.  The plots on the right show
    the measurement of the angle $\gamma$ in the $r^{(*)}_b-\gamma$
    (top, center) and $\rhobar-\etabar$ (bottom) planes.}}
\label{fig:gammadir}
\end{figure}

All measurements in Table \ref{tab:DK} are used to extract $\gamma$.
The p.d.f.'s of $\gamma$, $r_B$ ($r^*_B$) and the selected region in
the $\gamma$ vs $r_B$ plane are shown in Figure~\ref{fig:gammadir},
where also the effect of this measurement in the $\rhobar-\etabar$
plane is shown.

The comparison between the direct and the indirect determination is:
\begin{eqnarray}
\gamma[^{\circ}] & = & 61.5  \pm 7.0 ~[47.5;76.6]~{\rm at} ~95\%~ C.L.   ~~\rm
{indirect}     \nonumber \\
\gamma[^{\circ}] & = & 73 \pm 27 ~[4;133]~~{\rm at} ~95\%~ C.L. ( -107 \pm 27
~~[-176;-47])
~\rm {direct}  
\label{eq:gamma}
\end{eqnarray}
An important result of this analysis is also the p.d.f. for $r_B$ from
which the following result can be given:
\begin{eqnarray}
r_B   & = & 0.105  \pm 0.065 ~[~<~0.22]~{\rm at} ~95\%~ C.L.   ~~\rm
{direct~from~}DK.
\label{eq:rbrb}
\end{eqnarray}

\subsection{Determination of sin2$\alpha$ using $\pi \pi$ and $\rho \rho$
events}
\label{ref:alpha}
In the absence of contributions from penguin diagrams, the
measurements of the parameter S of the time-dependent CP asymmetry for
$B^0 \to \pi^+ \pi^-$ and $B^0 \to \rho^+ \rho^-$ give measurements of
the quantity sin(2$\alpha$). Even in presence of penguins, one can use
the SU(2) flavour symmetry to connect the measured value of S$\equiv
\sin(2\alpha_{eff})$ to the value of sin(2$\alpha$), constraining the
contribution from penguin diagrams using the Branching Fractions and
the direct CP asymmetry measurements of all the $B \to \pi \pi$ ($B
\to \rho \rho$) decays~\cite{gronaulondon}. The decay amplitudes in
the $SU(2)$ limit and neglecting electroweak penguins can be written
as:
\begin{eqnarray}
A^{+-} &=& -T e^{-i\alpha} + P e^{i\delta_P} \nonumber \\
A^{+0} &=& -\frac{1}{\sqrt{2}} \left [e^{-i\alpha} (T +
T_c~e^{i\delta_{T_c}})\right ] \nonumber \\
A^{00} &=& -\frac{1}{\sqrt{2}} \left [e^{-i\alpha}  T_c~e^{i\delta_{T_c}} +
P~e^{i\delta_P} \right ]\,.
\label{eq:su2analysis}
\end{eqnarray}
They can be expressed in terms of three independent hadronic
amplitudes, the absolute values of which are denoted as $T$, $T_c$ and
$P$.  Similarly, $\delta_P$ and $\delta_{T_c}$ are the strong phases
of $P$ and $T_c$, once the phase convention is chosen so that $T$ is
real. It should be noted that these parameters are different for $B
\to \pi \pi$ and $B \to \rho \rho$ decays. For the $B \to \rho \rho$
decays we have in principle to further double the parameters for the
longitudinal and the transverse polarization. On the other hand the
experimental measurements are compatible with decays which are fully
longitudinally polarized.  For this reason, in case of $B \to \rho
\rho$, we make the assumption of fitting the amplitude
with only one set of parameters.\\
Notice that the number of parameters exceeds the number of available
measurements. Nevertheless, one can still extract information on
$\alpha$,
in the same spirit of the bounds \`a la Grossman-Quinn.\\
Using the experimental measurements given in
Table~\ref{tab:alpha-exp}, we thus constrain all these parameters and
the value of the UT angle $\alpha$.

\begin{table}[tb]
  \centerline{\small
    \begin{tabular}{c|ccc|ccc}
     \hline\hline
                  &      \multicolumn{3}{c}{$\pi \pi$}  &
\multicolumn{3}{c}{$\rho \rho$} \\ 
     \hline     
     Observable   &      BaBar      &   Belle      &     Average    &    BaBar  
   &      Belle   &   Average \\
     \hline     
     C                &  -0.19  $\pm$ 0.20  & -0.58 $\pm$ 0.17 & -0.46  $\pm$
0.13  & -0.23 $\pm$ 0.28  &              & -0.23  $\pm$ 0.28 \\
     S                &  -0.40  $\pm$ 0.22  & -1.00 $\pm$ 0.22 & -0.74  $\pm$
0.16  & -0.19 $\pm$ 0.35  &              & -0.19  $\pm$ 0.35 \\
     $BR^{+-}(10^{-6})$ &   4.7   $\pm$ 0.6   &  4.4 $\pm$ 0.7   &  4.6   $\pm$
0.4   & 30.0  $\pm$ 6.0   &              &  30.0  $\pm$ 6.0 \\
     $BR^{+0}(10^{-6})$ &   5.5   $\pm$ 1.2   &  5.0 $\pm$ 1.3   &  5.2   $\pm$
0.8   & 22.5  $\pm$ 8.1   &   31.7 $\pm$ 9.8 &  26.4  $\pm$ 6.4 \\
     $BR^{00}(10^{-6})$ &   2.1   $\pm$ 0.7   &  1.6 $\pm$ 0.6   &  1.9   $\pm$
0.5   & 0.6   $\pm$ 0.8   &              &    0.6 $\pm$ 0.8 \\
     \hline\hline
    \end{tabular}}
  \caption{{\it Experimental inputs from Isospin Analysis in $B \to \pi \pi$ and
$B \to \rho \rho$
           decays~\cite{ref:hfag}. The $B \to \rho \rho$ decays are assumed to
be 
           fully polarized, in agreement with available measurements.}}
  \label{tab:alpha-exp}
\end{table}

\begin{figure}[htb!]
\begin{center}
\includegraphics[height=5.8cm]{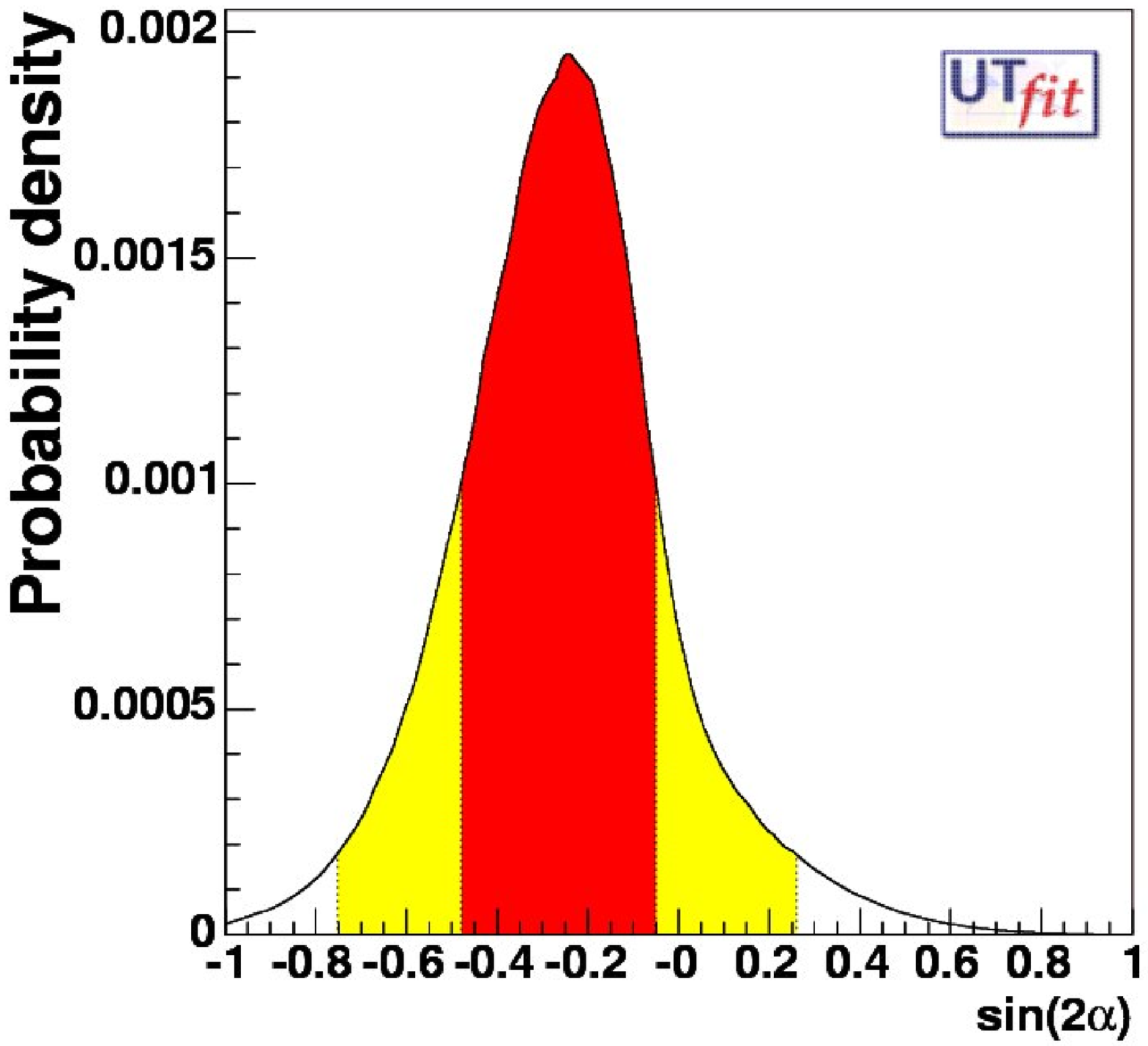}
\includegraphics[height=5.3cm]{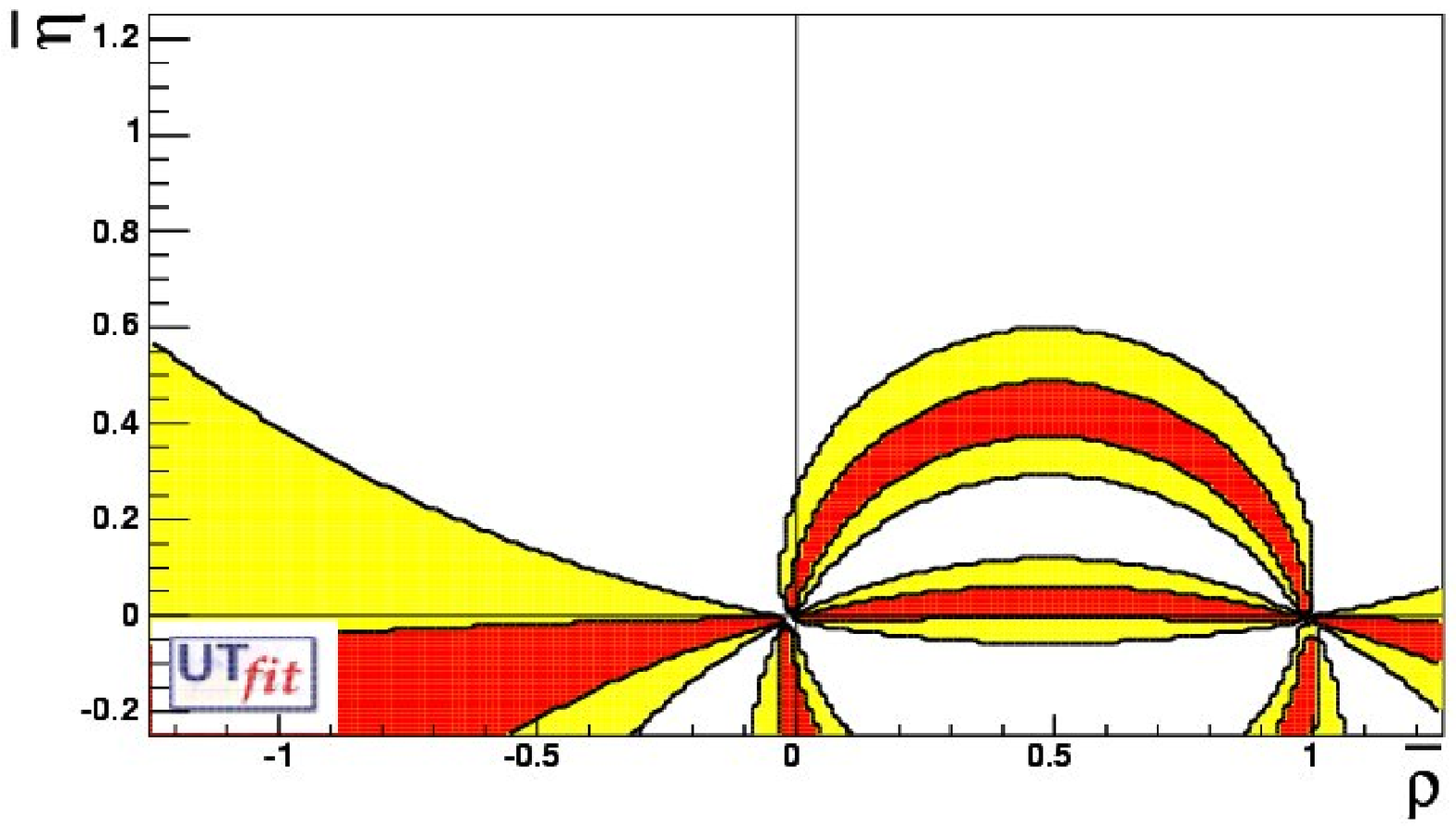} \\
\includegraphics[height=5.8cm]{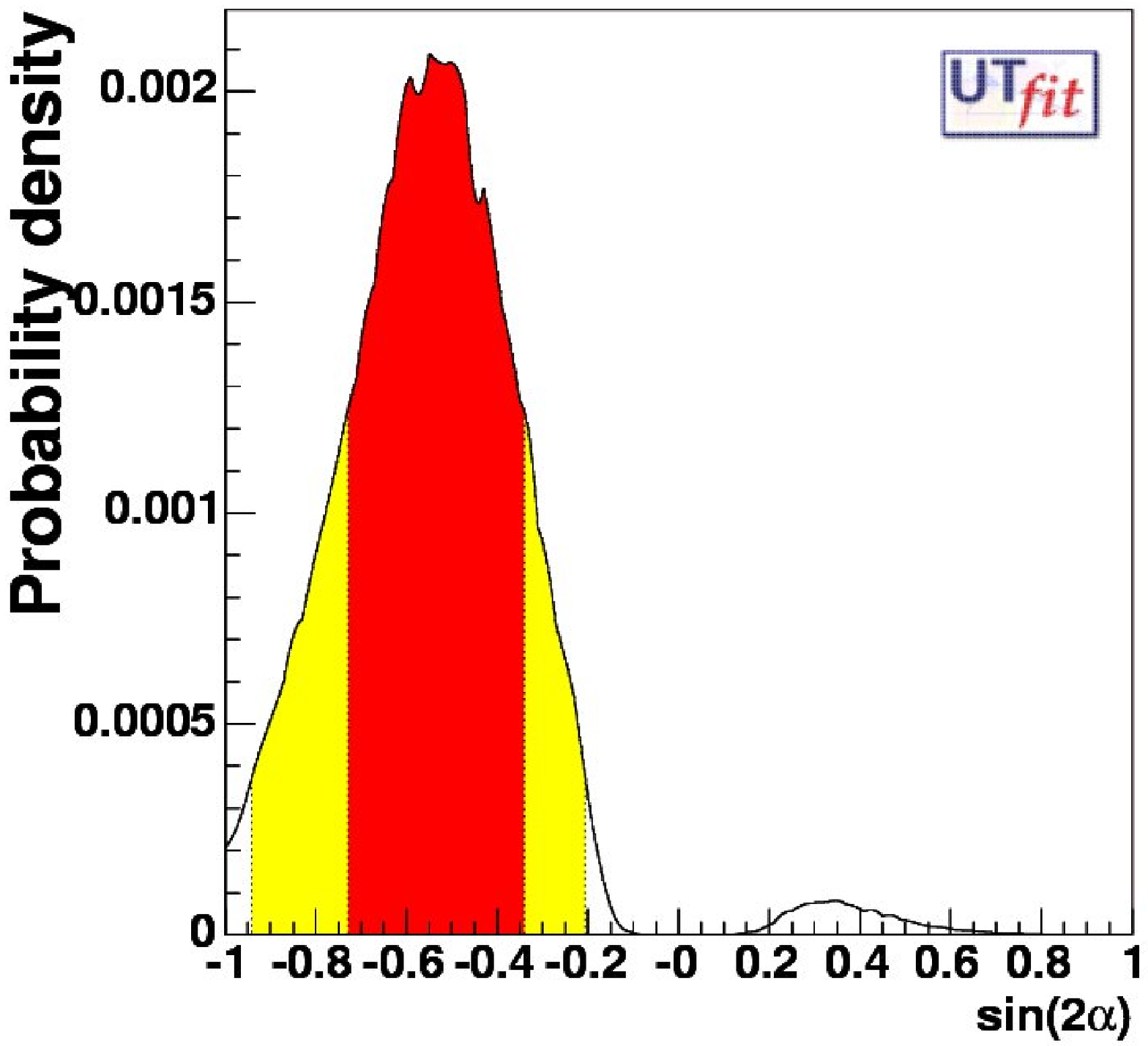}
\includegraphics[height=5.3cm]{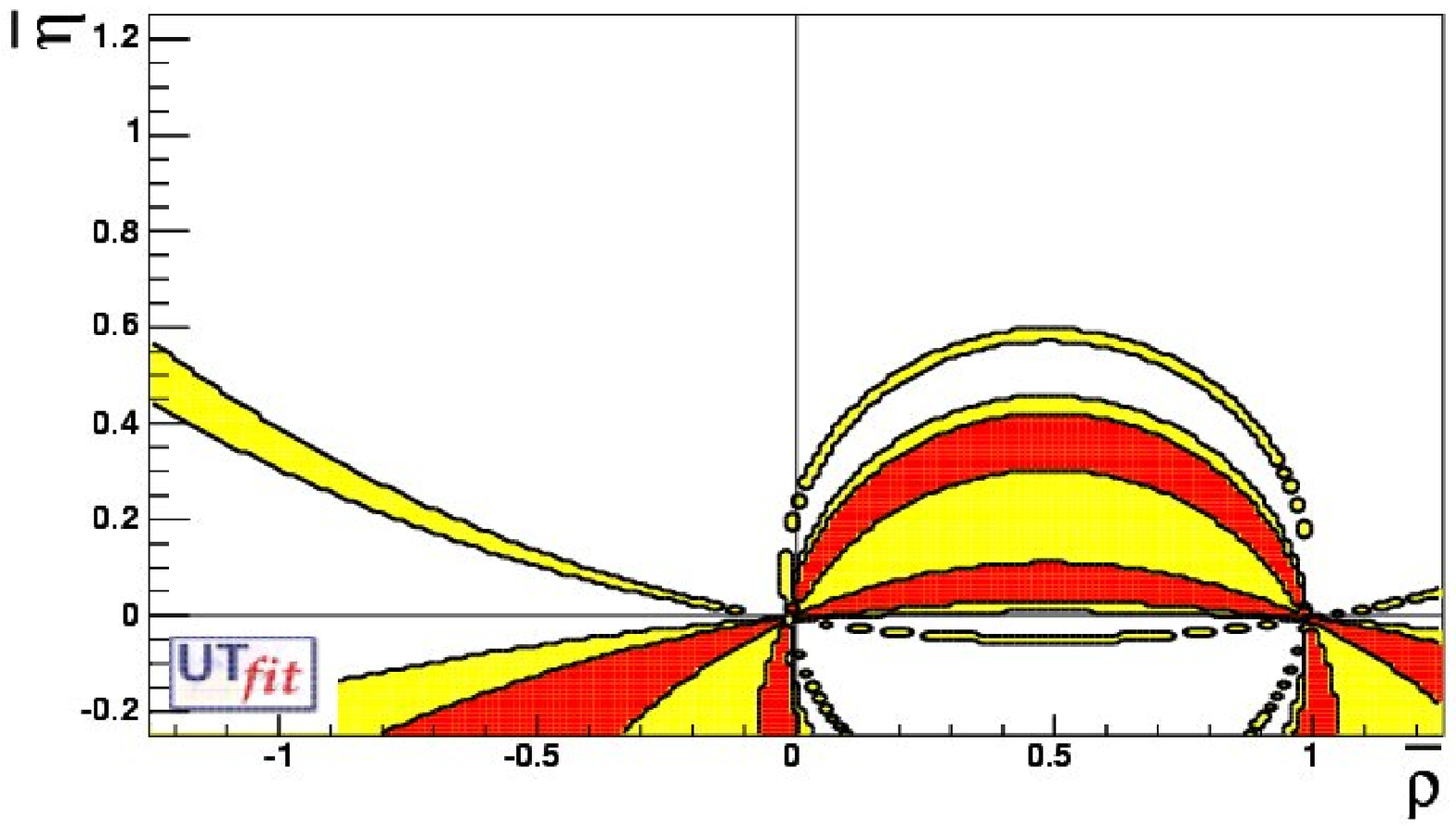}
\end{center}
\caption{{\it Distribution of sin(2$\alpha$) (left) and relative bound on
    $\rhobar-\etabar$ plane (right), using BaBar only (top) and World
    Average (bottom)~\cite{ref:hfag} values.}}
\label{fig:2alpha}
\end{figure}

In Figure~\ref{fig:2alpha} we show the results in terms of
$\sin{2\alpha}$ and of the allowed region in the $\rhobar-\etabar$
plane, from BaBar measurements only and using the world average.  It
has to be noted that the unphysical value found by Belle has a strong
impact on the selected area. On the other hand, the leading
contribution is given by $B \to \rho \rho$ decays.

The result we get can be compared to the indirect determination from the
standard fit:
\begin{eqnarray}
\sna & = & -0.16 \pm 0.26 ~[-0.62;0.35]~{\rm at} ~95\%~ C.L.   ~~\rm {indirect} 
   \nonumber \\
\sna & = & -0.24^{(+0.19)}_{(-0.24)} ~[-0.75;0.26]~{\rm at}  ~~95\%~ C.L.   
~~\rm {\pi\pi~and~\rho\rho~ BaBar} \nonumber  \\
\sna & = & -0.55^{(+0.21)}_{(-0.18)} ~[-0.94;-0.21]~{\rm at} ~~95\%~ C.L.   
~~\rm {\pi\pi ~and~\rho\rho~ (WA)}.
\label{eq:sin2alpha}
\end{eqnarray}

\begin{figure}[htb!]
\begin{center}
\includegraphics[height=4.2cm]{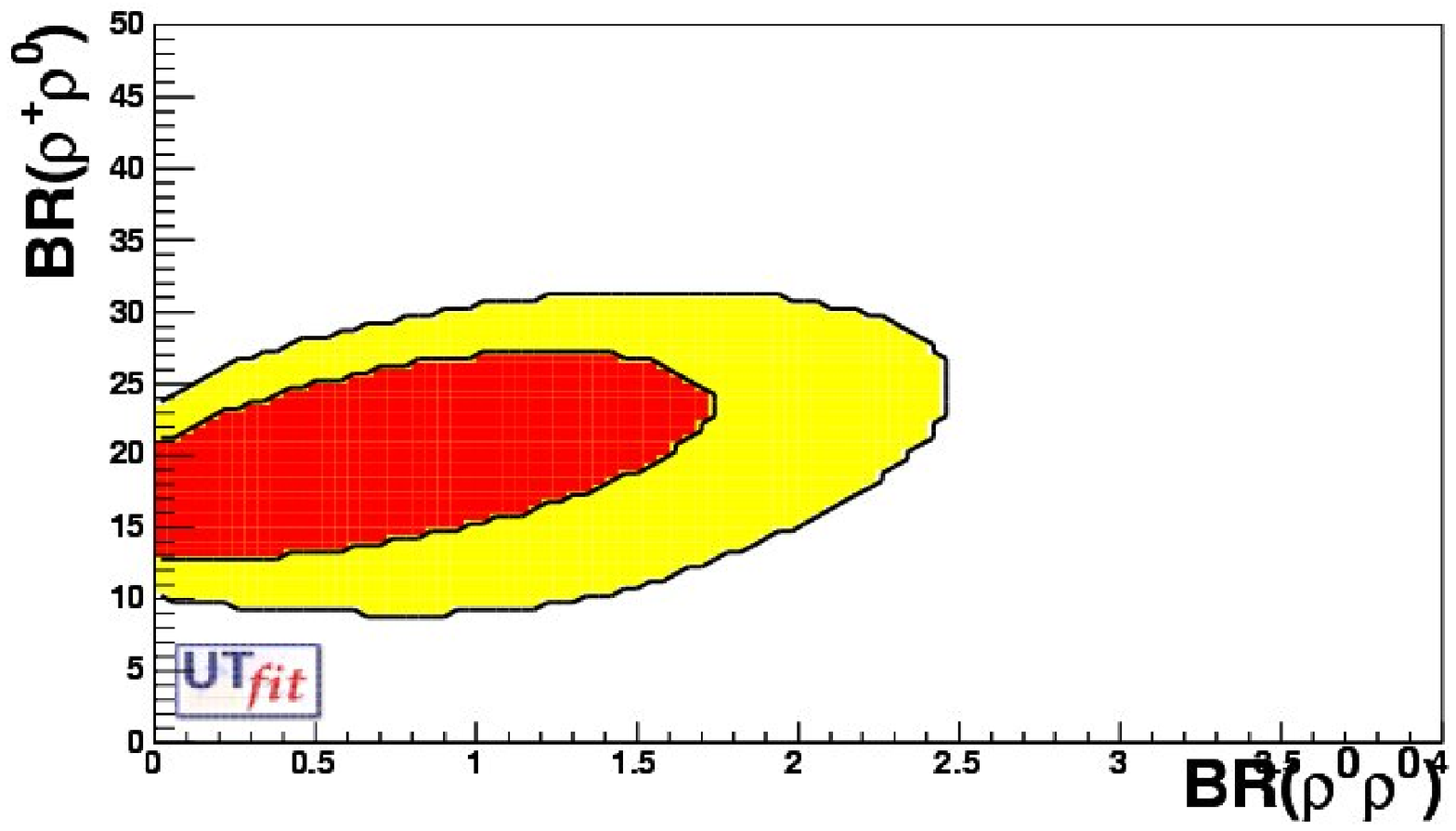}
\includegraphics[height=4.2cm]{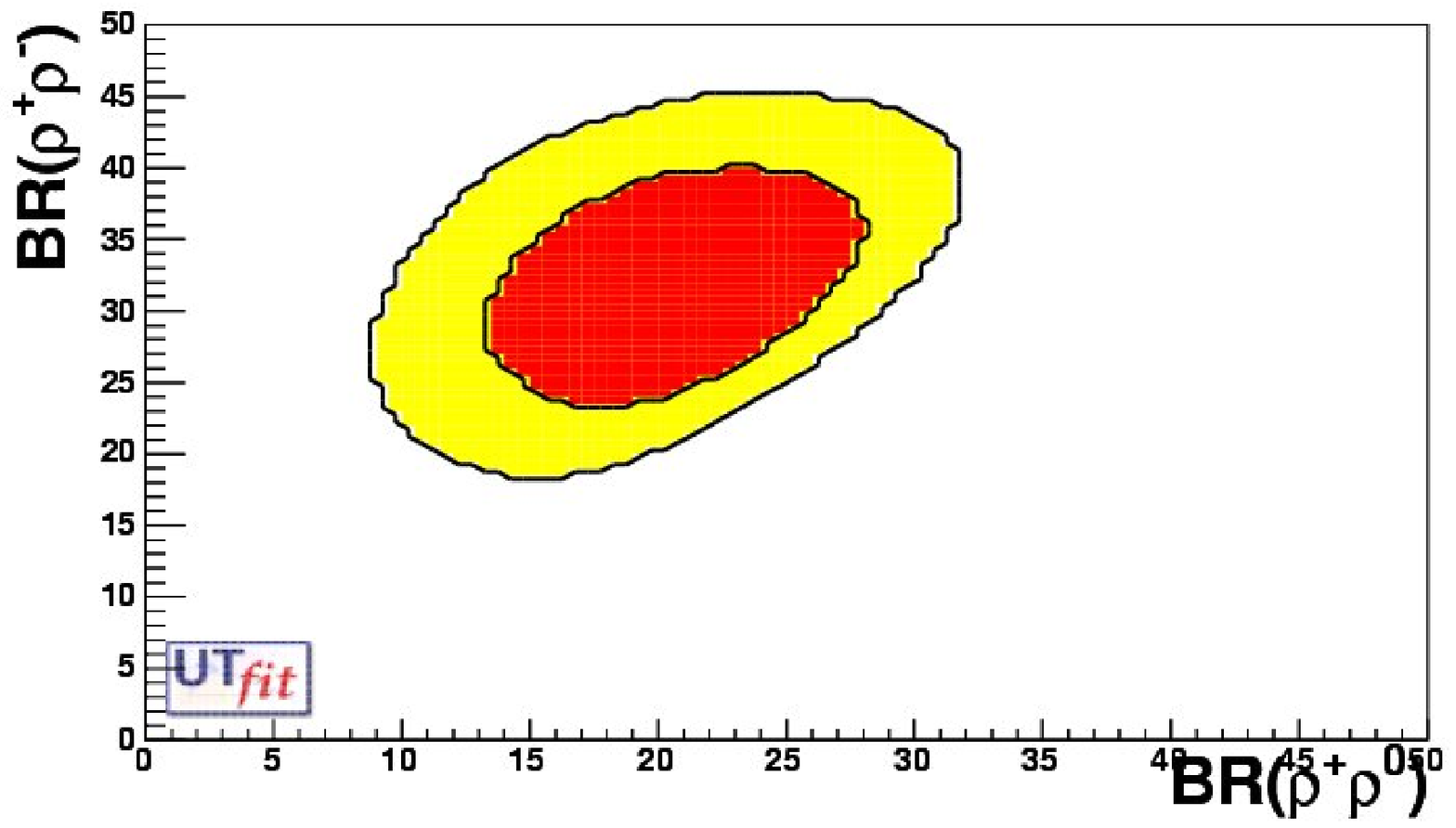} \\
\end{center}
\caption{{\it Plots showing the correlation of $BR(B^+ \to \rho^+ \rho^0)$ vs
    $BR(B^0 \to \rho^0 \rho^0)$(left) and $BR(B^0 \to \rho^+ \rho^-)$
    vs $BR(B^+ \to \rho^+ \rho^0)$(right) as obtained from SU(2)
    parameterizations in UTfit.}}
\label{fig:brrhorhp2D}
\end{figure}

It is important to stress the fact that the main assumptions we are
using here, i.e.  the validity of SU(2) flavour symmetry and the
absence of E.W. penguins.  can be directly tested in this framework
comparing the experimental and the fitted values of the Branching
Fractions (see Table~\ref{tab:2alpha-BRaposteriori}).  It is clear
that all experimental measurements of $B \to \pi \pi$ are in agreement
with the SU(2) assumption.  On the contrary, we observe from
Table~\ref{tab:2alpha-BRaposteriori} a disagreement between the fitted
and the experimental value of $BR(B^+ \to \rho^+ \rho^0)$.  This
discrepancy is shown in Figure~\ref{fig:brrhorhp2D} and, if confirmed
with increased accuracy, it would point towards a violation of the
assumptions on which the parameterization of
eqs.~(\ref{eq:su2analysis}) is based.

\begin{table}[htb]
  \centerline{\small
    \begin{tabular}{c|cc|cc}
     \hline\hline
                  &      \multicolumn{2}{c}{$\pi \pi$}  &
\multicolumn{2}{c}{$\rho \rho$} \\ 
     \hline     
     Observable   &      Average    &  UTfit &  Average &  UTfit \\
     \hline     
     $BR^{+-}(10^{-6})$ &  4.6   $\pm$ 0.4   &  4.6   $\pm$ 0.4   &  30.0  $\pm$
6.0  &  32.1  $\pm$ 5.5 \\
     $BR^{+0}(10^{-6})$ &  5.2   $\pm$ 0.8   &  5.2   $\pm$ 0.8   &  26.4  $\pm$
6.4  &  20.5  $\pm$ 4.8 \\
     $BR^{00}(10^{-6})$ &  1.9   $\pm$ 0.5   &  1.8   $\pm$ 0.5   &    0.6 $\pm$
0.8  &   0.7  $\pm$ 0.8 \\
     \hline\hline
    \end{tabular}}
  \caption{{\it Comparison of input and output values for the Branching
Fractions 
of $B \to \pi \pi$ and $B \to \rho \rho$ decay modes.}}
  \label{tab:2alpha-BRaposteriori}
\end{table}

\subsection{Determination of sin(2$\beta + \gamma$) using $D^{(*)} \pi$  events}
\label{ref:sin2bg}

$\sin(2\beta + \gamma)$ can be extracted from time-dependent
asymmetries in B decays to D$^{(*)}$ $\pi$ final states, looking at
the interference effects between the decay amplitudes implying $b
\rightarrow c$ and $b \rightarrow u$ transitions. The time-dependent
rates can be written as:
\begin{eqnarray}
R(B^0 \rightarrow D^{(*)-} \pi^+) = N~e^{-\Gamma t}~ {(1+C ~\cos(\Delta m_d t)
+S~ \sin(\Delta m_d t)~)}  \\ \nonumber
R(\overline{B}^0 \rightarrow D^{(*)-} \pi^+) = N~e^{-\Gamma t}~ {(1-C
~\cos(\Delta m_d t) -S~ \sin(\Delta m_d t)~)}   \\ \nonumber
R(B^0 \rightarrow D^{(*)+} \pi^-) = N~e^{-\Gamma t}~ {(1+C ~\cos(\Delta m_d t)
-\overline{S}~ \sin(\Delta m_d t)~)}   \\ \nonumber
R(\overline{B}^0 \rightarrow D^{(*)+} \pi^-) = N~e^{-\Gamma t}~ {(1-C
~\cos(\Delta m_d t) +\overline{S}~ \sin(\Delta m_d t)~)} 
\label{eq:betagamma}
\end{eqnarray}
where S and C parameters are defined as 
\begin{eqnarray}
S = 2~r/(1+r^2) \sin(2 \beta +\gamma-\delta)              \\ \nonumber
\overline{S} = 2~r/(1+r^2) \sin(2 \beta+\gamma+\delta)    \\ \nonumber
C = (1-r^2)/(1+r^2) 
\end{eqnarray}
and r and $\delta$ are the absolute value and the phase of the
amplitude ratio $A(\overline{B}^0 \rightarrow D^- \pi^+) / A(B^0
\rightarrow D^-\pi^+)$. This ratio $r$ is rather small being of the
order of $|V_{ub}/V_{cb}| \simeq$ 0.02.

There is a correlation between the tag side and the reconstruction
side in time dependent CP measurements at B-Factories~\cite{ref:owen}.
This is related to the fact that interference between $b \rightarrow
c$ and $b \rightarrow u$ transitions in $B \rightarrow D X$ decays can
occur also in the tag side.  $S$ and $\overline{S}$ entering the time
dependent rates can be replaced by
\begin{eqnarray}
a = 2 r \sin(2 \beta +\gamma) \cos(\delta)      \\ \nonumber
b = 2 r^{'} \sin(2 \beta +\gamma) \cos(\delta^{'})  \\ \nonumber
c = 2 \cos(2\beta +\gamma)(r \sin(\delta)-r^{'} \sin(\delta^{'})) 
\end{eqnarray}
where $r^{'}$ and $\delta^{'}$ are the analogue of $r$ and $\delta$
for the tag side.  It is important to stress that this interference on
the tag side cannot occur when B mesons are tagged using semileptonic
decays.  In other words, $r^{'}$= 0 when only semileptonic decays are
considered. In the following we will consider the observables $a$,
$c$(lepton), $a^*$ and $c^*$(lepton), which are functions of
$r^{(*)}$, $\delta^{(*)}$ and 2$\beta+\gamma$.

BaBar and Belle provided three different measurements of this channel,
with total (both)~\cite{dpi-tot} or partial (BaBar
only)~\cite{dpi-par} reconstruction of the final state, as summarized
in Table~\ref{tab:Dpi}.

\begin{table}[htbp!]
\begin{center}
\begin{tabular}{c|c} 
\hline\hline
 Parameters  &     HFAG average~\cite{ref:hfag}    \\ \hline
a            &  -0.038 $\pm$ 0.021  \\    
a$^*$        &   0.012 $\pm$ 0.030  \\  
c (lepton)          &  -0.041 $\pm$ 0.029  \\  
c$^*$   (lepton)      &  -0.015 $\pm$ 0.044  \\ \hline  \hline
\end{tabular}
\caption{ \it {Summary of the experimental results from BaBar and Belle, as
reported in~\cite{ref:hfag}.}}
\label{tab:Dpi}
\end{center}
\end{table}

\begin{figure}[htbp!]
\begin{center}
\includegraphics[width=5.cm]{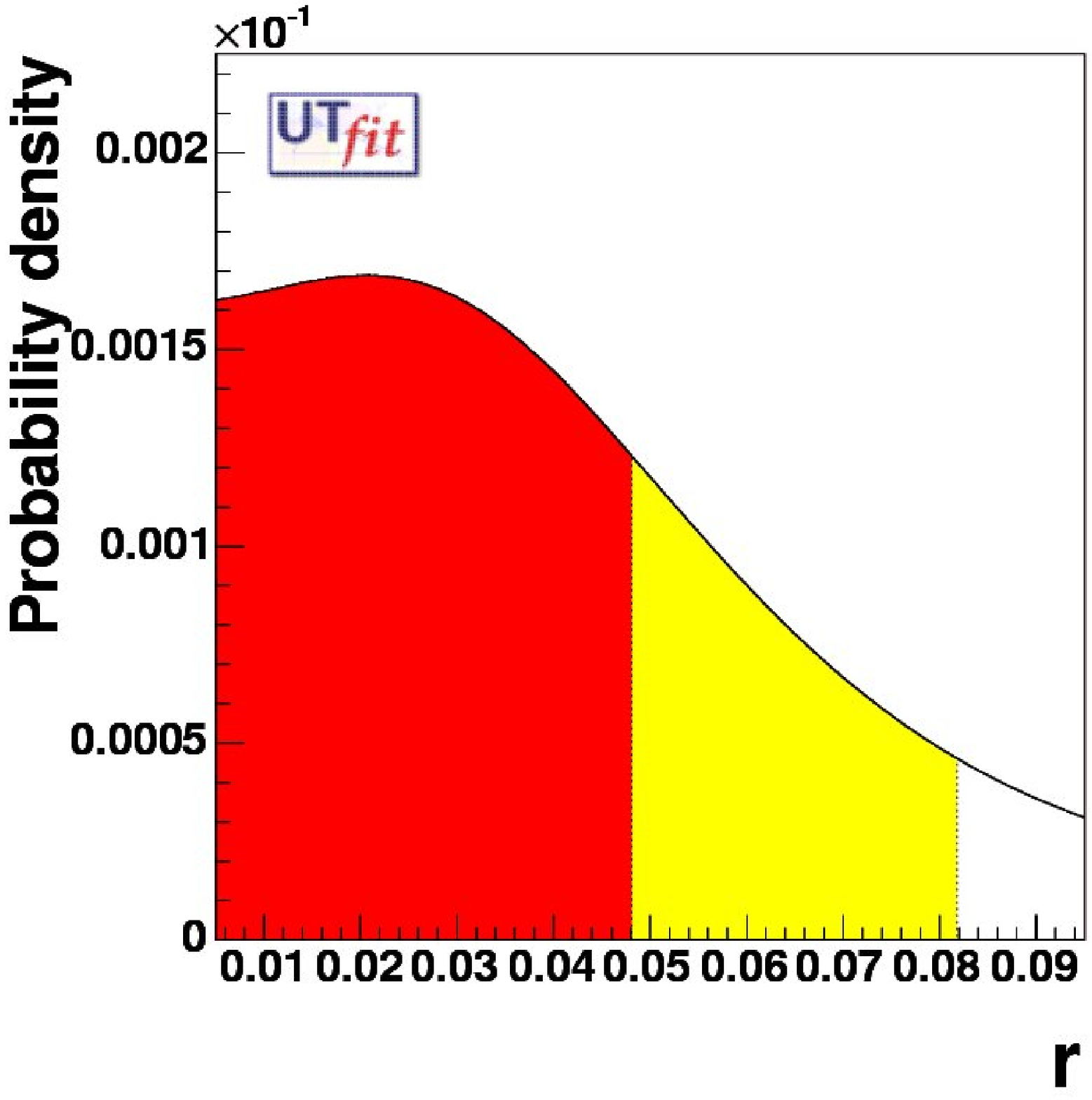}
\includegraphics[width=5.cm]{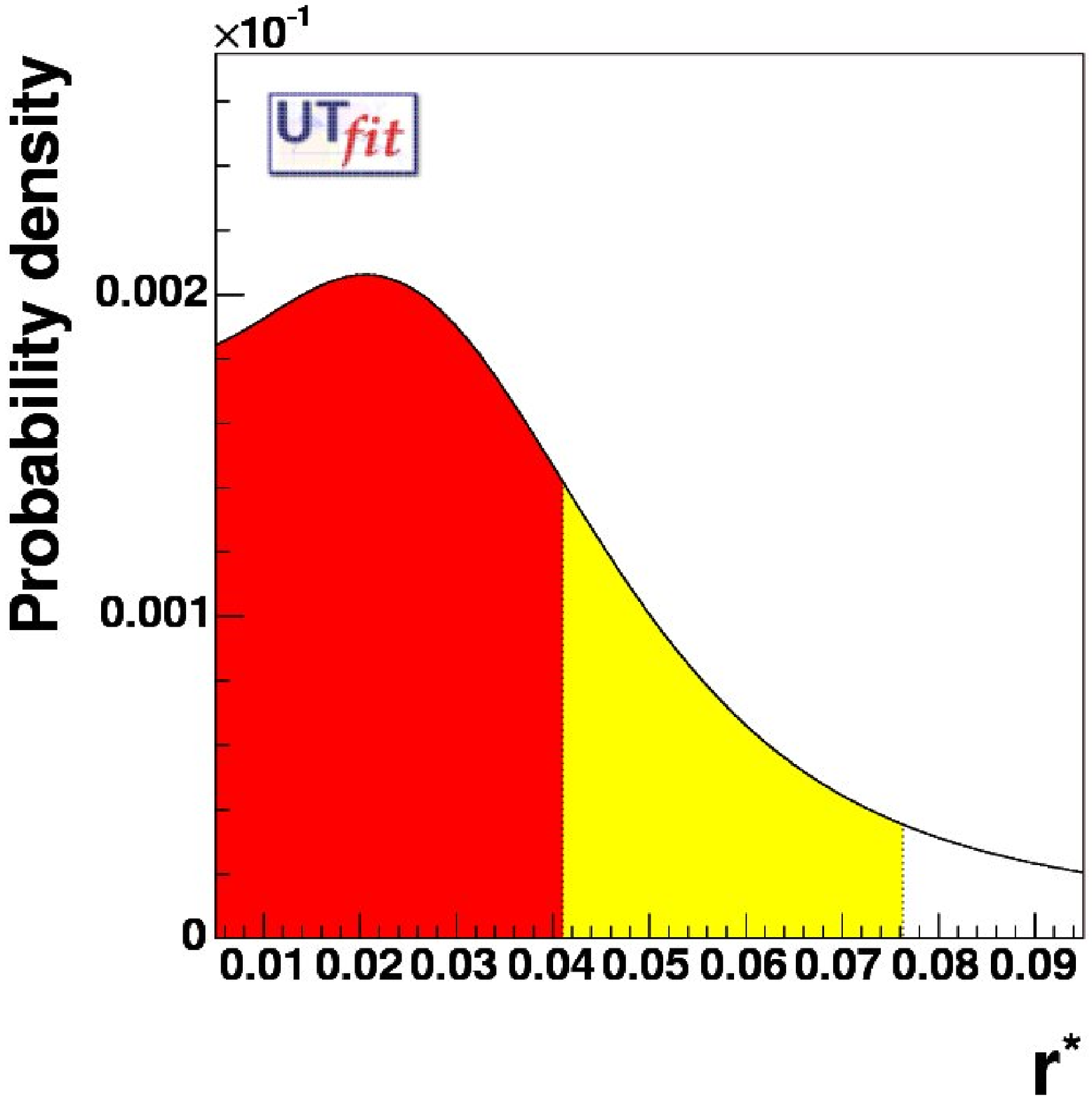}     \\
\includegraphics[width=5.cm]{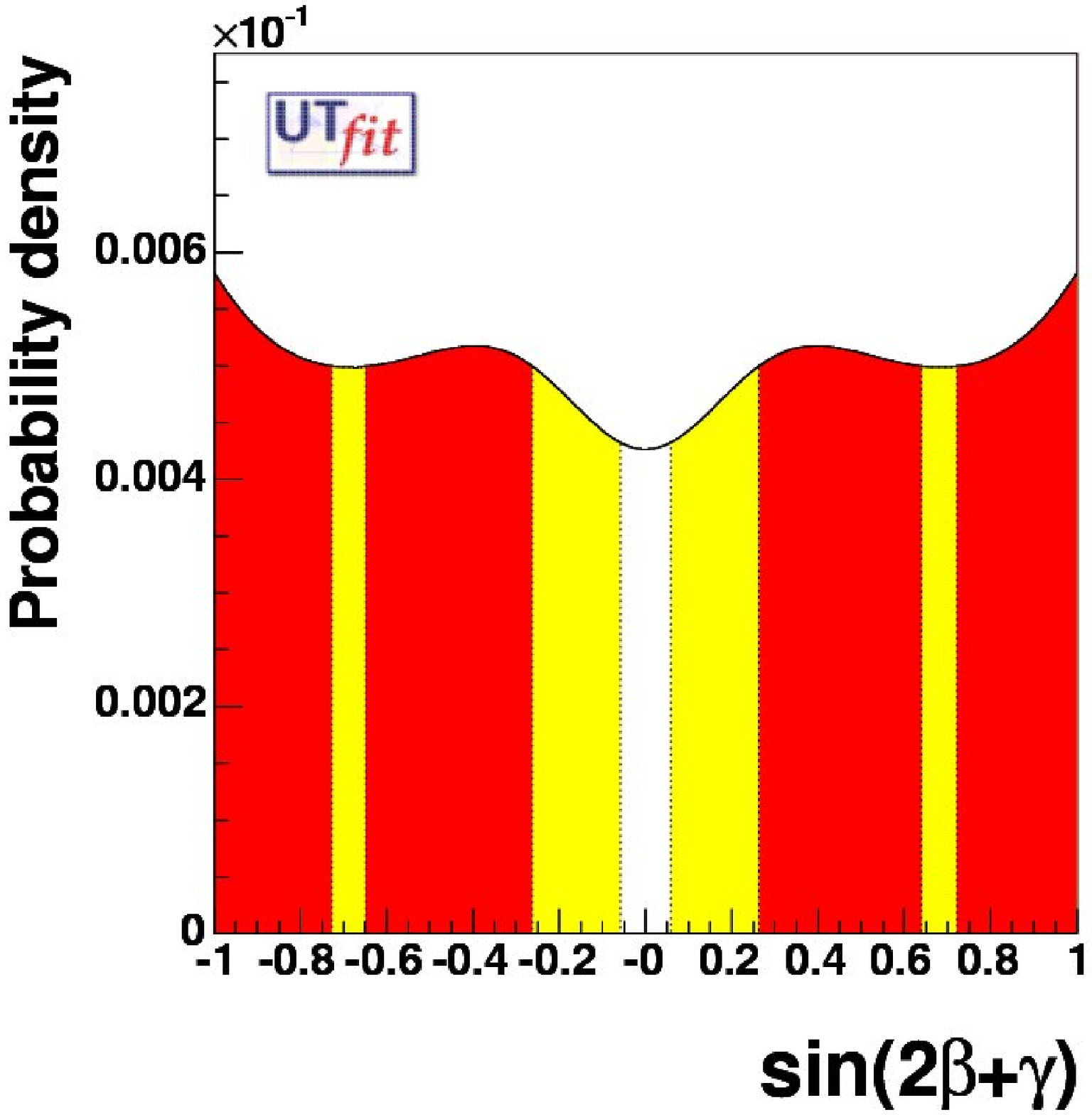}    
\includegraphics[width=5.cm]{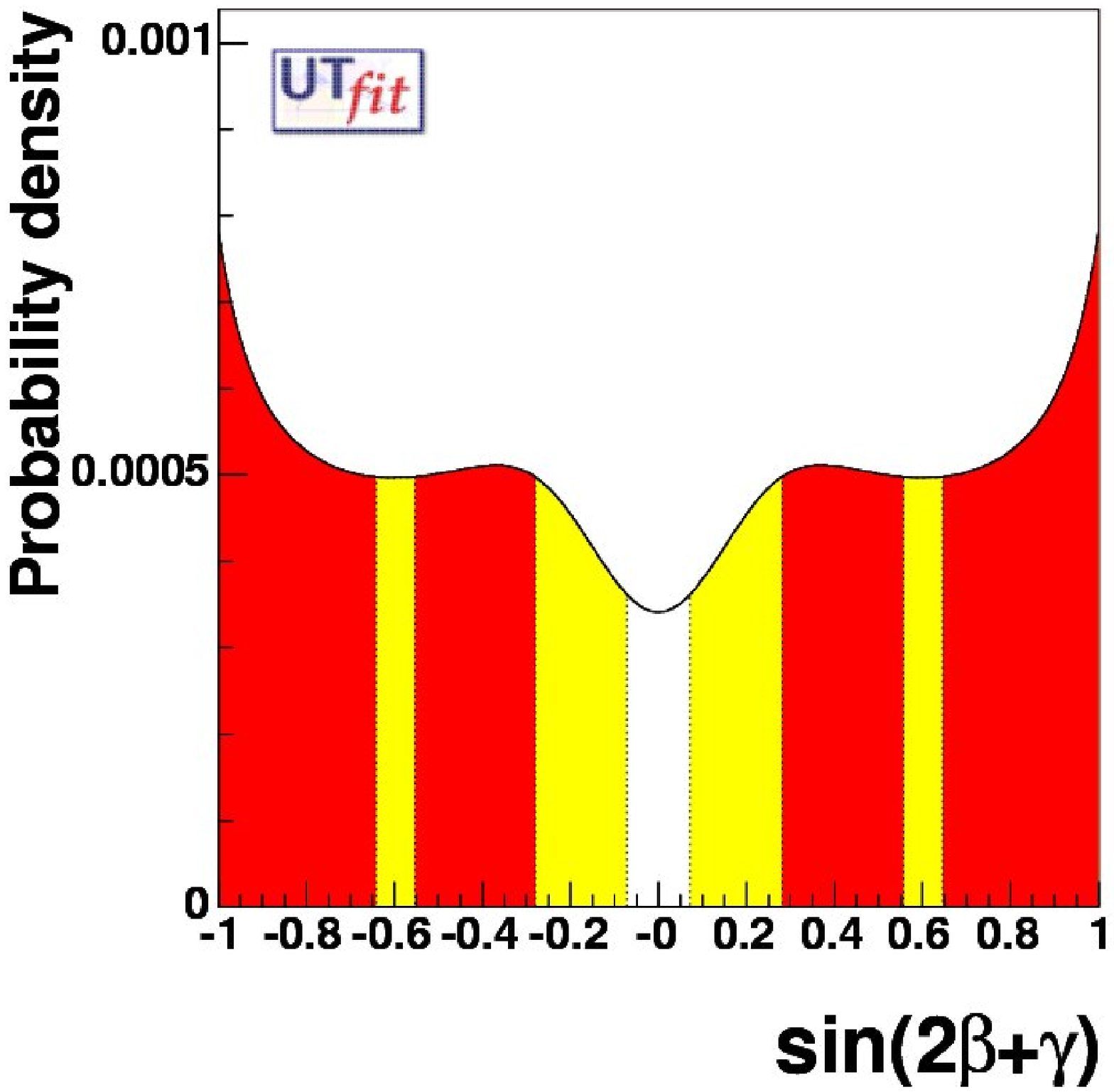} 
\includegraphics[width=5.cm]{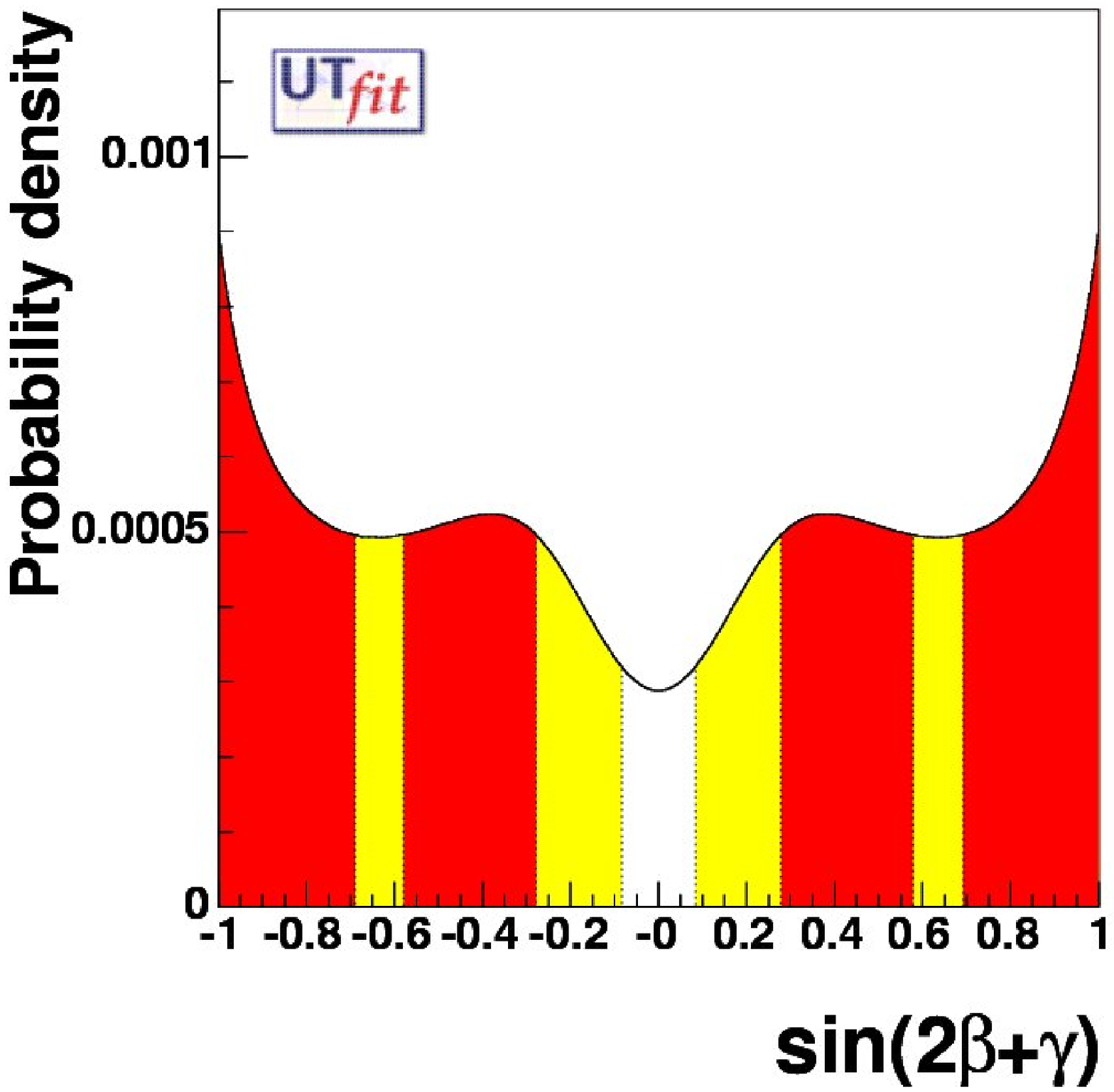} \\
\includegraphics[height=5.cm]{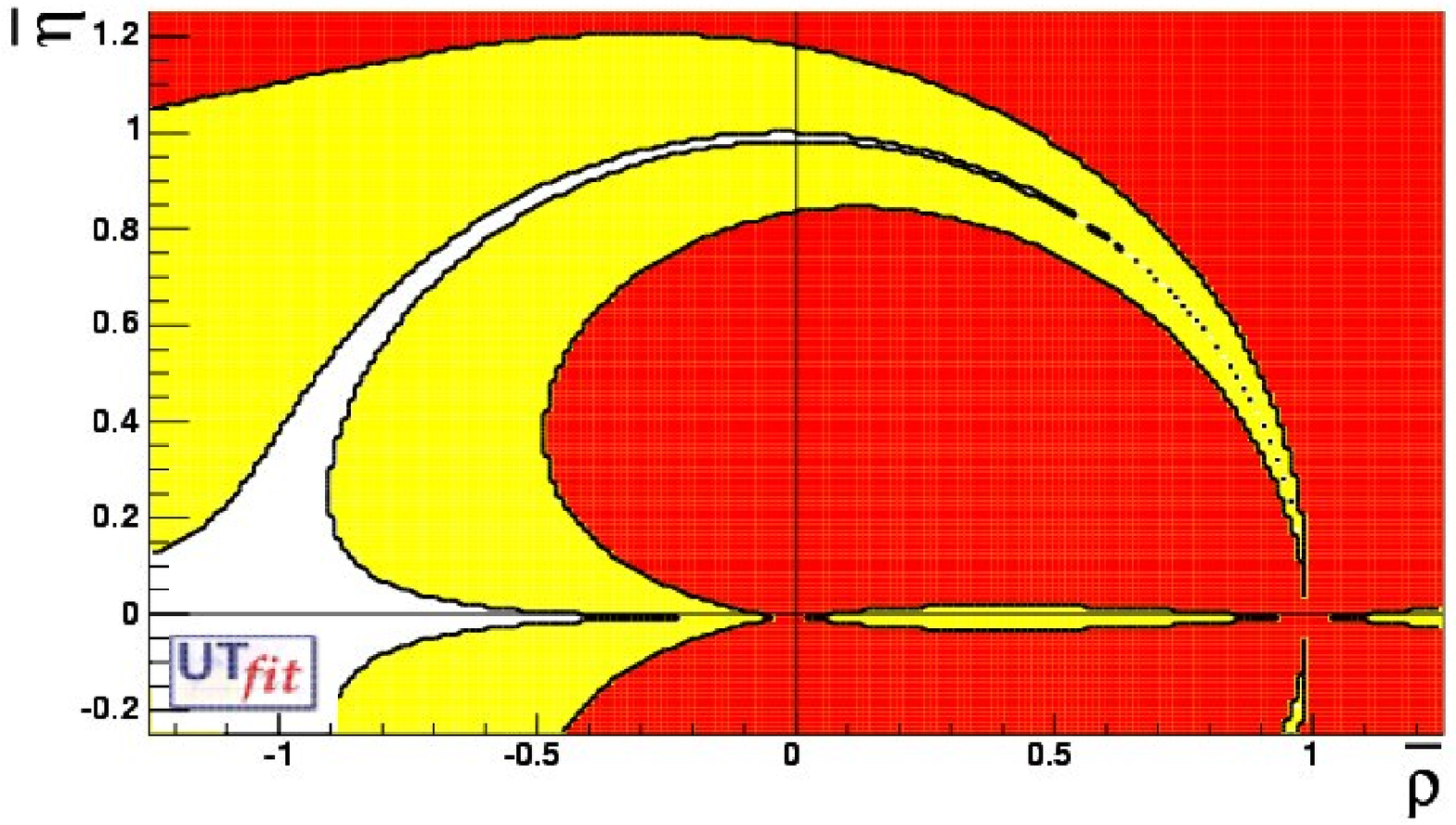}
\end{center}
\caption{ {\it Distributions for $r$ for $D\pi$ (top-left), 
    $r^*$ for $D^* \pi$ (top-right), $\sin(2\beta +\gamma)$ for $D\pi$
    (middle-left), $\sin(2\beta +\gamma)$ for $D^*\pi$ (middle-center)
    and $\sin(2\beta +\gamma)$ combined (middle-right) and bound from
    $\sin(2\beta +\gamma)$ on the $\rhobar-\etabar$ plane (bottom).}}
\label{fig:2bgdir}
\end{figure}

We use directly the four experimental quantities: $a^{(*)}$ and
$c^{(*)}$(lepton) defined before and we build a global p.d.f. as the
product of the p.d.f.'s of these four quantities.  We do not make any
assumption on $r$ and $r^*$ which are extracted in the range
[0.0,0.1].  The results on $r$, $r^*$, sin(2$\beta +\gamma$) and the
impact of this measurement in the $\rhobar-\etabar$ plane are shown in
Figure~\ref{fig:2bgdir}.

The comparison between the direct and the indirect determination of
sin$(2\beta + \gamma)$ is given below:
\begin{eqnarray}
\snbg &>&  ~~0.94 ~~{\rm at} ~68\%~ C.L. (> 0.88~{\rm at} ~95\%~ C.L.)~~\rm
{indirect}     \nonumber \\
\snbg &>&  ~~0.28 ~~{\rm at} ~68\%~ C.L. (> 0.08~{\rm at} ~95\%~ C.L.)~~ \rm
D^{(*)} \pi.
\label{eq:sin2bg}
\end{eqnarray}

\subsection{Determination of the Unitarity Triangle parameters using also the
new UT angle measurements}
\label{sec:newresults}

\begin{figure}[tbp]
\begin{center}
\includegraphics[width=14cm]{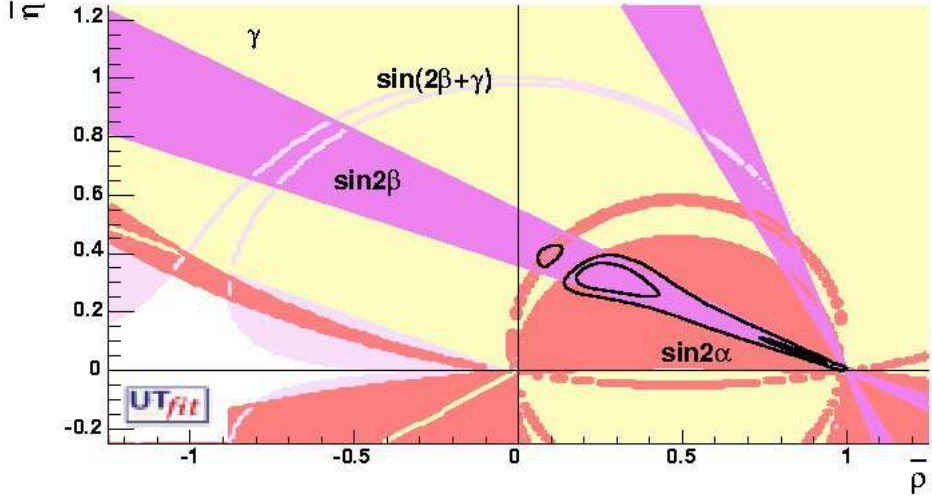}
\caption{ \it {Allowed regions for $\rhobar$ and $\etabar$ obtained using the
    measurements of the UT angles in the B sector: $\snb$, $\sna$,
    $\snbg$ and $\gamma$. The closed contours at $68\%$ and $95\%$
    probability are shown.  The full zones correspond to $95\%$
    probability regions from individual constraints.}}
\label{fig:soloangoli}
\end{center}
\end{figure}

It is interesting to see the selected region in $\rhobar-\etabar$
plane from the measurements of the UT angles in the B sector. The plot
is shown in Figure~\ref{fig:soloangoli}. In Table~\ref{tab-bfactory}
we report the results we get using these constraints.

\begin{figure}[htbp!]
\begin{center}
\includegraphics[width=14cm]{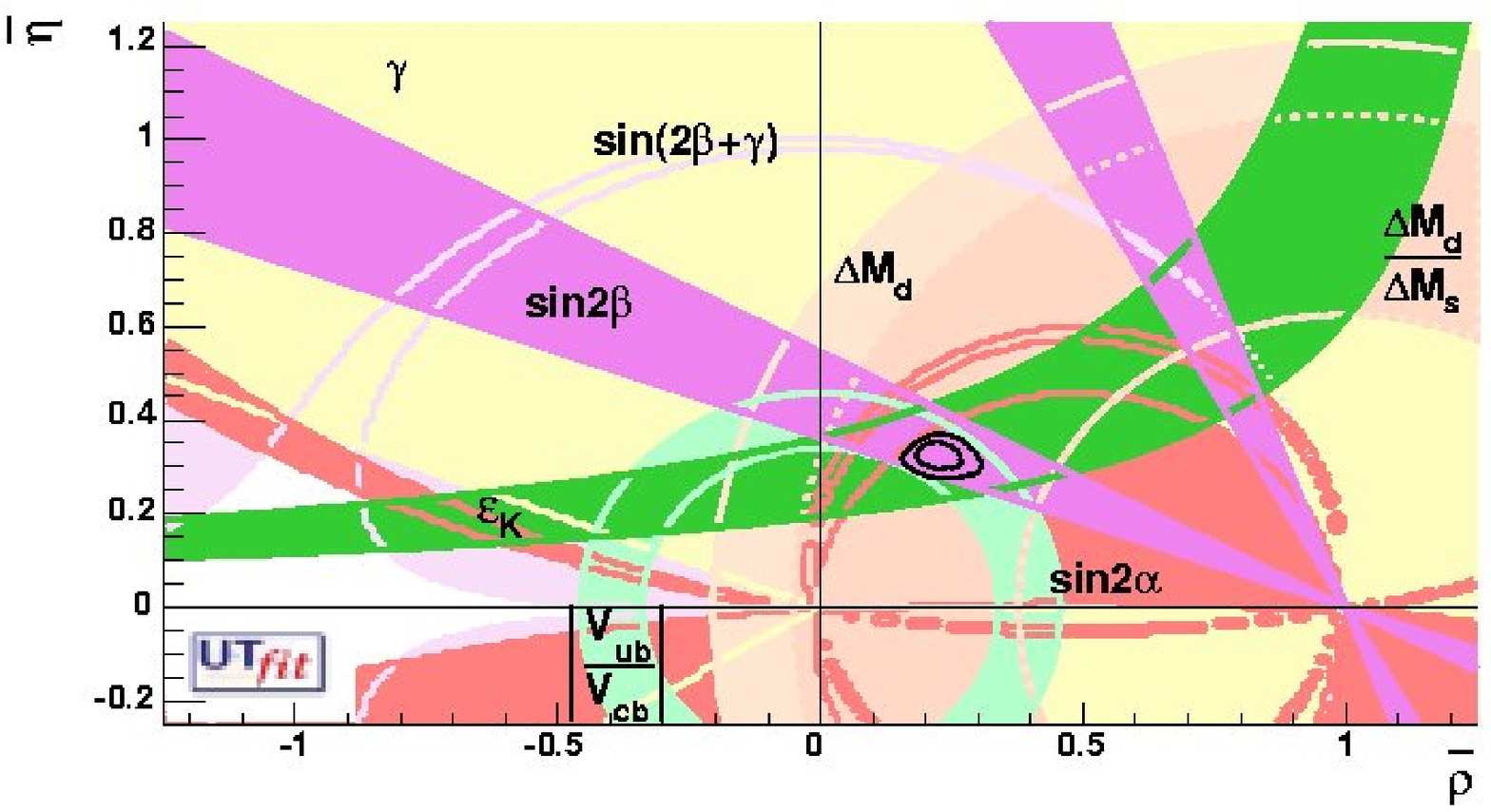}
\caption{ \it {Allowed regions for $\rhobar$ and $\etabar$ using the parameters
    listed in Table~\ref{tab:inputs}.  The closed contours at $68\%$
    and $95\%$ probability are shown. The full lines correspond to
    $95\%$ probability regions for the constraints, given by the
    measurements of $\left | V_{ub} \right |/\left | V_{cb} \right |$,
    $\epsilonk$, $\Delta m_d$, $\Delta m_s$ and $\snb$,$\gamma$,
    $\snbg$ and $\sna$.}}
\label{fig:allall}
\end{center}
\end{figure}

\begin{table*}[h]
\begin{center}
\begin{tabular}{@{}llllll}
\hline\hline  
    Parameter  &     ~~~~~68$\%$   &      ~~~~~95$\%$     &    ~~~~~99$\%$     
\\ 
\hline 
~~~~~$\overline {\eta}$  & 0.265$^{+0.120}_{-0.070}$   & [0.165;0.869]   &
[0.052;0.980] \\
~~~~~$\overline {\rho}$  & 0.315$^{+0.40}_{-0.51}$     & [0.040;0.378]   &
[0.004;0.437] \\
          ~$\snb$        & 0.733  $\pm$ 0.049          & [0.636;0.828]   &
[0.606;0.858] \\
          ~$\sna$        & -0.66  $\pm$ 0.26           &    $<$0.48      &    
$<$0.59  \\
~~~~$\gamma[^{\circ}$]   & 50.0$^{+9.2}_{-14.9}$       &    $<$61.5      &    
$<$79.6  \\
$Im {\lambda}_t$[$10^{-5}$] &12.1   $\pm$ 1.6          & [5.8,14.3]      &  
[5.1,16.0]   \\
\hline
\hline
\end{tabular} 
\end{center}
\caption {\it Values and probability ranges for the Unitarity Triangle
parameters obtained by using 
$\snb$, $\gamma$, $\snbg$ and $\sna$.}
\label{tab-bfactory} 
\end{table*}

The results given in Table~\ref{tab:1dimangoli} are obtained using all
the available constraints: $\left | V_{ub} \right |/\left | V_{cb}
\right |$, $\Delta {m_d}$, $\Delta {m_s}/\Delta {m_d} $, $\epsilonk$,
$\snb$, $\gamma$, $\snbg$ and $\sna$.  Figure~\ref{fig:allall} shows
the corresponding selected region in the $\rhobar-\etabar$ plane.

\begin{table*}[h]
\begin{center}
\begin{tabular}{@{}llllll}
\hline\hline  
    Parameter  &     ~~~~~68$\%$   &      ~~~~~95$\%$     &    ~~~~~99$\%$     
\\ 
\hline
~~~~~$\overline {\eta}$     & 0.324  $\pm$ 0.020          &[0.283,0.359]
&[0.268,0.373] \\
~~~~~$\overline {\rho}$     & 0.225  $\pm$ 0.030      &[0.171,0.288]
&[0.151,0.326] \\
          ~$\snb$           & 0.710  $\pm$ 0.032      &[0.645,0.773]
&[0.624,0.792] \\
          ~$\sna$           &-0.44$^{+0.17}_{-0.09}$  &[-0.68,-0.18]
&[-0.81,-0.13] \\
~~~~$\gamma[^{\circ}$]      & 53.9$^{+5.5}_{-2.4}$    &[46.6,62.4] &[41.1,64.6]
\\
Im ${\lambda}_t$[$10^{-5}$] & 12.4   $\pm$ 0.8        &[11.0,14.0] &[10.5,14.5]
\\
\hline
\hline
\end{tabular} 
\end{center}
\caption {\it Values and probability ranges for the Unitarity Triangle
parameters obtained by using all 
the available constraints: $\left | V_{ub} \right |/\left | V_{cb} \right |$, 
$\Delta {m_d}$, $\Delta {m_s}/\Delta {m_d} $, $\epsilonk$ and $\snb$, $\gamma$,
$\snbg$ and $\sna$.}
\label{tab:1dimangoli} 
\end{table*}

\section {Compatibility plots, or how to discover New Physics in the flavour
sector}
\label{ref:pull}

In this section we discuss the interest of measuring the various
physical quantities entering the UT analysis with a better precision.
We investigate, in particular, to which extent future and improved
determinations of the experimental constraints, such as
$\sin{2\beta}$, $\dms$ and $\gamma$, could allow us to possibly
invalidate the SM, thus signaling the presence of NP effects.

\subsection{Compatibility between individual constraints. The pull
distributions.}
\label{sec:compa}

In CKM fits based on a $\chi^2$ minimization, a conventional
evaluation of compatibility stems automatically from the value of the
$\chi^2$ at its minimum.  The compatibility between constraints in the
Bayesian approach is simply done by comparing two different p.d.f.'s.
For example, compare the value for $\snb$ obtained from the
measurement of the sides of the Unitarity Triangle (the random
variable $\bf x_1$) with the one obtained from the direct measurement
of the CP violation asymmetry (the random variable $\bf x_2$).  In
this case the distribution of the random variable
$\bf{y}=\bf{x_1}-\bf{x_2}$ has to be constructed and the integral of
this distribution above (or below) zero gives the one side probability
of compatibility\footnote{In the Gaussian case it coincides with the
  pull which is defined as the difference between the central values
  of the two distributions divided by the sum in quadrature of the
  r.m.s of the distributions themselves.}.  The advantage of this
approach is that the full overlap between the p.d.f.'s
is evaluated instead of a single number.\\
If two constraints turn out to be incompatible, further investigation
is necessary to tell if this originates from a ``wrong'' evaluation of
the input parameters or from a New Physics contribution.

\subsection{Pull distribution for $\sin{2\beta}$. Role of $\snb$ from Penguin
  processes.}  
\begin{figure}[htbp!]
\hbox to\hsize{\hss
\includegraphics[width=0.5\hsize]{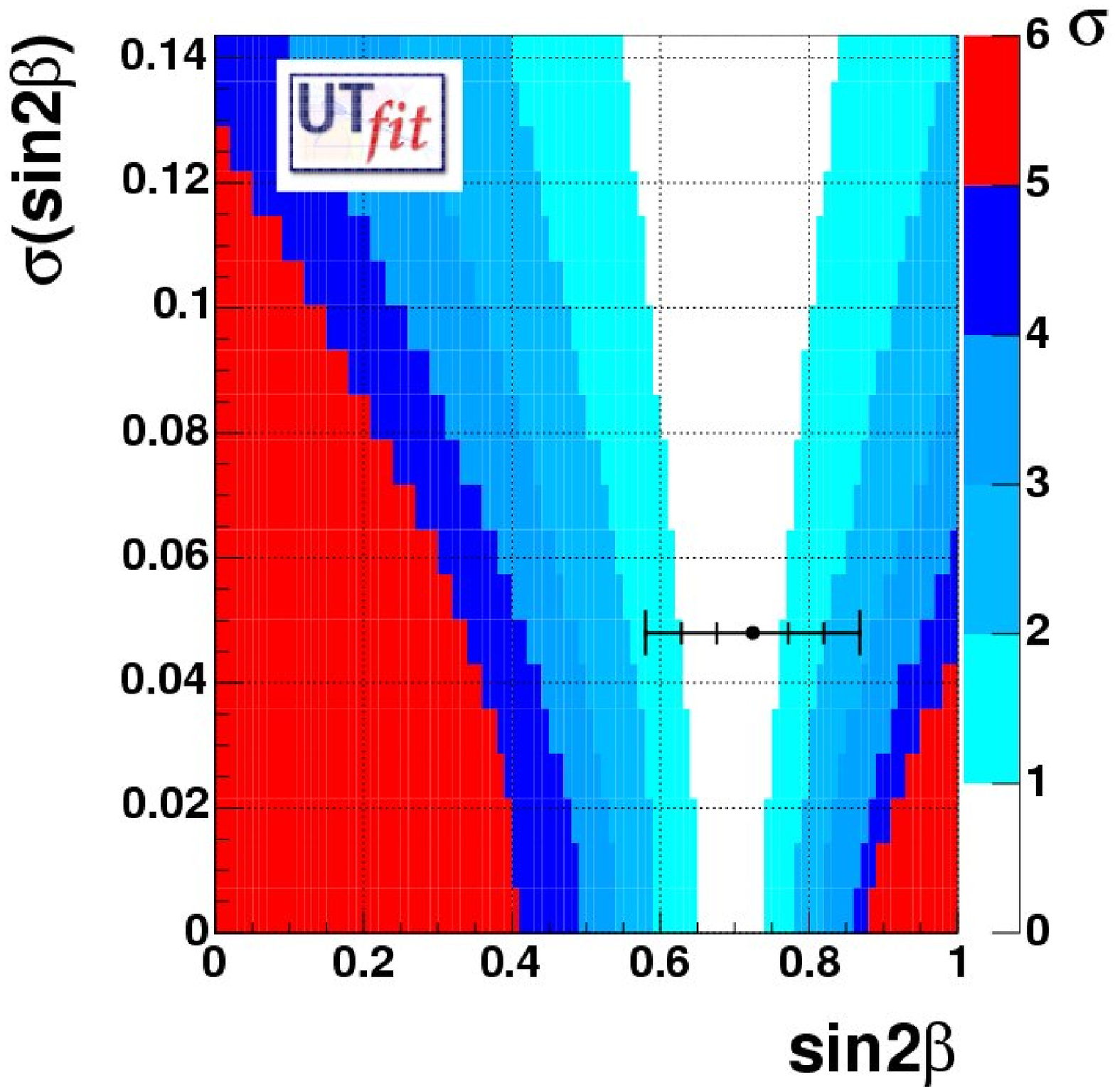}
\includegraphics[width=0.5\hsize]{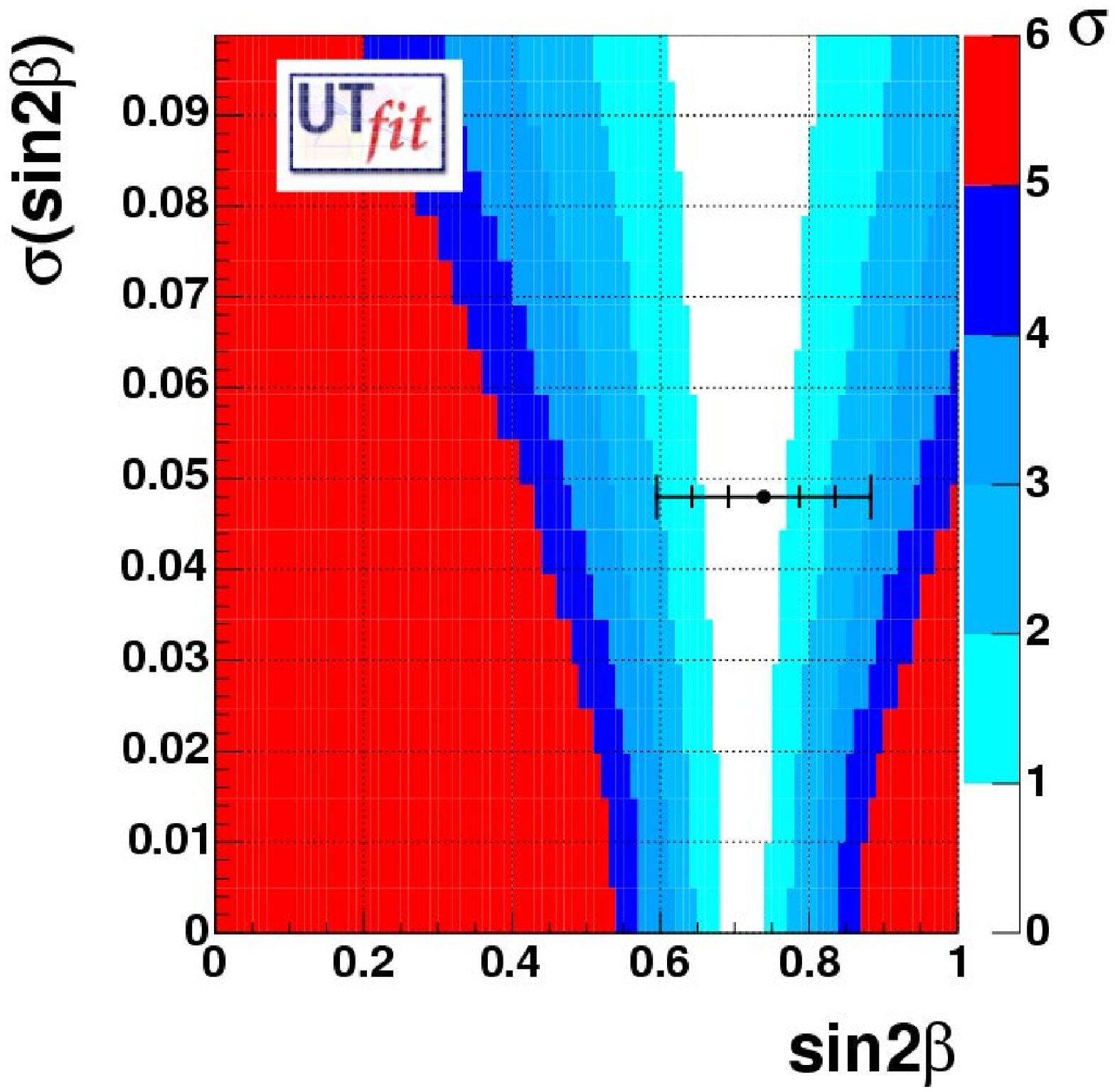}
\hss}
\caption{\it {The compatibility (``pull'') between the direct and indirect 
    determination of $\sin{2\beta}$ as a function of the value and
    error of $\sin{2\beta}$ measured from CP asymmetry in $J/\psi K^0$
    decays. The indirect distribution of $\sin{2\beta}$ is computed
    without using the direct measurement (left plot) or using the
    measurement of $\sin{2\beta}$ from CP asymmetry in $J/\psi K^0$
    decays (right plot).  The compatibility regions from 1 to 6
    $\sigma$ are also displayed.}}
\label{fig:pull_sin2b}
\end{figure}

We start this analysis by considering the measurement
of sin2$\beta$. The plots in Figure~\ref{fig:pull_sin2b} show the
compatibility (``pull'') between the direct and indirect distributions
of $\sin{2\beta}$, in the SM, as a function of the measured value
(x-axis) and error (y-axis) of $\sin{2\beta}$.

>From the left plot in Figure~\ref{fig:pull_sin2b}, it can be seen
that, considering the actual precision of about 0.05 on the measured
value of $\sin{2\beta}$, the 3$\sigma$ compatibility region is between
[0.49-0.87].  Values outside this range would be, therefore, not
compatible with the SM prediction at more than $3\sigma$ level. To get
these values, however, the presently measured central value should
shift by more than $4\sigma$.

The conclusion that can be derived from Figure~\ref{fig:pull_sin2b} is
the following: although the improvement of the error on sin2$\beta$
has an important impact on the accuracy of the UT parameter
determination, it is very unlikely that in the near future we will be
able to use this measurement to detect any failure of the SM, unless
the other constraints entering the fit improve substantially or, of
course, in case the central value of the direct measurement move away
from the present one by several standard deviations.

The right plot in Figure~\ref{fig:pull_sin2b} shows the compatibility
of the direct and indirect distributions of $\sin{2\beta}$ as a
function of the measured value and error of $\sin{2\beta}$. The
difference with respect to the left plot is that, in this case, all
the available constraints have been used to obtain the indirect
distribution of $\sin{2\beta}$, including the direct measurement of
$\sin{2\beta}$ from $J/\psi K^0$.

It was pointed out some time ago that the comparison of the
time-dependent CP asymmetries in various $B$ decay modes could provide
evidence of NP in B decay amplitudes~\cite{thphi}. Since
$\sin{2\beta}$ is known from $J/\psi K^0$, a significant deviation of
the time-dependent asymmetry parameters of penguin dominated channels
$B^0 \to \phi K^0$ and $B^0 \to K^0 \pi^0$ from their expected values
would indicate the presence of NP.

These asymmetries have been recently measured at
B-Factories~\cite{ref:phik01} and are reported in
Table~\ref{tab:phiks}.

\begin{table}[tb]
  \centerline{\small
    \begin{tabular}{c|ccc}
     \hline\hline
     Observable   &                         BaBar           &                   
Belle           &     Average\\
     \hline     
     $S_{\phi K^0_S}$               &   0.47 $\pm$ 0.34$^{+0.08}_{-0.06}$ &
-0.96 $\pm$ 0.50$^{+0.09}_{-0.11}$ & 0.02  $\pm$ 0.29    \\    
     $C_{\phi K^0_S}$               &   0.01 $\pm$ 0.33 $\pm$ 0.10        & 
0.15 $\pm$ 0.29 $\pm$ 0.08        & 0.09  $\pm$ 023     \\   
        \hline
     $S_{K^0_S \pi^0}$              &   0.48$^{+0.38}_{-0.47}$ $\pm$ 0.11 & - & 
 0.48$^{+0.38}_{-0.47}$ $\pm$ 0.11 \\            
     $C_{K^0_S \pi^0}$              &   0.40$^{+0.27}_{-0.28}$ $\pm$ 0.10 & - & 
 0.40$^{+0.27}_{-0.28}$ $\pm$ 0.10 \\       
     \hline\hline
    \end{tabular}}
  \caption{{\it Experimental inputs for S and C  in $B^0 \to \phi K^0_S$ and
$B^0 \to K^0_S \pi^0$
           decays~\cite{ref:hfag}.}}
  \label{tab:phiks}
\end{table}

\begin{figure}[htb!]
\begin{center}
\includegraphics[width=6.5cm]{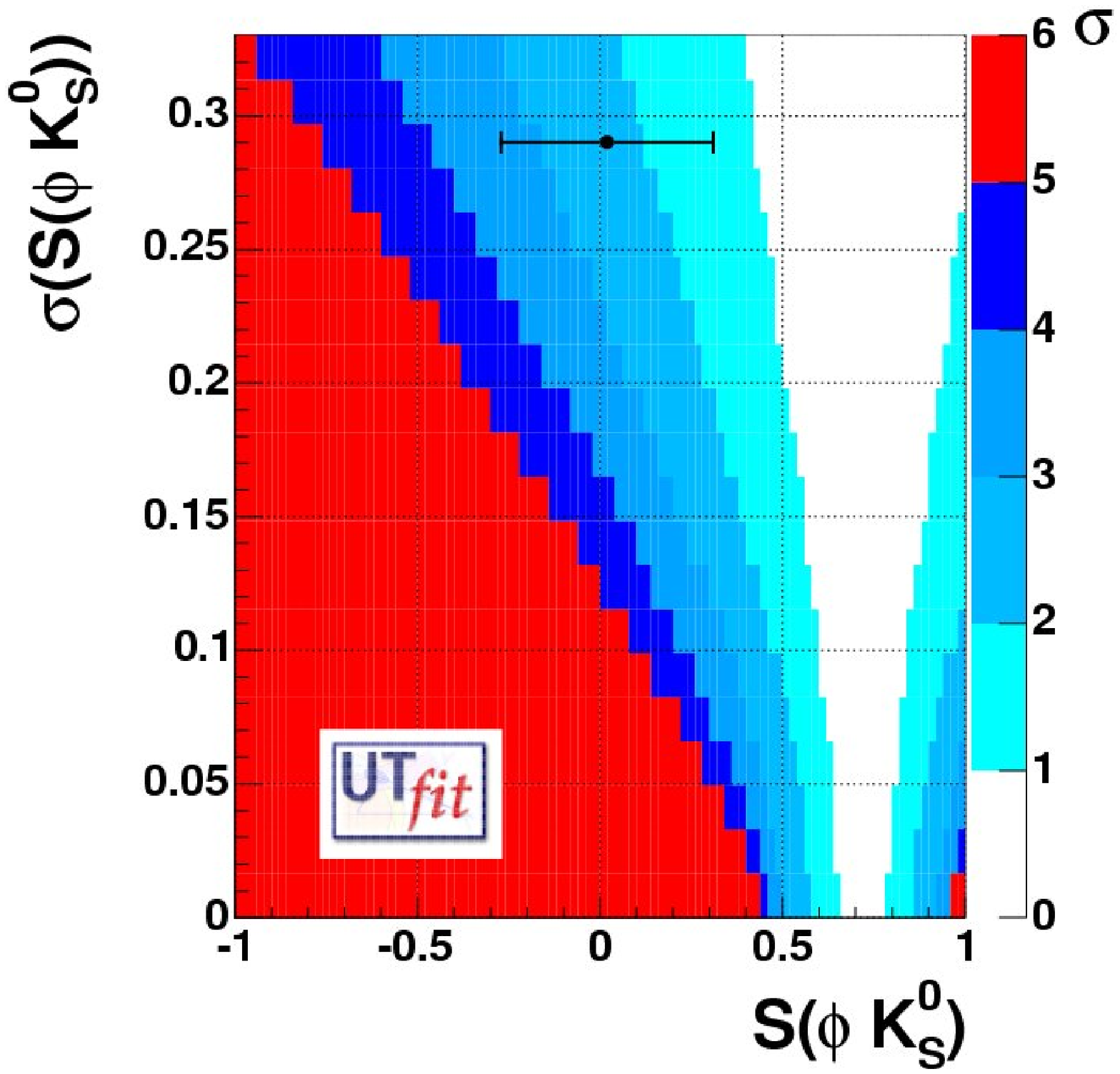}
\includegraphics[width=6.5cm]{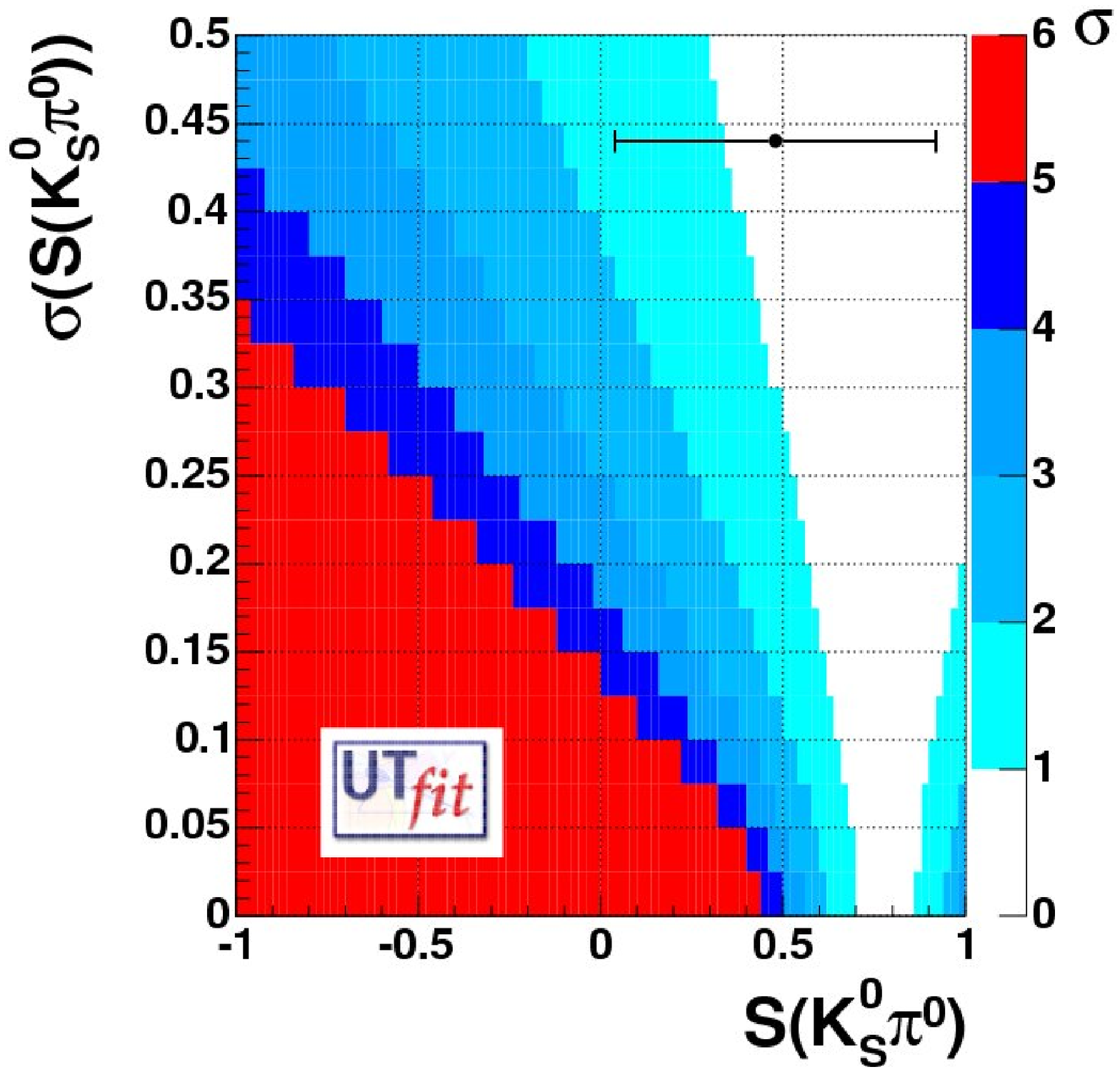}\\
\includegraphics[width=6.5cm]{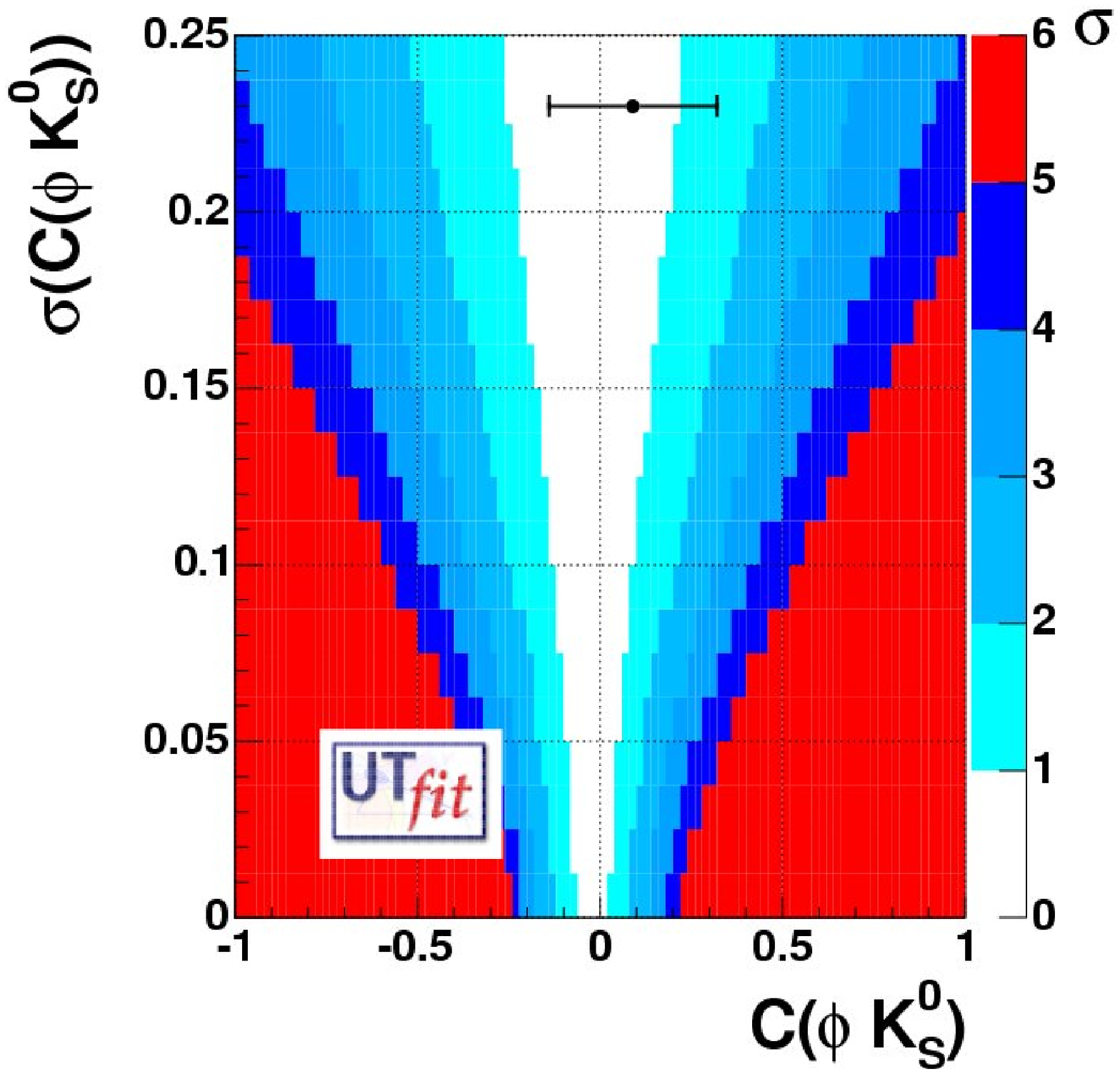}
\includegraphics[width=6.5cm]{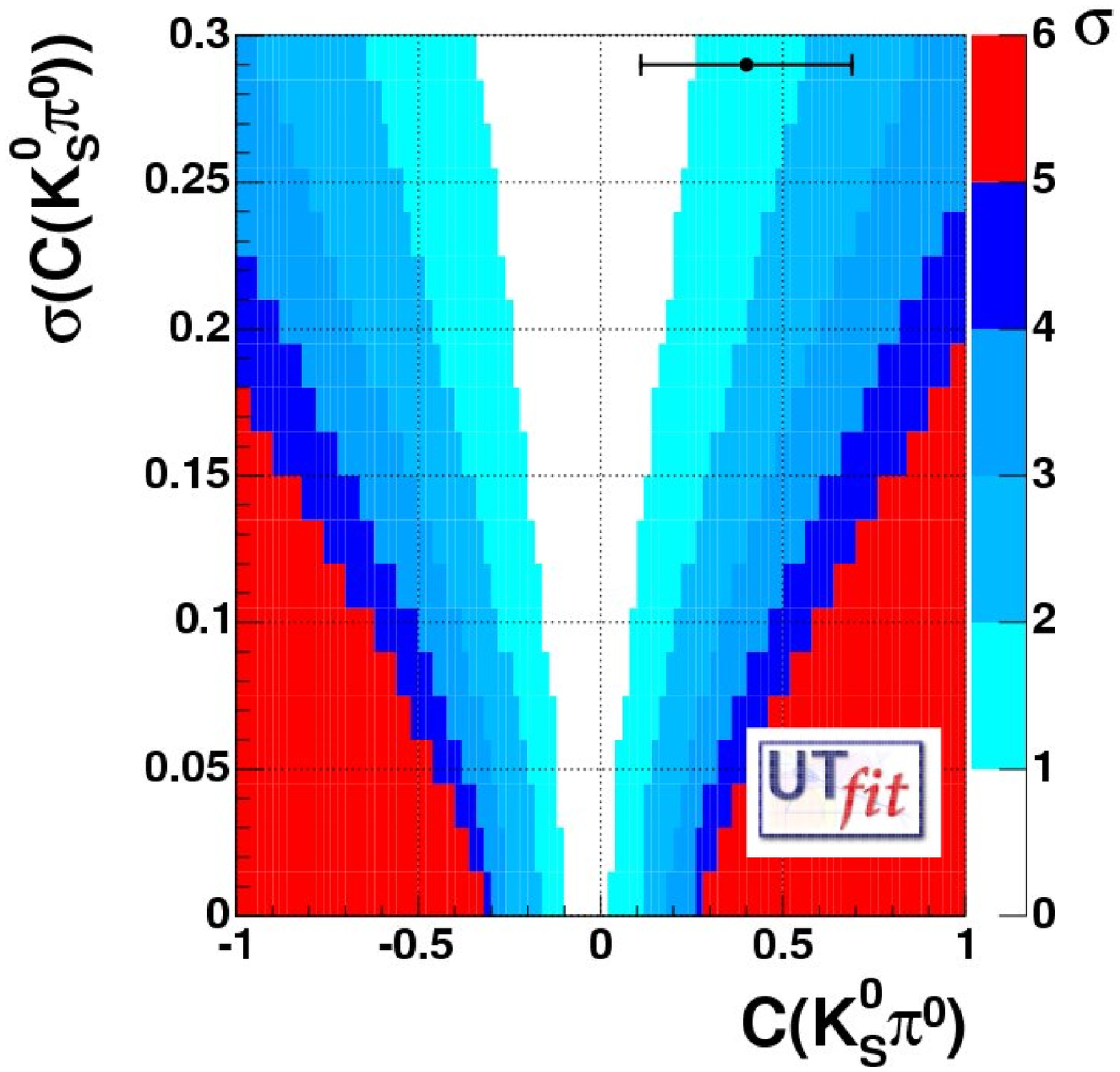}\\
\end{center}
\caption{ {\it Compatibility plot for S (top) and C (bottom) 
for $\phi K^0_S$ (left) and $K^0_S \pi^0$ (right), using UTfit result and
Charming Penguins model. Experimental measurements are superimposed.}}
\label{fig:pull_phik}
\end{figure}

In a na\"{\i}ve approach, one expects S $\sim \sin 2\beta$ and C
$\sim$ 0, but the theoretical uncertainties related to hadronic
physics can change this expectation.  Starting from the value of
sin2$\beta$ of the standard analysis, we used the Charming Penguins
model~\cite{strike} to take into account these hadronic uncertainties
and quantify the sensitivity of future measurements with the
compatibility plots shown in Figure~\ref{fig:pull_phik}.  It has to be
noted that, including these hadronic uncertainties, the theoretical
predictions of $S$ and $C$ in Table~\ref{tab:phiks} have a typical
uncertainty of $\sim 0.09$~\cite{morion04}.

\subsection{Pull distribution for $\dms$}
\begin{figure}[htbp!]
\hbox to\hsize{\hss
\includegraphics[width=0.5\hsize]{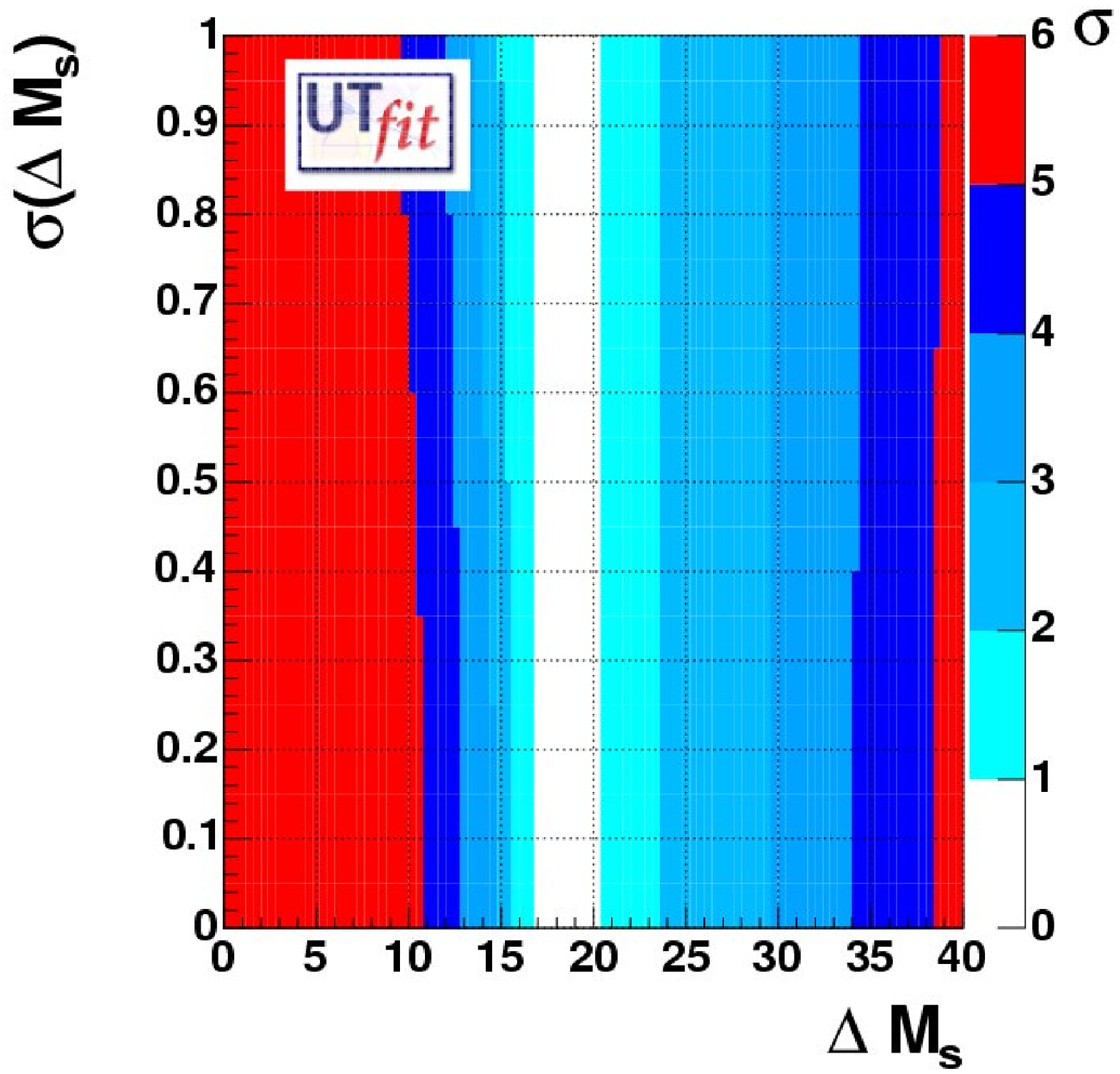}
\includegraphics[width=0.5\hsize]{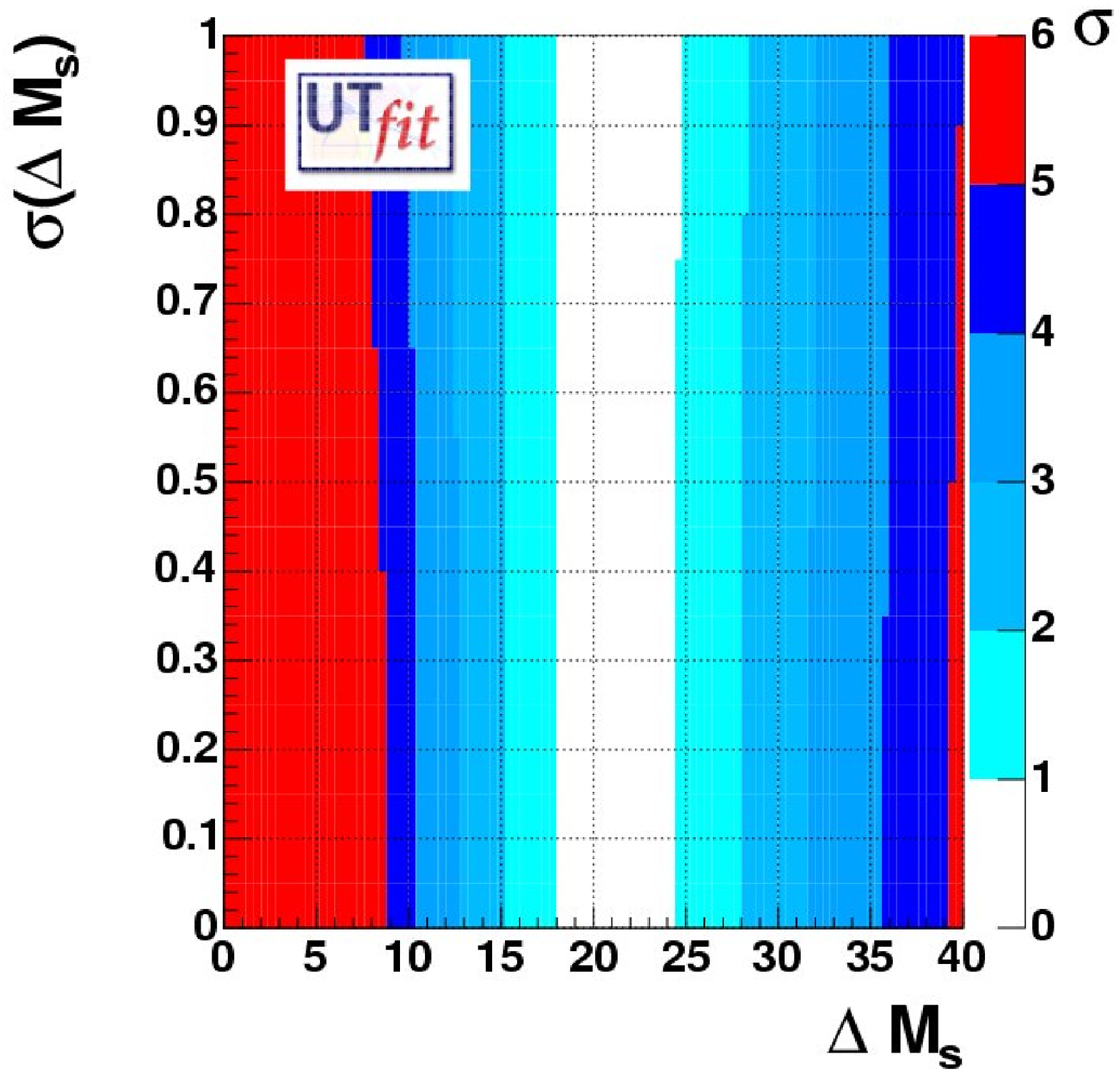}
\hss}
\caption{\it {The compatibility of the direct and indirect determination of
    $\dms$, as a function of the value of $\dms$, using (left) or
    ignoring (right) the present experimental bound.}}
\label{fig:pull_dms}
\end{figure}

The plot in Figure~\ref{fig:pull_dms} shows the compatibility of the
indirect determination of $\dms$ with a future determination of the
same quantity, obtained using or ignoring the experimental information
coming from the present bound.

>From the plot in Figure~\ref{fig:pull_dms} it can be concluded that,
once a measurement of $\dms$ with the expected accuracy of $\sim 1$
ps$^{-1}$ is available, a value of $\dms$ greater than $32$ ps$^{-1}$
would imply New Physics at $3~\sigma$ level.

\subsection{Pull distribution for the angle $\gamma$}

\begin{figure}[htbp!]
\hbox to\hsize{\hss
\includegraphics[width=0.5\hsize]{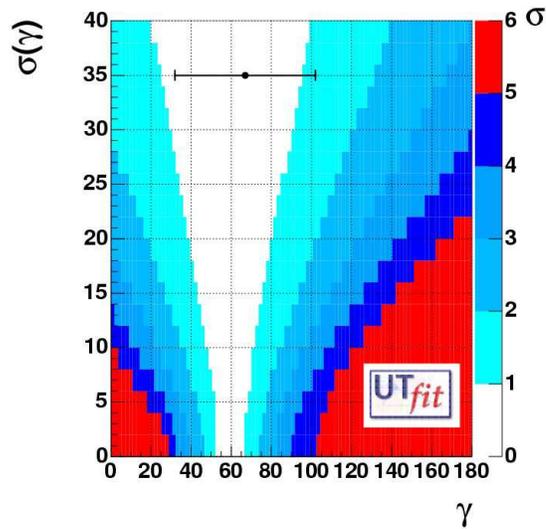}
\hss}
\caption{\it {The compatibility of the direct and indirect determination of
    $\gamma$, as a function of the value and the error of $\gamma$,
    using the UT fit results.}}
\label{fig:pull_gamma}
\end{figure}

The plot in Figure~\ref{fig:pull_gamma} shows the compatibility of the
indirect determination of $\gamma$ with a future determination of the
same angle obtained from B decays. It can be noted that even in case
the angle $\gamma$ can be measured with a precision of 10$^{\circ}$
from B decays, the predicted 3$\sigma$ region is still rather large,
corresponding to the interval [25-100]$^{\circ}$.  Values beyond
100$^{\circ}$ would clearly indicate physics beyond the Standard
Model. The actual determination of the angle $\gamma$ is not yet
precise enough to test the validity of the Standard Model as shown by
the point with the error bar in Figure~\ref{fig:pull_gamma} and given
in equation (\ref{eq:gamma}).  Nevertheless, a direct determination of
$\gamma$ is of crucial importance to test NP models~\cite{CKMNP}.

\section{Conclusions}

Flavour physics in the quark sector has entered its mature age.  Today
the Unitarity Triangle parameters are known with good precision.  A
crucial test has been already done: the comparison between the
Unitarity Triangle parameters, as determined with quantities sensitive
to the sides of the triangle (semileptonic B decays and oscillations),
and the measurements of CP violation in the kaon ($\epsilon_K$) and in
the B (sin2$\beta$) sectors. The agreement is ``unfortunately''
excellent. The Standard Model is ``Standardissimo'': it is also
working in the flavour sector. This agreement is also due to the
impressive improvements achieved on OPE, HQET and LQCD theories which
have been used to extract the CKM parameters.

Many B decay Branching Fractions and relative CP asymmetries have been
measured at B-Factories.  The outstanding result is the determination
of sin 2$\beta$ from B hadronic decays into charmonium-$K^0$ final
states. On the other hand many other exclusive hadronic rare B decays
have been measured and constitute a gold mine for weak and hadronic
physics, allowing in principle to extract different combinations of
the Unitarity Triangle angles.

Besides presenting an update of the standard UT analysis, we have
shown in this paper that new measurements at B-Factories start to have
an impact on the overall picture of the Unitarity Triangle
determination. In the following years they will provide further tests
of the Standard Model in the flavour sector to an accuracy up to the
per cent level.

Finally, introducing the compatibility plots, we have studied the impact
of future measurements on testing the SM and looking for New Physics.

\section{Acknowledgements}

We would like to warmly thank people which provide us the experimental
and theoretical inputs which are an essential part of this work and
helped us with useful suggestions for the correct use of the
experimental information. We thank: A.~Bevan, T.~Browder,
C.~Campagnari, G.~Cavoto, M.~Danielson, R.~Faccini, F.~Ferroni,
M.~Legendre, O.~Long, F.~Martinez, L.~Roos, A.~Poulenkov, M.~Rama,
Y.~Sakai, M.-H.~Schune, W.~Verkerke, M.~Zito.  We also thank
P.~Gambino and A.~Soni for useful discussions.  Finally, we thank
M.~Baldessari, C.~Bulfon, and all the BaBar Rome group for help in the
realization and for hosting the web site.


\begin{thebibliography}{99}

  
\bibitem{ref:pageweb} 
{\bf UT}{\it fit} Collaboration : M. Bona, M.
  Ciuchini, G. D'Agostini, E. Franco, V. Lubicz, G. Martinelli,
  F.Parodi, M. Pierini, P. Roudeau, C.
  Schiavi, L. Silvestrini, A. Stocchi. \\
  {\tt http://www.utfit.org}
  
\bibitem{ref:noi} M. Ciuchini, E. Franco, V. Lubicz, F. Parodi, L.
  Silvestrini and A. Stocchi, ``Unitarity Triangle in the Standard
  Model and sensitivity
  to new physics'' (hep-ph/0307195);\\
  M. Ciuchini, G. D'Agostini, E. Franco, V. Lubicz, G.  Martinelli, F.
  Parodi, P. Roudeau, A. Stocchi, ``2000 CKM-Triangle Analysis A
  Critical Review with Updated Experimental Inputs and Theoretical
  Parameters'', {\bf JHEP 0107} (2001) 013. (hep-ph/0012308); A. J.
  Buras, F. Parodi, A. Stocchi, ``The CKM Matrix and The Unitarity
  Triangle: another Look'' {\bf JHEP 0301} (2003) 029
  (hep-ph/0207101).

\bibitem{ref:loro}
M.~Lusignoli, L.~Maiani, G.~Martinelli and L.~Reina,
Nucl.\ Phys.\ B {\bf 369}, 139 (1992);
A.~Ali and D.~London,
%
arXiv:hep-ph/9405283;
arXiv:hep-ph/9409399;
Z.\ Phys.\ C {\bf 65}, 431 (1995);
S.~Herrlich and U.~Nierste,
%
Phys.\ Rev.\ D {\bf 52}, 6505 (1995);
M.~Ciuchini, E.~Franco, G.~Martinelli, L.~Reina and L.~Silvestrini,
%
Z.\ Phys.\ C {\bf 68}, 239 (1995);
A.~Ali and D.~London,
Nuovo Cim.\  {\bf 109A}, 957 (1996);
A.~Ali,
Acta Phys.\ Polon.\ B {\bf 27}, 3529 (1996);
A.~J.~Buras,
arXiv:hep-ph/9711217;
A.~J.~Buras and R.~Fleischer,
Adv.\ Ser.\ Direct.\ High Energy Phys.\  {\bf 15}, 65 (1998);
R.~Barbieri, L.~J.~Hall, S.~Raby and A.~Romanino,
Nucl.\ Phys.\ B {\bf 493}, 3 (1997);
A.~Ali and B.~Kayser,
arXiv:hep-ph/9806230;
P.~Paganini, F.~Parodi, P.~Roudeau and A.~Stocchi,
Phys.\ Scripta {\bf 58}, 556 (1998);
F.~Parodi, P.~Roudeau and A.~Stocchi,
Nuovo Cim.\ A {\bf 112}, 833 (1999);
F.~Caravaglios, F.~Parodi, P.~Roudeau and A.~Stocchi,
arXiv:hep-ph/0002171;
S.~Mele,
Phys.\ Rev.\ D {\bf 59}, 113011 (1999);
A.~Ali and D.~London,
Eur.\ Phys.\ J.\ C {\bf 9}, 687 (1999);
M.~Ciuchini, E.~Franco, L.~Giusti, V.~Lubicz and G.~Martinelli,
Nucl.\ Phys.\ B {\bf 573}, 201 (2000);
M.~Bargiotti {\it et al.}, {\it La Rivista del Nuovo Cimento} {\bf
    Vol.  23N3} (2000) 1;
S.~Plaszczynski and M.~H.~Schune,
arXiv:hep-ph/9911280;
S.~Mele,
in {\it Proc. of the 5th International Symposium on Radiative Corrections (RADCOR 2000) } ed. Howard E. Haber,
arXiv:hep-ph/0103040;
A.~Hocker, H.~Lacker, S.~Laplace and F.~Le Diberder,
Eur.\ Phys.\ J.\ C {\bf 21} (2001) 225;
M.~Ciuchini,
Nucl.\ Phys.\ Proc.\ Suppl.\  {\bf 109B} (2002) 307;
A.~Hocker, H.~Lacker, S.~Laplace and F.~Le Diberder,
AIP Conf.\ Proc.\  {\bf 618} (2002) 27;
F.~Caravaglios, P.~Roudeau and A.~Stocchi,
Nucl.\ Phys.\ B {\bf 633} (2002) 193;
A.~J.~Buras,
arXiv:hep-ph/0210291;
G.~P.~Dubois-Felsmann, D.~G.~Hitlin, F.~C.~Porter and G.~Eigen,
arXiv:hep-ph/0308262;
arXiv:hep-ex/0312062;
A.~Stocchi,
arXiv:hep-ph/0405038;
J.~Charles {\it et al.}  [CKMfitter Group Collaboration],
arXiv:hep-ph/0406184.

\bibitem{ref:hfag} HFAG: {\tt http://www.slac.stanford.edu/xorg/hfag/}

\bibitem{ref:ckm1} ``THE CKM MATRIX AND THE UNITARITY TRIANGLE''.
  {\it Based on the First Workshop on ``CKM Unitarity Triangle'' held
    at CERN, 13-16 February 2002}.  Edited by: M. Battaglia, A. Buras,
  P. Gambino and A.  Stocchi. To be published as {\bf CERN Yellow
    Book} (hep-ph/0304132).

\bibitem{ref:ckm2} Proceedings of the Second Workshop on ``CKM
  Unitarity Triangle'' {\it held in Durham, 5-9 April 2003.}  Edited
  by: P. Ball, P. Kluit, J. Flynn and A. Stocchi.

\bibitem{ref:allrhoeta} C. Dib, I. Dunietz, F. Gilman, Y. Nir, {\it
    Phys. Rev.} {\bf D41}
  (1990) 1522;
  G. Buchalla, A.J. Buras and M.E. Lautenbacher, {\it Rev. Mod. Phys.}
  {\bf 68} (1996)  1125;
  A. Ali and D. London, in Proceeding of ``ECFA Workshop on the
  Physics of a
  $B$ Meson Factory'', Ed. R. Aleksan, A. Ali (1993);
  A. Ali and D. London, {\it Nucl. Phys.} {\bf 54A} (1997) 297.

\bibitem{ref:DKmethods} M. Gronau and D. London, {\it Phys. Lett.}
  {\bf B253}, 483 (1991);~ M. Gronau and D. Wyler, {\it Phys. Lett.}
  {\bf B265}, 172 (1991);~ I. Dunietz, {\it Phys. Lett.} {\bf B270},
  75 (1991);~ I. Dunietz, {\it Z. Phys.} {\bf C56}, 129 (1992);~ D.
  Atwood, G. Eilam, M. Gronau and A. Soni, {\it Phys.  Lett.} {\bf
    B341}, 372 (1995);~ D. Atwood, I. Dunietz and A. Soni, {\it Phys.
    Rev.  Lett.} {\bf 78}, 3257 (1997).

\bibitem{ref:dalitz} A. Giri, Yu. Grossman, A. Soffer and J. Zupan,
  {\it Phys. Rev.} {\bf D68}, 054018 (2003).

\bibitem{ref:belle} A.~Poluektov {\it et al.}  [Belle Collaboration],
arXiv:hep-ex/0406067.

\bibitem{gronaulondon} M.~Gronau and D.~London,
Phys.\ Rev.\ Lett.\  {\bf 65} (1990) 3381.

\bibitem{ref:owen} O.~Long, M.~Baak, R.~N.~Cahn and D.~Kirkby,
Phys.\ Rev.\ D {\bf 68} (2003) 034010
[arXiv:hep-ex/0303030].
  
\bibitem{dpi-tot} B.~Aubert {\it et al.}  [BABAR Collaboration],
arXiv:hep-ex/0309017.
K.~Abe {\it et al.}  [BELLE Collaboration],
arXiv:hep-ex/0308048.

\bibitem{dpi-par} B.~Aubert {\it et al.}  [BABAR Collaboration],
Phys.\ Rev.\ Lett.\  {\bf 92} (2004) 251802
[arXiv:hep-ex/0310037].

\bibitem{thphi} Y.~Grossman and M.~P.~Worah,
Phys.\ Lett.\ B {\bf 395} (1997) 241
[arXiv:hep-ph/9612269];
M.~Ciuchini, E.~Franco, G.~Martinelli, A.~Masiero and L.~Silvestrini,
Phys.\ Rev.\ Lett.\  {\bf 79} (1997) 978
[arXiv:hep-ph/9704274];
Y.~Grossman, G.~Isidori and M.~P.~Worah,
Phys.\ Rev.\ D {\bf 58} (1998) 057504
[arXiv:hep-ph/9708305].

\bibitem{ref:phik01} B.~Aubert {\it et al.}  [BABAR Collaboration],
arXiv:hep-ex/0403026.
K.~Abe {\it et al.}  [Belle Collaboration],
Phys.\ Rev.\ Lett.\  {\bf 91} (2003) 261602
[arXiv:hep-ex/0308035].
B.~Aubert {\it et al.}  [BABAR Collaboration],
arXiv:hep-ex/0403001.

\bibitem{strike}
M.~Ciuchini, E.~Franco, G.~Martinelli, M.~Pierini and L.~Silvestrini,
Phys.\ Lett.\ B {\bf 515} (2001) 33 [arXiv:hep-ph/0104126].

\bibitem{morion04} M.~Ciuchini, E.~Franco, G.~Martinelli, A.~Masiero,
  M.~Pierini and L.~Silvestrini,
arXiv:hep-ph/0407073.

\bibitem{CKMNP} M.~Ciuchini, E.~Franco, F.~Parodi, V.~Lubicz,
  L.~Silvestrini and A.~Stocchi,
eConf {\bf C0304052} (2003) WG306
[arXiv:hep-ph/0307195].

\end{thebibliography}
\end{document}